\documentstyle[12pt]{article}
\begin{document}
\baselineskip6mm
\def\be{\begin{equation}\label}\def\ee{\end{equation}}
\def\ban{\begin{eqnarray}\label}{}
\def\ean{\end{eqnarray}}\def\nn{\nonumber}
\def\bea{\begin{eqnarray*}}\def\eea{\end{eqnarray*}}
\def\bp{\begin{picture}}\def\ep{\end{picture}}
\def\ba{\begin{array}}\def\ea{\end{array}}
\def\s{\scriptstyle}
\def\la{\langle\,}\def\ra{\,\rangle}\def\mi{\,|\,}
\def\O{{\cal O}}
\def\F{F^{\O}}
\def\N{N^{\O}}
\def\f{f^{\O}}
\def\a{\alpha}
\def\b{\beta}
\def\t{\theta}
\def\ua{{\underline\a}}
\def\ut{{\underline\t}}
\def\u{{\underline u}}
\def\Res{\mathop{\rm Res}}
\def\ds{\displaystyle}
\def\S{\dot S}
\def\da{\dot a}
\def\db{\dot b}
\def\h#1{\mbox{$\frac#12$}}

\newtheorem{theo}{Theorem}[section]
\newtheorem{defi}[theo]{Definition}
\newtheorem{lemm}[theo]{Lemma}
\renewcommand{\theequation}{\mbox{\arabic{section}.\arabic{equation}}}

\title{Exact Form Factors in Integrable Quantum Field Theories:
the Sine-Gordon Model}

\author{\\
H. Babujian$^{1,2}$,
A. Fring$^3$,
M. Karowski$^4$
and A. Zapletal$^{5}$
\\{\small\it Institut f\"ur Theoretische Physik}\\
{\small\it Freie Universit\"at Berlin, Arnimallee 14, 14195 Berlin, Germany} 
}
\date{\small\it In memory of Harry Lehmann}

\maketitle

\footnotetext[1]{Permanent address: Yerevan Physics Institute,
Alikhanian Brothers 2, Yerevan, 375036 Armenia.}
\footnotetext[2]{e-mail: babujian@lx2.yerphi.am, babujian@physik.fu-berlin.de}
\footnotetext[3]{e-mail: fring@physik.fu-berlin.de}
\footnotetext[4]{e-mail: karowski@physik.fu-berlin.de}
\footnotetext[5]{e-mail: zapletal@physik.fu-berlin.de}

\begin{abstract}
We provide detailed arguments on how to derive properties of generalized form
factors,
originally proposed by one of the authors (M.K.) and Weisz twenty years ago,
solely based on the assumption of ``maximal analyticity" and the validity of 
the LSZ reduction formalism. These properties constitute consistency 
equations which allow the explicit evaluation of the n-particle form
factors once the scattering matrix is known. 
The equations give rise to a matrix Riemann-Hilbert problem. 
Exploiting the ``off-shell" Bethe ansatz we
propose a general formula for form factors for an odd number of particles.
For the Sine-Gordon model alias the massive Thirring model 
we exemplify the general solution for several operators.
In particular we calculate the three particle form factor of the
soliton field, carry out a consistency check against the Thirring model
perturbation theory and thus confirm the general formalism.
\end{abstract}

\newpage

\section{Introduction}
More than fifty years ago, Heisenberg \cite{Heisen} pointed out the
importance of studying analytic continuations of scattering amplitudes into
the complex momentum plane. The first concrete investigations in this
direction were carried out by Jost \cite{Jost} and Bargmann \cite{Bargmann},
initially for non-relativistic scattering processes. The original ideas
turned out to be very fruitful and lead to interesting results on-shell,
i.e. for the S-matrix \cite{ELOP}, as well as off-shell, that is for the
two-particle form factors, see for instance \cite{Barton}.

Once one restricts ones attention to 1+1 dimensional integrable theories,
the n-particle scattering matrix factories into two particle S-matrices and
the approach, now usually referred to as the bootstrap program, reveals its
full strength. On-shell, it leads to the exact determination of the
scattering matrix \cite{KTTW,KT}, (for reviews see also  
[9-12]). The results obtained in this way agree with the S-matrix
obtained from the extrapolation of semi-classical expressions for the
Sine-Gordon model \cite{ZamS}.
The first off-shell
considerations were carried out about two decades ago by one of the 
authors (M.K.) et al. 
\cite{KW,BKW}, who introduced the concept of a generalized form factor and
formulated several consistency equations which are expected to be satisfied
by these objects. Thereafter this approach was mainly developed further and
studied in the context of several explicit models by Smirnov et al. [15-23]. 
Recently this program has seen some revival in relation to models which
arise as perturbations of certain conformal field theories \cite{BPZ}, 
particularly in the context of affine Toda theories \cite{Toda}
and closely related models [26-48].

An entirely different method, the  Bethe ansatz \cite{Bethe}, 
was initially formulated in order to solve 
the eigenvalue problem for certain integrable Hamiltonians.
The approach has found  applications in the context of numerous
models and  has led to a detailed study of various mass spectra 
and S-matrices (for reviews and an extensive list of references
see for instance \cite{Betherev}).
The original techniques have been refined into several directions,
of which in particular the so-called
``off-shell'' Bethe ansatz, which was originally formulated by 
one of the authors (H.B.) \cite{Hrachik1,Hrachik2}, will be exploited
for our purposes. This version of the Bethe ansatz paves the way
to extend the approach to the off-shell physics and opens up the 
intriguing  possibility to merge the two methods, that is the form factor
approach and the Bethe ansatz. The basis for this opportunity lies
in the observation  \cite{Resh,Pl,BKZ}, that the ``off-shell'' Bethe ansatz
captures the vectorial structure  of Watson's equations (see section \ref{2.2}
properties (i) and (ii)).  
These are matrix difference equations giving rise to a matrix
Riemann-Hilbert problem which is solved by an ''off-shell" Bethe ansatz.
Furthermore, there exist interesting speculations in order to
make contact with general concepts of algebraic quantum field theory \cite
{Niedermaier,Schroer}.

Conceptionally, the on and off-shell approaches are very similar. For
the on-shell situation one has certain constraints resulting from general
physical and in particular analytic properties (referred to as
``maximal analyticity assumption''), which lead to a set of
conditions which turned out to be  so restrictive that they allow to
construct the exact scattering matrix almost uniquely. This approach  is
adopted in order to determine the key off-shell quantities, i.e.~the form
factors. In the present manuscript we provide a detailed derivation of
the consistency equations solely based on the maximal analyticity 
assumption and the validity of the LSZ-formalism \cite{LSZ} 
(see also \cite{IZ}).
Form factors are vector valued functions, representing matrix
elements of some local operator ${\mathcal{O}}(x)$ at the origin between
an in-state and the vacuum, which we denote by (refer  equation (\ref{3.2})
for more details)
\begin{equation}
F_{\underline{\alpha }}^{\mathcal{O}}\left( ((p_i +p_j)^2   
+i\varepsilon )_{(1\leq
i<j\leq n)}\right) :=\left\langle 0\left| {\mathcal{O}}(0)
\right| p_{1},\ldots
,p_{n}\right\rangle _{\alpha _{1}\ldots \alpha _{n}}^{in}\,.
\label{1.1}
\end{equation}
Once all the $n$-particle form factors are known, one is in principle in a
position to compute all correlation functions. In particular the two point
function for an hermitian operator $\mathcal{O}$ in real Euclidian space
reads 
\begin{equation}
\langle {\mathcal{O}}(x)\,{\mathcal{O}}(0)\rangle\,=\,\sum_{n=0}^{\infty }
\int\frac{d\theta _{1}\ldots d\theta _{n}}{n!\,(4\pi )^{n}}\mid 
F_{\underline{\alpha }}^{\mathcal{O}}(\theta _{1}, \ldots , 
\theta _{n})\mid ^{2}\exp \left(
-r\sum_{i=1}^{n}  m_i  \cosh \theta _{i}\right) .  \label{1.2}
\end{equation}
Here  $r$  denotes the radial 
distance $r=\sqrt{x_{1}^{2}+x_{2}^{2}}$ and
$\theta$ is the rapidity related to the momentum via $p_i= m_i \sinh 
\theta_i$
(see section 3.2 for more details). The
explicit evaluation of all integrals and sums remains an open challenge for
almost all theories, except the Ising model\footnote{%
Of course one may also adopt a very practical point of view and resort to
the well-known fact that the series expansion of correlation functions in
terms of form factors (\ref{1.2}) converges very rapidly.
Consequently correlations functions may be approximated very often quite
well by simply including the two-particle form factor into the expansion. 
>From that point of
view the form factor program is completed, since the calculation of the
two-particle form factors is well understood.}. 
Important progress towards a
solution of this problem has  recently been achieved in \cite{Korepin}.

A commonly
used procedure which will yield expressions which satisfy all of the
consistency requirements is constituted out  of 
the following steps: First of all one 
has to have solved the on-shell system, that is one requires expressions for
the S-matrix. In the next step one usually makes an ansatz for the form 
factors of a type already
introduced in \cite{KW}, in which one extracts explicitly the expected
 singularity
structure. The nature of the ansatz guarantees by
construction that the generalized Watson's equations (properties (i) and (ii)) 
are satisfied once the
scattering matrix is diagonal. For generically non-diagonal scattering
matrices one may invoke also the techniques of the  ``off-shell''
 Bethe ansatz 
\cite{Resh,Pl,BKZ}  in order to capture the vectorial structure of the form
factors. The ansatz only involves the 
rapidity differences, apart from a possible
pre-factor, which takes the spin of the local field ${\mathcal{O}}$
into account, and has therefore the desired behavior under Lorentz 
transformations (refer property (v) in section 3.2).
General solutions for the so-called minimal form factors (the function which
satisfies the functional equation (4.10)) are always fairly easy to
find. Once the scattering matrix is non-diagonal one has also to 
encode the vectorial structure at this stage.
What is then left, is to determine  a general function which takes
the complete singularity structure into account. For this purpose one
may now invoke  properties (iii) and (iv) (equations  (3.12) and (3.13)
for the bosonic case), which lead to a set
of recursive equations. In principle these equations may now be solved step
by step, once the first non-vanishing form factor for a particular operator
is properly fixed. However, only after a few steps the expressions become
usually algebraically very complex and reveal very little insight.
Therefore, it is highly desirable to search for structures of a more general
nature, that is in particular to seek for closed expressions for all
n-particle form factors. Only such expressions may ultimately shed more
light on the analytic expressions for the correlation functions
(\ref{1.2}). Alternatively, one may try to construct directly 
a representation for the creation operators of the particles 
in the in-state in (\ref{1.1}) \cite{MJ,Na,TV,NPT}. Representing the local
operator $\mathcal{O}$ in the same space, one may in principle also compute the
form factors.

In the present manuscript we provide a general  expression (see theorem 4.1)
of a different kind, which solves all the consistency requirements. 
It is  very generic by
construction and, roughly speaking, captures the vectorial nature 
of the form factors by means
of ``off-shell'' Bethe ansatz states and the pole structure 
by particular contour integrals. We exemplify this general
expression for the form factors of the Sine-Gordon model involving an odd
number of states, which was hitherto unknown. For the even case similar
expressions may be found in \cite{nankai,Smirnov3}.
We present a detailed analysis 
of the three particle form factor.

Once solutions for the set of consistency equations are found, it is highly 
desirable to verify the solutions with some alternative method. 
Several different methods  have been
developed in recent years. Assuming that the theory under consideration
results from the perturbation of some conformal field theory, one may carry
out the following consistency checks. For instance one
may  take the operator in the form factor to be
the trace of the energy momentum tensor and  exploit the so-called 
c-theorem \cite{Zamc} in order to obtain a first indication about the result.
This check is not extremely restrictive what the higher n-particle form
factors concerns, since the expected value for c is usually already saturated
after the two-particle contribution. Alternatively one may also compare
with the perturbation theory around the conformal field theory, which is
possible for all operators of the model. The latter approach has turned
out to be very fruitful \cite{YLZam}. A further consistency check 
consist out of the comparison between the exact result obtained
from the form factors with the predictions of the renormalisation
group (that is asymptotic freedom etc.~\cite{Sash,BA}).
In the present manuscript we present a check
of our solutions against conventional perturbation 
theory in standard quantum field theory.

The manuscript is organized as follows: In section 2 we review the 
properties of the general scattering matrix and in particular the
Sine-Gordon S-matrix. In section 3 we motivate the general properties of
the generalized form factors, for simplicity initially only for the bosonic
case, which we thereafter extend to the general situation involving also
fermions. In section 4  we briefly explain the ``off-shell" Bethe ansatz
and state theorem 4.1, the main result of the manuscript. We present a general
 formula\footnote{Our formula is similar to an analogous one
of Smirnov \cite{Smirnov3} for even number of particles. This should be a
starting point for a comparison of both formulae.} (based on the ''off-shell"
Bethe ansatz) for form factors with an odd number of solitons or anti-solitons.
Furthermore, we 
provide an explicit analysis of several two- and three particle form
factors  and carry out various consistency checks relating different
form factors to each other. In section 5 we compare our solution for
a three particle form factor against perturbative perturbation theory.
Our conclusions are stated in section 6.
In appendix A we provide the proofs of the properties of the
generalized form factors. In appendix B we proof
theorem 4.1 and appendix C serves  as a depot for several 
useful formulae employed in the
working.

\setcounter{equation}{0}
\section{The $S$-matrix}\label{s2}
\subsection{General Properties}\label{s2.1}
In this section we briefly review some of the well known facts on the
general properties of the scattering matrices.
The  Fock space
is spanned by the in- or out-states of the particles 
\be{2.1}
\mi p_1,\dots,p_n\ra^{in/out}_{\a_1\dots\a_n}
=a^{in/out\,\dagger}_{\a_1}(p_1)\cdots a^{in/out\,\dagger}_{\a_n}(p_n)
\mi0\ra 
\ee
where the $a^\dagger$'s are creation operators.
The $p$'s denote the momenta and the $\a$'s the internal quantum
numbers of the particles, such as the particle type etc.
We choose the normalization
\be{2.2}
_{\a'}\la p'\mi p\ra_\a=\delta_{\a'\a}\,
2\omega\,2\pi\,\delta(p'-p)=\delta_{\a'\a}\,4\pi\,\delta(\t'-\t)
\ee
where the rapidity  $\theta$ is related to the momentum by $p=m\sinh\t$ and
$\omega=\sqrt{m^2+p^2}$.

In an integrable quantum 
field theory in 1+1-dimensions there exists an
infinite set of conservation laws. Therefore in a scattering process 
the sets of incoming and outgoing momenta are equal
$$
\{p_1,\dots,p_n\}=\{p'_1,\dots,p'_{n'}\}\,.
$$
The n-particle S-matrix is defined by
$$
\ba{rcl}
\mi p_1,\dots,p_n\ra^{in}_{\a_1\dots\a_n}
&=&\mi p_1,\dots,p_n\ra^{out}_{\a'_1\dots\a'_n}
{S^{(n)}}^{\a'_n\dots\a'_1}_{\a_1\dots\a_n}(p_1,\dots,p_n)\\[3mm]
&=&\mi p_n,\dots,p_1\ra^{out}_{\a'_n\dots\a'_1}
(\sigma^{(n)}{S^{(n)})}^{\a'_n\dots\a'_1}_{\a_1\dots\a_n}(p_1,\dots,p_n).
\ea
$$
The statistics of the particles has been taken into account by the diagonal
matrix $\sigma^{(n)}$. It is a product of all two particle matrices $\sigma$
with entries $-1$ if both particles are fermions and $+1$ otherwise
(see \cite{K1}). As a consequence of integrability , i.e.~the existence
of an infinite number of conserved quantities,
the n-particle S-matrix factorizes into $n(n-1)/2$ two-particle ones
$$
\sigma^{(n)} S^{(n)}(p_1,\dots,p_n)=\prod_{i<j}\sigma S^{(2)}(p_i,p_j)\,,
$$
where the product on the right hand side has to be taken in a specific order
(see e.g.~\cite{KT}).
For this reason it is sufficient to investigate
the properties of the two-particle scattering matrix.
As is usual in integrable quantum field theories in 1+1-dimensions it is most
convenient to regard the two-particle
S-matrix as a function of the rapidity differences $\t=|\t_i-\t_j|$
rather than as a function of the Mandelstam variables $s_{ij}=(p_i+p_j)^2$.
In order to establish the analytic properties of the two-particle S-matrix
one may employ
the relations $s_{ij} = m_i^2 + m_j^2 + 2 m_i m_j \cosh \t_{ij}$,
$t_{ij} = (p_i-p_j)^2=2 m_i^2 + 2 m_j^2 - s_{ij}$.
Considering the scattering matrix as
a function in the complex $s_{ij}$-plane, there will be two branch
cuts present, the s-channel one for $s_{ij} > (m_i+m_j)^2 $ and the
t-channel one for $s_{ij} < (m_i-m_j)^2 $.
\begin{figure}[hbt]
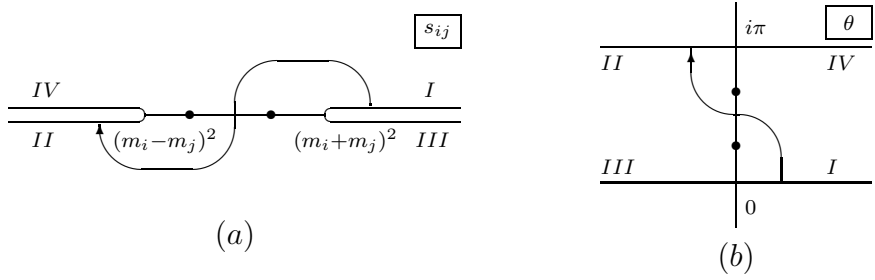

$$
\ba{c}
\unitlength6mm
\bp(10,4)
\put(3,2){\line(1,0){4}}
\put(10,2){\oval(6,.3)[l]}
\put(0,2){\oval(6,.3)[r]}
\put(2.3,1.3){$\s (m_i-m_j)^2$}
\put(6.3,1.3){$\s (m_i+m_j)^2$}
\put(.5,2.4){$\s IV$}
\put(.5,1.3){$\s II$}
\put(9.2,2.4){$\s I$}
\put(9,1.3){$\s III$}
\put(4,2){\makebox(0,0){$\s\bullet$}}
\put(5.8,2){\makebox(0,0){$\s\bullet$}}
\put(6.5,2.2){\oval(3,2)[t]}
\put(3.5,1.8){\oval(3,2)[b]}
\put(5,2.3){\line(0,-1){.6}}
\put(2.01,1.6){\vector(0,1){.2}}
\put(9,3.5){\framebox(1,.7){$\s s_{ij}$}}
\ep
\\(a)
\ea~~~~~~~~~~~
\ba{c}
\unitlength6mm
\bp(6,5)
\put(0,1){\line(1,0){6}}
\put(0,4){\line(1,0){6}}
\put(3,0){\line(0,1){5}}
\put(3,1){\oval(2,3)[rt]}
\put(3,4){\oval(2,3)[lb]}
\put(2,3.6){\vector(0,1){.3}}
\put(3.2,.3){$\s0$}
\put(3.2,4.3){$\s i\pi$}
\put(5,1.2){$\s I$}
\put(0,1.2){$\s III$}
\put(5,3.5){$\s IV$}
\put(0,3.5){$\s II$}
\put(3,1.8){\makebox(0,0){$\s\bullet$}}
\put(3,3){\makebox(0,0){$\s\bullet$}}
\put(5,4.2){\framebox(1,.7){$\s\t$}}
\ep
\\(b)
\ea~~~~~~~~
$$
\caption[The analyt..]{\label{f31}\it
The analyticity domains in the complex planes of $(a)$ the Mandelstam
variable $s_{ij}=(p_i+p_j)^2$ and $(b)$ the rapidity difference variable
$\t=|\t_1-\t_2|$. The physical regimes in the s- and t-channels are
denoted by I
and II, respectively. The crossing transition from the s- to the t-channel is
indicated by the arrow. As explained in the main text, the interchange of
{\em in} and {\em out} means transition from I to III for the s-channel,
and II to IV for the t-channel. The dots denote the possible
positions of poles corresponding to one particle intermediate states.}
\end{figure}
In figure 1 the physical s-channel and t-channel regions in the complex
$(a)$ $s$-- and $(b)$ $\theta$--plains are labeled by I and II, respectively.
The crossing transition is depicted by an arrow. This and the transitions
corresponding to the exchange of in-- and out--going waves are given by:
$$
\begin{tabular}{rll}
$I \leftrightarrow II:$&$\qquad s_{ij}+i\epsilon\leftrightarrow
t_{ij}-i\epsilon$&$\qquad\Leftrightarrow\qquad\t
\leftrightarrow i\pi-\t$\\
$ I \leftrightarrow III:$&$\qquad s_{ij}+i\epsilon\leftrightarrow 
s_{ij}-i\epsilon$&$\qquad\Leftrightarrow\qquad\t
\leftrightarrow-\t$\\
$II\leftrightarrow VI:$&$\qquad t_{ij}-i\epsilon\leftrightarrow 
t_{ij}+i\epsilon $&$\qquad\Leftrightarrow\qquad i\pi-\t
\leftrightarrow i\pi+\t$
\end{tabular}
$$
(It will be important in the
following to notice that the $t$-channel cut (II-IV) is not present
for form factors as a function defined in the complex 
$s_{ij}$-plane.)

Let $V$ be a finite dimensional vector space, whose basis vectors label
all types of particles of the model. Then one considers the 
S-Matrix as an intertwining operator acting  on the tensor product
of two of these spaces
$$
S_{12}(\t)\,:\,V_1\otimes V_2\to V_2\otimes V_1\,.
$$
The unitarity of the S-matrix reads
\be{2.3}
\sum_{\a'\b'}\Big(S^{\a''\b''}_{\b'\a'}(\t)\Big)^*\,S^{\b'\a'}_{\a\b}(\t)=
\delta_{\a''\a}\delta_{\b''\b}~~~{\rm or}~~
S_{21}(-\t)\,S_{12}(\t)=1
\ee
since by analytic continuation from positive to negative variable
one has $S^\dagger_{12}(\t)=S_{21}(-\t)$. The crossing relations are
\be{2.4}
S_{\alpha\beta}^{\delta\gamma}(\t)
=S_{\bar\delta\alpha}^{\gamma\bar\beta}(i\pi-\t)
=S_{\beta\bar\gamma}^{\bar\alpha\delta}(i\pi-\t)
\ee
where the bar refers to the anti-particles.
The Yang-Baxter equation which follows from the higher conservation laws is
\be{2.5}
(\sigma S)_{12}(|\t_{12}|)(\sigma S)_{13}(|\t_{13}|)(\sigma S)_{23}(|\t_{23}|)
=(\sigma S)_{23}(|\t_{23}|)(\sigma S)_{13}(|\t_{13}|)(\sigma S)_{12}(|\t_{12}|)
\ee
where $\t_{ij}=\t_i-\t_j$. When there are no  transitions of the sort that
two bosons change into two fermions, the signs given by the statistics cancel.

As usual we use here and in the following the notation for a vector with
components $v^{\a_1\dots\a_n}$ and a matrix with elements
$A^{\b_1\dots\b_n}_{\a_1\dots\a_n}$ acting on these vector
\be{2.6}
v^{1\dots n}\in V_{1\dots n}=V_1\otimes\cdots\otimes V_n~,~~
A_{1\dots n}:~V_{1\dots n}\to V_{1\dots n}
\ee
where all vector spaces $V_i$ are isomorphic to $V$ and whose basis vectors
label all kinds of particles. An S-matrix as $S_{ij}$ acts nontrivial
only on the factors $V_i\otimes V_j$ and in addition exchanges these factors.
If we want to express the fact that a particle
belongs to a multiplet of a specific type of particles, we also write
$v^a\in V_a,v^b\in V_b$ etc. and consider $V=\bigoplus_aV_a$ as the direct
sum of all these spaces. Usually these spaces $V_a$ are the representation
spaces of a symmetry group or quantum group of the model.

The physical S-matrix in the formulas above is given for
positive values of the rapidity parameter $\t$. For later convenience
we will also consider an auxiliary matrix $\S$ regarded  as a function
depending on the individual rapidities of both particles $\t_1,~\t_2$ or
$\t_{12}=\t_1-\t_2$
\be{2.7}
\S_{12}(\t_1,\t_2)=\S_{12}(\t_1-\t_2)=\cases{
(\sigma S)_{12}(|\t_1-\t_2|) & for $\t_1>\t_2$\cr
(S\sigma)_{21}^{-1}(|\t_1-\t_2|) & for $\t_1<\t_2$}
\ee
with $\sigma$ taking into account the statistics
of the particles.  Up to these statistics factors $\S$
is obviously the analytic extension of the physical S-matrix $S$ from positive
to negative values of $\t$, due to the unitarity (\ref{2.3}).

It appears convenient to introduce a graphical representation
for several of the amplitudes, which will allow us to develop
a more direct graphical intuition for the derivation of several
relations.
The auxiliary matrix $\S$ may be depicted as
$$
\S_{12}(\t_1,\t_2)~~=~~
\begin{array}{c}
\setlength{\unitlength}{3mm}
\begin{picture}(5,4)
\put(1,0){\line(1,1){4}}
\put(5,0){\line(-1,1){4}}
\put(0,.6){$\t_1$}
\put(4.8,.6){$\t_2$}
\end{picture}
\end{array}
$$
Here and in the following we associate  a rapidity variable
$\t_i\in\mathbf{C}$ to each space $V_i$ which is graphically represented by a
line  labeled by $\t_i$ or simply by $i$.
In terms of the components of the S-matrix we have
$$
\S_{\alpha\beta}^{\delta\gamma}(\t_1,\t_2)~~=~~
\begin{array}{c}
\setlength{\unitlength}{2.5mm}
\begin{picture}(6,6)
\put(1,1){\line(1,1){4}}
\put(5,1){\line(-1,1){4}}
\put(.3,0){$\alpha$}
\put(5.2,0){$\beta$}
\put(5,5.6){$\gamma$}
\put(.5,5.6){$\delta$}
\put(0,2.1){$\t_1$}
\put(4.8,2.1){$\t_2$}
\end{picture}~.
\end{array}
$$
In terms of the auxiliary S-matrix the Yang-Baxter equation has the general
form
$$
\S_{12}(\t_{12})\,\S_{13}(\t_{13})\,\S_{23}(\t_{23})
=\S_{23}(\t_{23})\,\S_{13}(\t_{13})\,\S_{12}(\t_{12})
$$
which grphically simply reads
\be{2.8}
\ba{c}
\unitlength6mm
\bp(9,4)
\put(0,1){\line(1,1){3}}
\put(0,3){\line(1,-1){3}}
\put(2,0){\line(0,1){4}}
\put(4.3,2){$=$}
\put(6,0){\line(1,1){3}}
\put(6,4){\line(1,-1){3}}
\put(7,0){\line(0,1){4}}
\put(.2,.5){$1$}
\put(1.3,0){$2$}
\put(3,.2){$3$}
\put(5.5,.2){$1$}
\put(7.3,0){$2$}
\put(8.4,.4){$3$}
\ep~~~.
\ea
\ee
Unitarity and crossing may be written and depicted as
$$
\S_{21}(\t_{21})\S_{12}(\t_{12})=1~:~~~~~
\ba{c}
\unitlength3mm
\bp(8,5)
\put(0,1){\line(1,1){2}}
\put(2,1){\line(-1,1){2}}
\put(0,3){\line(1,1){2}}
\put(2,3){\line(-1,1){2}}
\put(6,1){\line(0,1){4}}
\put(8,1){\line(0,1){4}}
\put(3.7,2.7){$=$}
\put(0,-.5){$1$}
\put(1.5,-.5){$2$}
\put(6,-.5){$1$}
\put(7.5,-.5){$2$}
\ep
\ea
$$
$$
\ba{rcccl}
\S_{12}(\t_1-\t_2)
&=&{\bf C}^{2\bar2}\,\S_{\bar21}(\t_2+i\pi-\t_1)\,{\bf C}_{\bar22}
&=&{\bf C}^{1\bar1}\,\S_{2\bar1}(\t_2-(\t_1-i\pi)\,{\bf C}^{\bar11}:
\\[5mm]
\ba{c}
\unitlength3mm
\bp(4,5)
\put(0,1){\line(1,1){4}}
\put(4,1){\line(-1,1){4}}
\put(0,-.5){$1$}
\put(3.7,-.5){$2$}
\ep
\ea
&=&
\ba{c}
\unitlength3mm
\bp(6,5)
\put(1,1){\line(1,1){4}}
\put(4,1){\line(-1,2){2}}
\put(1,5){\oval(2,8)[lb]}
\put(5,1){\oval(2,8)[tr]}
\put(3.5,-.5){$1$}
\put(5.7,-.5){$2$}
\ep
\ea
&=&
\ba{c}
\unitlength3mm
\bp(6,5)
\put(2,1){\line(1,2){2}}
\put(5,1){\line(-1,1){4}}
\put(1,1){\oval(2,8)[lt]}
\put(5,5){\oval(2,8)[br]}
\put(0,-.5){$1$}
\put(2,-.5){$2$}
\ep
\ea
\ea
$$
where ${\bf C}^{1\bar1}$ and ${\bf C}_{1\bar1}$ are charge conjugation
operators with components ${\bf C}^{\a\bar\b}={\bf C}_{\a\bar\b}=
\delta_{\a\b}$.
We have introduced the graphical rule, that a line changing the
``time direction'' also interchanges particles and antiparticles and
changes the rapidity as $\t\to\t\pm i\pi$, as follows
\be{2.9}
{\bf C}_{\alpha\bar\beta}=\delta_{\alpha\beta}=
\ba{c}
\unitlength4mm
\bp(7,3)
\put(2,1){\oval(2,4)[t]}
\put(0,1){$\t$}
\put(3.3,1){$\t-i\pi$}
\put(.7,0){$\alpha$}
\put(2.7,0){$\bar\beta$}
\ep
\ea
,~~~~
{\bf C}^{\alpha\bar\beta}=\delta_{\alpha\beta}=
\ba{c}
\unitlength4mm
\bp(7,3)
\put(2,2){\oval(2,4)[b]}
\put(0,1){$\t$}
\put(3.3,1){$\t+i\pi$}
\put(.7,2.2){$\alpha$}
\put(2.7,2.2){$\bar\beta$}
\ep
\ea
\ee
Similar crossing relations will be used below to investigate the 
properties of form factors.

\subsection{Bound states}\label{s2.2}
Let the two particles labeled by 1 and 2 of mass $m_1$ and
$m_2$, respectively form a bound state labeled by $(12)$ of mass $m_{(12)}$.
If the mass of the bound state is
$$m_{(12)}=\sqrt{m_1^2+m_2^2+2m_1m_2\cosh\t_{12}^{(12)}}
~~,~~~({\rm Re}\,\t_{12}^{(12)}=0,~0<{\rm Im}\,\t_{12}^{(12)}<\pi)$$
the corresponding eigenvalue of the S-matrix  $S_e(\t)$ will have a pole at
$\t=\t_{12}^{(12)}$
such that
\be{2.10}
S_e(\t)\approx\frac{R_e}{\t-\t_{12}^{(12)}}~~,~~
{\rm for}~~\t\to\t_{12}^{(12)},
\ee
giving rise to a residue $ R_e$.
The eigenvalues are given by the diagonalization of the S-matrix
\be{2.11}
S_{12}(\t)=\sum_e\varphi^{21}_e\,S_e(\t)\,\varphi_{12}^e
\ee
where the projections onto the eigenspaces are given by the intertwiners
(Clebsch-Gordan cofficients) $\varphi_{12}^e$ with
$$
\sum_e\varphi^{12}_e\varphi_{12}^e={\bf1}_{12}~~,~~~
\varphi_{12}^{e'}\varphi^{12}_e=\delta_{e'e}.
$$
Formula (\ref{2.10}) may also be written as
\be{2.12}
\Res_{\t=\t_{12}^{(12)}}S_{12}(\t)=
\varphi^{21}_{(12)}R_{(12)}\varphi_{12}^{(12)}
\ee
where a matrix product with respect to the space of bound states $V_{(12)}$ is
assumed.\\[4mm]
{\bf Remark:}
In general an eigenvalue of $S$ may have several poles
corresponding to bound states of different masses. On the 
other hand several eigenvalues may have poles
at the same point,
which means that there are several types of bound states (12) of the particles
$1$ and $2$ with the same mass.
The space of the bound states $V_{(12)}$ is then a direct sum of spaces
belonging to these types of particles.

\bigskip

The corresponding fields are related by a normal-product relation like
$$
\Psi_{(12)}(x)={\cal N}[\Psi_1\Psi_2](x)\varphi^{12}_{(12)}\,.
$$
The bound state S-matrix which describes the scattering of a bound state with
another particle is given by \cite{K1}
\be{2.13}
\S_{(12)3}(\t_{(12)3})=\sqrt{R_{(12)}}\varphi^{(12)}_{12}\S_{13}(\t_{13})
\S_{23}(\t_{23})\varphi^{12}_{(12)}/\sqrt{R_{(12)}}~\Big|_{\t_1-\t_2=
\t^{(12)}_{12}}
\ee
where the rapidity $\t_{(12)}$ is fixed by $p_1+p_2=p_{(12)}$.
Here and below we use the phase convention that $\sqrt{R_e}=i\sqrt{-R_e}$ if
$R_e<0$.

In integrable  quantum field theories there exist different types of bound
state
spectra which may be characterized by the absence or presence of solitons or
kinks. Of course, in quantum field theory the bootstrap picture means that all
particles are to be considered on the same footing. The names 'solitons',
'kinks' and 'breathers' are motivated by the classical non-linear equations
associated with the quantum model. These equations may possess soliton or
kink solutions, i.e localized non-singular solutions with a localized energy
density. Special solutions consisting of a soliton and an antisoliton
are called 'breathers' because of their oscillatory behaviour.
In the quantum case we call a particle a soliton if it is a bound state of
itself and another particle.
Similarly (and more general), we call a particle a kink if it is a bound
state of a particle with the same mass and another particle.
The mass spectra of integrable quantum field theories characterized by the
absence or presence of solitons are given as follows:
\begin{itemize}
\item[i)] There are particles labeled by $a$ with mass \cite{KT,K2}
$$
m_a=m_1\frac{\sin\frac\pi2\nu a}{\sin\frac\pi2\nu}
~,~~a=1,2,\dots<2/\nu.
$$
This means that two particles of mass $m_a$ and $m_b$ form a bound state
of mass $m_{c=a+b}$.
The corresponding poles of the 
two particle S-matrix element and the rapidities
in the bound state formula (\ref{2.13}) are given by
$$
\t_{ab}^c=i\h\pi\nu(a+b)~~,~~\t_a=\t_c+i\h\pi\nu b~~,~~\t_b=\t_c-i\h\pi\nu a.
$$
The chiral $SU(N)$-Gross-Neveu model \cite{SUN}, the
$Z(N)$ invariant  Ising models \cite{Roland} or
the $SU(N)$-affine Toda field theories
are examples for the above spectrum  with $\nu=2/N$.
In general the mass spectrum is more involved,  
for instance for affine Toda field theories
(with real coupling constant)
related to simply laced algebras the masses
constitute the entries of the Perron-Frobenius
eigenvector of the Cartan matrix \cite{TodaS,Mass}
and for theories related to non-simply laced 
algebras they do not even renormalise
uniformly \cite{nonsim}. 

\item[ii)] If there exist kinks (solitons) of mass $M$ labeled by $A$
then there are three types of bound states:
\begin{itemize}
\item[a)] Particles (breathers) labeled by $a$ are kink-antikink bound
states with
\be{2.14}
m_a=2M\sin\h\pi\nu a~,~~~a=1,2,\dots< 1 /\nu .
\ee
Here the corresponding poles of the kink-antikink S-matrix and the rapidities
in the bound state formula (\ref{2.13}) are given by
$$
\t_{AB}^a=i\pi(1-a\nu)~~,~~
\t_A=\t_c+\frac12\t_{AB}^a~~,~~~\t_B=\t_c-\frac12\t_{AB}^a.
$$
\item[b)] The kink $B$ may be considered as a bound state of a particle $a$
and a kink $A$ such that the pole of the ($a$-$A$)-S-matrix and the rapidities
in the bound state formula (\ref{2.13}) are
$$
\t_{aA}^B=\frac{i\pi}2\,(1+a\nu)~~,~~
\t_a=\t_B+\frac{i\pi}2\,(1-a\nu)~~,~~~\t_A=\t_B-i\pi a\nu.
$$
\item[c)] In addition as in i) two particles of mass $m_a$ and $m_b$ form
a bound state of mass $m_{c=a+b}$, however, here $a<1/\nu$.
\end{itemize}
\end{itemize}
Examples for the latter case
are the sine-Gordon (SU(2)-affine Toda theory)
alias the massive Thirring model
with $\nu=\b^2/(8\pi-\b^2)=\pi/(\pi+2g)$
and the $O(2N)$-Gross-Neveu model with $\nu=1/(N-1)$. Also in this case
the mass spectrum is in general more complicated,
for example all affine Toda field theories with purely 
imaginary coupling fall into this category \cite{TodaIS}. 

The bound state formulae above may be depicted as follows:
For $\t_{12}=\t_1-\t_2=\t^{(12)}_{12}$ with Im $\t_{12}>0$ we introduce
$$
\sqrt{R_{(12)}}\,\varphi^{(12)}_{12}~=
\ba{c}
\unitlength3mm
\bp(4,4)
\put(2,2){\line(0,1){2}}
\put(2,0){\oval(2,4)[t]}
\put(0,0){1}
\put(3.3,0){2}
\put(2.3,3){(12)}
\ep
\ea~~,~~~
(\sigma\varphi)_{(12)}^{21}\sqrt{R_{(12)}}~=
\ba{c}
\unitlength3mm
\bp(4,4)
\put(2,0){\line(0,1){2}}
\put(2,4){\oval(2,4)[b]}
\put(0,3){2}
\put(3.3,3){1}
\put(2.3,0){(12)}
\ep
\ea
$$
$$
1/\sqrt{R_{(12)}}\,(\varphi\sigma)^{(12)}_{21}~=
\ba{c}
\unitlength3mm
\bp(4,4)
\put(2,2){\line(0,1){2}}
\put(2,0){\oval(2,4)[t]}
\put(0,0){2}
\put(3.3,0){1}
\put(2.3,3){(12)}
\ep
\ea~~,~~~
\varphi_{(12)}^{12}\,/\sqrt{R_{(12)}}~=
\ba{c}
\unitlength3mm
\bp(4,4)
\put(2,0){\line(0,1){2}}
\put(2,4){\oval(2,4)[b]}
\put(0,3){1}
\put(3.3,3){2}
\put(2.3,0){(12)}
\ep~~~.
\ea
$$
Then we have the relations
$$
\ba{c}
\unitlength3mm
\bp(4,6)
\put(2,0){\line(0,1){1.5}}
\put(2,3){\oval(2,3)}
\put(2,4.5){\line(0,1){1.5}}
\put(.9,0){$e$}
\put(.8,5){$e'$}
\put(0,2.7){1}
\put(3.3,2.7){2}
\ep
\ea~~=~\delta_{e'e}~
\ba{c}
\unitlength3mm
\bp(1,4)
\put(1,0){\line(0,1){4}}
\put(0,0){$e$}
\ep
\ea
~~~,~~~
\sum_e~
\ba{c}
\unitlength3mm
\bp(4,6)
\put(2,0){\oval(2,4)[t]}
\put(2,2){\line(0,1){2}}
\put(2,6){\oval(2,4)[b]}
\put(0,0){1}
\put(3.3,0){2}
\put(2.3,2.7){$e$}
\put(0,5){1}
\put(3.3,5){2}
\ep
\ea~~=~~
\ba{c}
\unitlength3mm
\bp(4,4)
\put(1,0){\line(0,1){4}}
\put(3,0){\line(0,1){4}}
\put(0,0){1}
\put(3.3,0){2}
\ep
\ea
$$
where formally we have put $R_e/R_e=1$ even if $R_e=0$ in case that $e$ does
not correspond to a bound state.
The sum in the last formula is over all eigenspaces
$V_e\subset V_1\otimes V_2$ of the S-matrix.
If we would sum only over those $e=(12)$ which correspond to bound states,
we would get the projector onto the subspace of bound states in
$V_1\otimes V_2$. Moreover formula (\ref{2.12}) is depicted as
$$
\Res_{\t_{12}=\t^{(12)}_{12}}
\ba{c}
\unitlength3mm
\bp(4,4)
\put(1,0){\line(1,2){2}}
\put(3,0){\line(-1,2){2}}
\put(0,0){1}
\put(3.3,0){2}
\ep
\ea
~~=~~
\ba{c}
\unitlength3mm
\bp(4,6)
\put(2,0){\oval(2,4)[t]}
\put(2,2){\line(0,1){2}}
\put(2,6){\oval(2,4)[b]}
\put(0,0){1}
\put(3.3,0){2}
\put(2.3,2.7){(12)}
\put(0,5){2}
\put(3.3,5){1}
\ep
\ea
$$
The bound state formula (\ref{2.13}) may be depicted as
$$
\ba{c}
\unitlength3mm
\bp(6,4)
\put(1,0){\line(1,1){4}}
\put(5,0){\line(-1,1){4}}
\put(-1.5,.3){(12)}
\put(5.3,.3){3}
\ep
\ea
~~~=~~~~~~
\ba{c}
\unitlength3mm
\bp(8,6)
\put(1,0){\line(1,1){1.7}}
\put(4,3){\oval(3.5,3.5)}
\put(7,6){\line(-1,-1){1.7}}
\put(7,0){\line(-1,1){6}}
\put(-1.5,.5){(12)}
\put(7.3,.5){3}
\put(.8,2.6){1}
\put(6.4,2.6){2}
\ep
\ea~.
$$
It implies relations of two-particle S-matrices \cite{K2} called
'pentagon equations' (also referred to as bootstrap equations) like
$$
S_{(12)3}\,\sqrt{R_{(12)}}\,\varphi^{(12)}_{12}~=
\sqrt{R_{(12)}}\,\varphi^{(12)}_{12}~
S_{13}S_{23}~~~:~~~~
\ba{c}
\unitlength3mm
\bp(5,5)
\put(2,2){\line(1,1){3}}
\put(5,2){\line(-1,1){3}}
\put(0,2){\line(1,0){2}}
\put(2,0){\line(0,1){2}}
\put(0,.7){1}
\put(1,0){2}
\put(4,1.2){3}
\ep
\ea~=~
\ba{c}
\unitlength3mm
\bp(5,5)
\put(3,3){\line(1,1){2}}
\put(4,0){\line(-1,1){4}}
\put(0,2){\line(3,1){3}}
\put(2,0){\line(1,3){1}}
\put(0,.8){1}
\put(1,0){2}
\put(4,.4){3}
\ep
\ea~~.
$$

\subsection{The Sine-Gordon model S-matrix}\label{s2.3}
The Sine-Gordon model alias the massive Thirring model is defined by the
Lagrangians
$${\cal L}^{SG}
=\frac12(\partial_\mu\phi)^2+\frac\alpha{\beta^2}(\cos\beta\phi-1),$$
$${\cal L}^{MTM}
=\bar\psi(i\gamma\partial-M)\psi-\frac12g(\bar\psi\gamma^\mu\psi)^2,$$
respectively.

The Fermi field $\psi$ correspond to the soliton and antisoliton and the
bose field $\phi$ to the lowest `breather' which is the lowest soliton
antisoliton bound state. The precise relation between the related
coupling constants was found by Coleman  \cite{Co} within the
framework of perturbation theory
$$\nu=\frac{\b^2}{8\pi-\b^2}=\frac{\pi}{\pi+2g}$$
where the parameter $\nu$ is introduced for later
convenience. The two-particle S-matrix is
\be{2.15}
S(\t,\nu)=\left(\matrix{a&&&&&&\cr&b&c&&&&\cr&c&b&&&&\cr&&&a&&&
\cr&&&&S_{sb}&&\cr&&&&&S_{bb}&\cr&&&&&&\ddots}\right)
\ee
where the soliton-soliton amplitude $a(\t)$ and the soliton-antisoliton
forward and backward amplitudes $b(\t)$ and $c(\t)$
$$
\def\pS#1#2#3#4#5#6#7#8{~\ba{c}\unitlength2mm\bp(4,4)
\put(0,0){\line(1,1){4}}\put(4,0){\line(-1,1){4}}
\put(1,1){\vector(#1 1,#2 1){.3}}\put(3,1){\vector(#3 1,#4 1){.3}}
\put(3,3){\vector(#5 1,#6 1){.3}}\put(1,3){\vector(#7 1,#8 1){.3}}\ep\ea}
a=\pS{}{}{-}{}{}{}{-}{}=\pS{-}{-}{}{-}{-}{-}{}{-}~,~~
b=\pS{-}{-}{-}{}{-}{-}{-}{}=\pS{}{}{}{-}{}{}{}{-}~,~~
c=\pS{-}{-}{-}{}{}{}{}{-}=\pS{}{}{}{-}{-}{-}{-}{}
$$
are given by \cite{ZamS}
\be{2.16}
\ba{rcl}
b(\t)&=&\ds\frac{\sinh\t/\nu}{\sinh(i\pi-\t)/\nu}\,a(\t)\,,~~~~
c(\t)=\frac{\sinh i\pi/\nu}{\sinh(i\pi-\t)/\nu}\,a(\t)\,,
\\[3mm]
a(\t)&=&\ds\exp\int_0^\infty\frac{dt}t\,\frac{\sinh\frac12(1-\nu)t}
{\sinh\frac12\nu t\,\cosh\frac12t}\,\sinh t\frac\t{i\pi}\,.
\ea
\ee
These amplitudes fulfill 'crossing'
\be{2.17}
a(i\pi-\t)=b(\t)~,~~c(i\pi-\t)=c(\t)
\ee
and unitarity
\be{2.18}
a(-\t)a(\t)=1~,~~b(-\t)b(\t)+c(-\t)c(\t)=1.
\ee
The intertwiners $\varphi^e_{ab}$ of section \ref{s2.1} are given by the
non-vanishing components
\be{2.19}
\varphi^0_{ss}=\varphi^{\bar0}_{\bar s\bar s}=1~,~~~
\varphi^\pm_{s\bar s}=1/\sqrt2~,~~~\varphi^\pm_{\bar ss}=\pm1/\sqrt2
\ee
and the corresponding  S-matrix eigenvalues are
\be{2.20}
S_0=S_{\bar0}=a~,~~~S_\pm=b\pm c\,.
\ee
The amplitudes $S_0=S_{\bar0}$ have no poles corresponding to bound states.
The amplitudes $S_\pm(\t)$ have poles at $\t=i\pi(1-k\nu)$ for even/odd
$k<1/\nu$ corresponding to the k-th breather as soliton-antisoliton bound
states.

As examples of soliton-breather and breather-breather amplitudes those
for the lowest breather are \cite{KT}
\be{2.21}
\ba{rcl}
S_{sb}(\t)&=&\ds\frac{\sinh\t+i\sin\frac12\pi(1+\nu)}
{\sinh\t-i\sin\frac12\pi(1+\nu)}
=-\exp\int_0^\infty\frac{dt}t\,2\frac{\cosh\frac12\nu t}
{\cosh\frac12t}\,\sinh t\frac\t{i\pi}\,,
\\[3mm]
S_{bb}(\t)&=&\ds\frac{\sinh\t+i\sin\pi\nu}{\sinh\t-i\sin\pi\nu}
=-\exp\int_0^\infty\frac{dt}t\,2\frac{\cosh(\frac12-\nu)t}
{\cosh\frac12t}\,\sinh t\frac\t{i\pi}\,.
\ea
\ee
The S-matrix element $S_{bb}$ has been dicussed before in \cite{AK}.
The pole of $S_{sb}(\t)$ at $\t=i\pi(1+\nu)/2$ belongs to the soliton as a
soliton-breather bound state and the pole of $S_{bb}(\t)$ at $\t=i\pi\nu$
to the second breather $b_2$ as a breather-breather bound state.
The intertwiners $\varphi^e_{ab}$ of section \ref{s2.1} are given by the
non-vanishing components
\be{2.22}
\varphi^s_{sb}=\varphi^s_{bs}=\varphi^{b_2}_{bb}=1.
\ee
The formulae involving  higher breather may be found in \cite{KT}, e.g.
\be{2.23}
S_{sb_k}(\t)
=(-1)^k\exp\int_0^\infty\frac{dt}t\,2\frac{\cosh\frac12\nu t\sinh\frac12\nu kt}
{\cosh\frac12t\sinh\frac12\nu t}\,\sinh t\frac\t{i\pi}
\ee
for $k<l$
\be{2.24}
S_{b_kb_l}(\t)
=\exp\int_0^\infty\frac{dt}t\,4
\frac{\cosh\frac12\nu t\sinh\frac12\nu kt\cosh\frac12(1-\nu l)t}
{\cosh\frac12t\sinh\frac12\nu t}\,\sinh t\frac\t{i\pi}
\ee
and
\be{2.25}
S_{b_kb_k}(\t)
=-\exp\int_0^\infty\frac{dt}t\,2
\frac{\cosh\frac12\nu t\sinh\frac12(2k\nu-1) kt+\sinh\frac12(1-\nu)t}
{\cosh\frac12t\sinh\frac12\nu t}\,\sinh t\frac\t{i\pi}\,.
\ee

\setcounter{equation}{0}
\section{Properties of generalized form factors}\label{s3}
We investigate the properties of generalized form factors, in particular
for integrable quantum field theories in 1+1 dimensions.
Some formulae, originally proposed in \cite{KW}, are recalled and 
the physical arguments on how to
derive them are provided in appendix \ref{sb}. All arguments are solely based
on the validity LSZ reduction formalism \cite{LSZ} 
(see also \cite{IZ}) and the  additional assumption of
``maximal analyticity" which means, roughly speaking,
that the S-matrix and the form factors
are analytic functions everywhere except at those points where they posses
singularities due to physical intermediate states. In other words the
entire  pole structure is of physical origin and in the following we
investigate it employing the arguments of  \cite{KW,BKW,K2} 
(see also \cite{nankai,Smirnov3,YZ}).
\subsection{Form factors in momentum space and rapidity space}
For simplicity we first consider the case of bosonic charged particles.
The extension to the general situation will be provided below.
The corresponding  Fock space
is spanned by the in- or out-states of particles and anti-particles
given by (\ref{2.1}) and (\ref{2.2}).
In addition to the notation of vectors and matrices of (\ref{2.6}) we denote
co-vectors by
\be{3.1}
v_{1\dots n}\in V_{1\dots n}^\dagger
\ee
with components $v_\ua=v_{\a_1\dots\a_n}$.

Let now  $\O(x)$ be a local scalar operator, the generalized form factors
are defined as the co-vector valued functions given by
\be{3.2}
\la0\mi\O(x)\mi
p_1,\dots,p_n\ra^{in}_{\a_1\dots\a_n}=e^{-ix(p_1+\dots+p_n)}\,
\F_\ua\Big((s_{ij}+i\epsilon)_{(1\le i<j\le n)}\Big)
\ee
where $\ua=\{\a_1,\dots,\a_n\}$ and where $s_{ij}=(p_i+p_j)^2$ is one of the
Mandelstam variables, as in the previous section. 
There may also be anti-particles in the state.
As is well known these functions are boundary values of analytic functions
as indicated by the $\epsilon$-prescription. We assume that the domain of
analyticity is much larger than could be proven by means of general
principles. Similar as for the scattering matrix
we  assume in addition at this point ``maximal analyticity" meaning 
that there should be no redundant
poles, but all singularities should be of physical origin as particle states
etc.  Since the $x$-dependence of the form factors is trivial, in the sense
that we may always carry out a translation as in eq.~(\ref{3.2}), we consider
in the following the operator always at the origin, i.e. $\O=\O(0)$.

Under the assumption that $F$ is an analytic function, an 
interchange in eq.~(\ref{3.2}) of the $in$ and $out$ states leads
to the replacement of $s+i\epsilon$ by
$s-i\epsilon$. This means in particular that
\be{3.3}
\la0\mi\O\mi
p_1,\dots,p_n\ra^{out}_{\a_1\dots\a_n}=
\F_\ua\Big((s_{ij}-i\epsilon)_{(1\le i<j\le n)}\Big).
\ee
The crossing property for the connected part of the matrix element yields
\be{3.4}
^{~~~~out}_{\a_1\dots\a_m}\la p_1,\dots,p_m\mi\O\mi
p_{m+1},\dots,p_n\ra^{in~{\rm conn.}}_{\a_{m+1}\dots\a_n}
=\F_\ua(s_{ij}+i\epsilon,t_{rs}-i\epsilon,s_{kl}+i\epsilon)
\ee
where $1\le i<j\le m,~1\le r\le m<s\le n,~m\le k<l\le n$
and $t_{rs}=(p_r-p_s)^2$ is another Mandelstam variable. See
appendix~\ref{sb} for a proper derivation  of this claim.

The most basic properties of the form factors are usually refered to as
Watson's equations \cite{Wat}, which have been already known in the fifties.  
It is instructive at this point to discuss them first for the case $n=2$. 
Using the completeness of the $out$-states we have
\be{3.5}
\F_{\a_1\a_2}(s_{12}+i\epsilon)=\la0\mi\O\mi p_1,p_2\ra^{in}_{\a_1\a_2}
=\sum_{out}\la0\mi\O\mi out\ra\la out\mi p_1,p_2\ra^{in}_{\a_1\a_2}.
\ee
For $4m^2\le s_{12}<$ 'lowest inelastic threshold' only the two particle
S-matrix contributes
\be{3.6}
\F_{\a_1\a_2}(s_{12}+i\epsilon)=
\F_{\a'_1\a'_2}(s_{12}-i\epsilon)\,S^{\a'_2\a'_1}_{\a_1\a_2}(s_{12})
\ee
and analogously starting with $_{\a_1}\la p_1\mi\O\mi p_2\ra_{\a_2}$ in
(\ref{3.5}) we obtain
\be{3.7}
\F_{\a_1\a_2}(t_{12}-i\epsilon)=\F_{\a_1\a_2}(t_{12}+i\epsilon)\,,
\ee
where the fact has been used that the one-particle S-matrix is always 
trivial.
In integrable theories there are no inelastic transitions,
therefore eq.~(\ref{3.6}) holds for all
$s\ge 4m^2$.
The generalized Watson's equations for $1\le m\le n$ read as (see \cite{KW})
\be{3.8}
\F_\ua(s_{ij}+i\epsilon,t_{rs}-i\epsilon,s_{kl}+i\epsilon)=
S^{\a_m\dots\a_1}_{\a'_1\dots\a'_m}(s_{ij})\,
\F_{\ua'}(s_{ij}-i\epsilon,t_{rs}+i\epsilon,s_{kl}-i\epsilon)\,
S^{\a'_n\dots\a'_{m+1}}_{\a_{m+1}\dots\a_n}(s_{lk}).
\ee
For a diagonal S-matrix these equations have been discussed before in
\cite{VG}.

\begin{figure}[hbt]
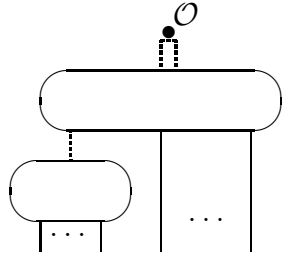

$$
\unitlength4mm
\bp(9,8)
\put(2,2){\oval(4,2)}
\put(5,5){\oval(8,2)}
\put(1,0){\line(0,1){1}}
\put(3,0){\line(0,1){1}}
\put(5,0){\line(0,1){4}}
\put(8,0){\line(0,1){4}}
\thicklines
\put(2,3){\dashbox{.1}(0,1){}}
\put(5,6){\dashbox{.1}(.5,1){}}
\put(5,7){$\bullet$}
\put(1.3,.5){$\dots$}
\put(5.9,1){$\dots$}
\put(5.4,7.5){$\O$}
\ep
$$
\caption{\label{f30}\it
A singular contribution to the n-particle form factor diagram corresponding to
a sub-channel. The dashed lines belong to off-shell lines.
}
\end{figure}

The generalized form factors also contain singularities  
\cite{KW,BKW,K} which are determined
by the one-particle states in all sub-channels
$(\a_i,\dots,\a_j)\subset(\a_1,\dots,\a_n)$ (see figure \ref{f30}).
Poles occur if the square of the total momentum in the sub-channel equals
the one-particle mass squared.
In particular there are poles, if for instance 
particle 1 is the anti-particle of
particle 2 and particle 1 is crossed to the out-state together with 
$p_2\to p_1$,
which means $(p_2+p_3-p_1)^2\to m_3^2$. Alternatively, if particle 3 is 
a bound state
of particle 1 and 2, in which case  $(p_2+p_1)^2\to m_3^2$. 
The residues of the form factors at these poles are 
related to form factors with fewer legs,
as indicated in figure \ref{f30}. We will discuss these facts later in detail.

Similarly as for the S-matrix we may also write the form factors 
(\ref{3.4}) as co-vector valued analytic functions of the
rapidity differences $\t_{ij}=\t_i-\t_j$
\bea
\F_\ua(s_{ij}+i\epsilon,t_{rs}-i\epsilon,s_{kl}+i\epsilon)
&=&\F_\ua(|\t_{ij}|,i\pi-|\t_{rs}|,|\t_{kl}|)\\
\F_\ua(s_{ij}-i\epsilon,t_{rs}+i\epsilon,s_{kl}-i\epsilon)
&=&\F_\ua(-|\t_{ij}|,i\pi+|\t_{rs}|,-|\t_{kl}|)
\eea
The domains of analyticity and the physical regimes in the complex planes of
the Mandelstam variables and the rapidity difference variables are depicted
in figure~\ref{f31}. However, now the branch cut between between region 
II and IV is absent (c.f. eqs.~(\ref{3.7} and (\ref{3.8})).

\subsection{The auxiliary form factor function}\label{s3.2}
\begin{figure}[hbt]
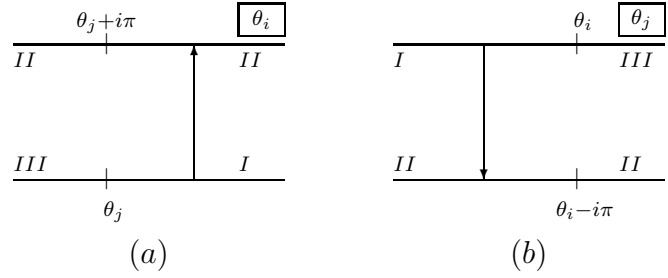

$$
\ba{c}
\unitlength6mm
\bp(6,5)
\put(0,1){\line(1,0){6}}
\put(0,4){\line(1,0){6}}
\put(4,1){\vector(0,1){3}}
\put(2,.9){$\s|$}
\put(2,.2){$\s\t_j$}
\put(2,3.9){$\s|$}
\put(1.4,4.4){$\s\t_j+i\pi$}
\put(5,1.2){$\s I$}
\put(0,1.2){$\s III$}
\put(0,3.5){$\s II$}
\put(5,3.5){$\s II$}
\put(5,4.2){\framebox(1,.7){$\s\t_i$}}
\ep
\\(a)
\ea~~~~~~~~
\ba{c}
\unitlength6mm
\bp(6,5)
\put(0,1){\line(1,0){6}}
\put(0,4){\line(1,0){6}}
\put(2,4){\vector(0,-1){3}}
\put(4,.9){$\s|$}
\put(3.6,.2){$\s\t_i-i\pi$}
\put(4,3.9){$\s|$}
\put(4,4.4){$\s\t_i$}
\put(0,3.5){$\s I$}
\put(5,3.5){$\s III$}
\put(0,1.2){$\s II$}
\put(5,1.2){$\s II$}
\put(5,4.2){\framebox(1,.7){$\s\t_j$}}
\ep
\\(b)
\ea
$$
\caption{\label{f32}\it
The physical regimes in the complex planes of the rapidity variables
$(a)~~\t_i$ and $(b)~~\t_j$ for $\t_i>\t_k>\t_j~(k=1,\dots,n)$. Again the
crossing transitions (see appendix B) are indicated by the arrows.
}
\end{figure}

Furthermore, it is convenient to introduce a new co-vector valued auxiliary
function $\f_\ua(\ut)$ which is considered as an analytic function of the
individual rapidities of
the particles, instead of analytic functions of all rapidity differences
(see also \cite{nankai}). It coincides with the generalized form factor for a
particular order of the rapidities
\be{3.9}
\f_\ua(\t_1,\dots,\t_n)=\F_\ua(|\t_{ij}|)=
\la0\mi\O\mi p_1,\dots,p_n\ra^{in}_\ua~,~~~{\rm for}~~\t_1>\dots>\t_n.
\ee
For all other arrangements of the rapidities the functions 
$\f_\ua(\ut)$ are given by analytic continuation.
The domains of analyticity, the physical regimes and the transitions to the
crossed regions in the complex planes of $\t_i$ and $\t_j$ for $\t_i>\t_j$
are depicted in figure~\ref{f32}.

Now we formulate the main properties of generalized form factors in
terms of the auxiliary functions $\f_{1\dots n}$
under the assumptions of ``maximal analyticity''.\\[3mm]
{\bf Properties:} 
The co-vector valued auxiliary function $\f_{1\dots n}(\ut)$ is meromorphic in
all variables $\t_1,\dots,\t_n$ and
\begin{itemize}
\item[(i)] fulfills the symmetry property under the permutation of
both, the variables $\t_i,\t_j$ and the spaces $i,j$ at the same time
\be{3.10}
\f_{\dots ij\dots}(\dots,\t_i,\t_j,\dots)
=\f_{\dots ji\dots}(\dots,\t_j,\t_i,\dots)\,
S_{ij}(\t_i-\t_j)
\ee
for all possible arrangements of the $\t$'s,
\item[(ii)] fulfills the periodicity property under the cyclic permutation
of the rapidity variables and spaces 
\be{3.11}
\f_{1\dots n}(\t_1,\t_2,\dots,\t_n,)=
\f_{2\dots n1}(\t_2,\dots,\t_n,\t_1-2\pi i),
\ee
\item[(iii)] and has poles determined by one-particle states in each
sub-channel (see figure~\ref{f30}).
In particular the function $\f_\ua(\ut)$ has a pole at $\t_{12}=i\pi$
such that
\be{3.12}
\Res_{\t_{12}=i\pi}\f_{1\dots n}(\t_1,\dots,\t_n)=2i\,{\bf C}_{12}\,
\f_{3\dots n}(\t_3,\dots,\t_n)\Big({\bf1}-S_{2n}\dots S_{23}\Big)
\ee
where ${\bf C}_{12}$ is the charge conjugation matrix with matrix elements
${\bf C}_{\a\a'}=\delta_{\bar\a\a'}$.
\item[(iv)]
If there are also bound states in the model 
the function $\f_\ua(\ut)$ has additional poles. If for instance
 the particles 1 and 2
form a bound state (12), there 
is a pole at $\t_{12}=\t^{(12)}_{12}$ such that  
\be{3.13}
\Res_{\t_{12}=\t^{(12)}_{12}}\f_{12\dots n}(\t_1,\t_2,\dots,\t_n)\,
\varphi^{12}_{(12)}=\f_{(12)\dots n}(\t_{(12)},\dots,\t_n)\,\sqrt{2iR_{(12)}}
\ee
where the quantities 
$\varphi^{12}_{(12)},~R_{(12)}$ and the values of $\t_1,~\t_2,~\t_{(12)}$ and
$\t^{(12)}_{12}$ were discussed in section \ref{s2.2}.
\end{itemize}
The property (i) - (iv) may be depicted as
$$
\ba{rrcl}
{\rm(i)}&
\ba{c}
\unitlength4mm
\bp(7,3)
\put(3.5,2){\oval(7,2)}
\put(3.5,2){\makebox(0,0){$\f$}}
\put(1,0){\line(0,1){1}}
\put(3,0){\line(0,1){1}}
\put(4,0){\line(0,1){1}}
\put(6,0){\line(0,1){1}}
\put(1.4,.5){$\dots$}
\put(4.4,.5){$\dots$}
\ep
\ea
&=&
\ba{c}
\unitlength4mm
\bp(7,4)
\put(3.5,3){\oval(7,2)}
\put(3.5,3){\makebox(0,0){$\f$}}
\put(1,0){\line(0,1){2}}
\put(3,0){\line(1,2){1}}
\put(4,0){\line(-1,2){1}}
\put(6,0){\line(0,1){2}}
\put(1.4,1){$\dots$}
\put(4.4,1){$\dots$}
\ep
\ea
\\
(ii)&
\ba{c}
\unitlength4mm
\bp(5,4)
\put(2.5,2){\oval(5,2)}
\put(2.5,2){\makebox(0,0){$\f$}}
\put(1,0){\line(0,1){1}}
\put(2,0){\line(0,1){1}}
\put(4,0){\line(0,1){1}}
\put(2.4,.5){$\dots$}
\ep
\ea
&=&
\ba{c}
\unitlength4mm
\bp(7,4)
\put(0,0){\line(0,1){1}}
\put(6,1){\oval(2,2)[b]}
\put(3.5,1){\oval(7,6)[t]}
\put(3.5,2){\oval(5,2)}
\put(3.5,2){\makebox(0,0){$\f$}}
\put(2,0){\line(0,1){1}}
\put(4,0){\line(0,1){1}}
\put(2.4,.5){$\dots$}
\ep
\ea
\\
(iii)&
\ds\frac1{2i}\,\Res_{\t_{12}=i\pi}~~~
\ba{c}
\unitlength4mm
\bp(6,4)
\put(3,2){\oval(6,2)}
\put(3,2){\makebox(0,0){$\f$}}
\put(1,0){\line(0,1){1}}
\put(2,0){\line(0,1){1}}
\put(3,0){\line(0,1){1}}
\put(5,0){\line(0,1){1}}
\put(3.4,.5){$\dots$}
\ep
\ea
&=&
\ba{c}
\unitlength4mm
\bp(5,4)
\put(.5,0){\oval(1,2)[t]}
\put(3,2){\oval(4,2)}
\put(3,2){\makebox(0,0){$\f$}}
\put(2,0){\line(0,1){1}}
\put(4,0){\line(0,1){1}}
\put(2.4,.5){$\dots$}
\ep
\ea~~-~~
\ba{c}
\unitlength4mm
\bp(6,5)
\put(0,0){\line(0,1){3}}
\put(3,3){\oval(6,4)[t]}
\put(3,3){\oval(6,4)[br]}
\put(3,0){\oval(4,2)[tl]}
\put(3,3){\oval(4,2)}
\put(3,3){\makebox(0,0){$\f$}}
\put(2,0){\line(0,1){2}}
\put(4,0){\line(0,1){2}}
\put(2.4,1.5){$\dots$}
\ep
\ea
\\
(iv)&
\ds\frac1{\sqrt{2i}}\,\Res_{\t_{12}=\t^{(12)}_{12}}~
\ba{c}
\unitlength4mm
\bp(5,4)
\put(2.5,3){\oval(5,2)}
\put(2.5,3){\makebox(0,0){$\f$}}
\put(1.5,2){\oval(1,2)[b]}
\put(1.5,0){\line(0,1){1}}
\put(4,0){\line(0,1){2}}
\put(2.4,1){$\dots$}
\ep
\ea
&=&
\ba{c}
\unitlength4mm
\bp(4,3)
\put(2,2){\oval(4,2)}
\put(2,2){\makebox(0,0){$\f$}}
\put(1,0){\line(0,1){1}}
\put(3,0){\line(0,1){1}}
\put(1.4,.5){$\dots$}
\ep
\ea
\ea
$$
Both properties (ii) and (iii) are consequences of the general
{\bf crossing formulae}
\be{3.14}\ba{ccl}
\lefteqn{_{\bar1}\la p_1\mi\O\mi p_2,\dots, p_n\ra^{in}_{2\dots n}}\\[2mm]
&=&\sum\limits_{j=2}^n\,{}_{\bar1}\la p_1\mi p_j\ra_j\,
\f_{2\dots\hat j\dots n}\,S_{2j}\cdots S_{j-1j}
+{\bf C}^{\bar11}\,\f_{12\dots n}(\t_1+i\pi_-,\dots,\t_n)\\[2mm]
&=&\sum\limits_{j=2}^n\,{}_{\bar1}\la p_1\mi p_j\ra_j\,
\f_{2\dots\hat j\dots n}\,S_{jn}\cdots S_{jj+1}
+\f_{2\dots n1}(\dots,\t_n,\t_1-i\pi_-)\,{\bf C}^{1\bar1}.
\ea\ee
where we introduced the notation
$\f_{2\dots\hat j\dots n}$, meaning that the space $j$ and the corresponding
variable $\t_j$ are missing. In terms of the components,
$_{\bar1}\la p_1\mi p_j\ra_j$ means
$\delta_{\bar\a_1\a_j}\,4\pi\,\delta(\t_1-\t_j)$ and $\delta^{1\bar1}$ means
$\delta_{\a_1\bar\a_1}$.
These are equations for distributions where on the right hand side the second
terms are understood as boundary values of analytic functions with $\pi_-=
\pi-\epsilon$.
The crossing formulae may be depicted as
\bea
\ba{c}
\unitlength4mm
\bp(5,6)
\put(2.5,3){\oval(5,2)}
\put(2.5,3){\makebox(0,0){$\f$}}
\put(1,1){\line(0,1){1}}
\put(4,1){\line(0,1){1}}
\put(2.5,4){\line(0,1){1}}
\put(1.9,1.5){$\dots$}
\put(.8,0){$\s2$}
\put(3.8,0){$\s n$}
\put(2.3,5.5){$\s\bar1$}
\ep
\ea~~&=&\sum_{j=2}^n~~
\ba{c}
\unitlength4mm
\bp(6,6)
\put(2,5){\oval(4,6)[bl]}
\put(2,1){\oval(3,2)[tr]}
\put(3.5,4){\oval(5,2)}
\put(3.5,4){\makebox(0,0){$\f$}}
\put(2,1){\line(0,1){2}}
\put(5,1){\line(0,1){2}}
\put(2.8,2.5){$\dots$}
\put(1.8,0){$\s2$}
\put(3.3,0){$\s j$}
\put(4.8,0){$\s n$}
\put(-.1,5.5){$\s\bar1$}
\ep
\ea~~+~~
\ba{c}
\unitlength4mm
\bp(6,6)
\put(0,2){\line(0,1){3}}
\put(1,2){\oval(2,2)[b]}
\put(3.5,3){\oval(5,2)}
\put(3.5,3){\makebox(0,0){$\f$}}
\put(3,1){\line(0,1){1}}
\put(5,1){\line(0,1){1}}
\put(3.4,1.5){$\dots$}
\put(2.8,0){$\s2$}
\put(4.8,0){$\s n$}
\put(-.1,5.5){$\s\bar1$}
\ep
\ea\\
&=&\sum_{j=2}^n~~
\ba{c}
\unitlength4mm
\bp(6,6)
\put(4,5){\oval(4,6)[br]}
\put(4,1){\oval(3,2)[tl]}
\put(2.5,4){\oval(5,2)}
\put(2.5,4){\makebox(0,0){$\f$}}
\put(1,1){\line(0,1){2}}
\put(4,1){\line(0,1){2}}
\put(1.8,2.5){$\dots$}
\put(.8,0){$\s2$}
\put(2.3,0){$\s j$}
\put(3.8,0){$\s n$}
\put(5.8,5.5){$\s\bar1$}
\ep
\ea~~+~~
\ba{c}
\unitlength4mm
\bp(6,6)
\put(6,2){\line(0,1){3}}
\put(5,2){\oval(2,2)[b]}
\put(2.5,3){\oval(5,2)}
\put(2.5,3){\makebox(0,0){$\f$}}
\put(1,1){\line(0,1){1}}
\put(3,1){\line(0,1){1}}
\put(1.4,1.5){$\dots$}
\put(.8,0){$\s2$}
\put(2.8,0){$\s n$}
\put(5.8,5.5){$\s\bar1$}
\ep
\ea
\eea
where we have again used the graphical rule (\ref{2.9}), which states
that a line
changing the ``time direction" also interchanges particles and anti-particles
and changes $\t\to\t\pm i\pi$.
Taking the analytic part of the crossing relation one obtains  property (ii)
and considering in addition the part with point like support one gets
property (iii). The proofs of the properties  (i)-(iv) 
and equation (\ref{3.14}) are provided in appendix \ref{sb}.
\begin{itemize}
\item[(v)]
Naturally, since we are dealing with relativistic
quantum field theories we finally have
\be{3.15}
\f_{1\dots n}(\t_1+u,\dots,\t_n+u)=e^{\pm s u}\,
\f_{1\dots n}(\t_1,\dots,\t_n)
\ee
if the local operator transforms under Lorentz transformations as
$\O\to e^{\pm s u}\O$ where $s$ is the ``spin" of $\O$.
\end{itemize}

\subsection{The general bosonic and fermionic case}
For the general case where the states involve also fermions and
where $\O(x)$ is a local bosonic or fermionic operator
with arbitrary spin we write the matrix elements of $\O(0)$ as
\be{3.16}
\la0\mi\O\mi p_1,\dots,p_n\ra^{in}_{\a_1\dots\bar\a_n}=
\sum_l\bar v(p_i)\cdots\Gamma^{(l)}_{\mu_1\dots\mu_k}\cdots u(p_j)\,
p^{\mu_1}_{i_1}\dots p^{\mu_k}_{i_k}\,G^{(l),\O}_\ua(s_{ij}+i\epsilon)
\ee
where the $\Gamma$ are matrices in spinor space. For the invariant form
factor functions $G^\O_\ua$, the Watson's equations look
quite analogously to those in the bosonic case. However, sometimes it is more
convenient to consider the full matrix elements and then we must take into
account sign factors due to the fermions.
Analogously to eq.~(\ref{3.9}) we introduce the co-vector valued auxiliary
function $\f$ which determines the form factors for a
specific order of the rapidities.
For the general case the three main properties of the
co-vector valued function $\f$ may be written as:
\be{3.17}\ba{lcl}
(\rm i)~~~~\f_{1\dots ij\dots n}(\t_1,\dots,\t_i,\t_j,\dots, \t_n)
&=&\f_{1\dots ji\dots n}(\t_1,\dots,\t_j,\t_i,\dots,\t_n)\,\S_{ij}
\\[2mm]
(\rm ii)&=&\f_{2\dots n1}(\t_2,\dots,\t_n,\t_1-2i\pi)\,\sigma_{\O1}\\[2mm]
(\rm iii)&\approx&\frac{2i}{\t_{12}-i\pi}\,
{\bf C}_{12}\,\f_{3\dots n}(\t_3,\dots,\t_n)\Big({\bf 1}-S_{2n}\dots
S_{23}\Big)
\ea\ee
The bound state formula (iv) is in general true for the invariant part
of the form factors. For the case of fermions, spinors have to be taken into
account (see the examples below). In the formulae (\ref{3.17})
the statistics of the operator $\O$ is taken into account by $\sigma_{\O1}=
-1$ if both $\O$ and particle 1 are fermionic and $\sigma_{\O1}=1$ otherwise.
The statistics of the particles is taken into account by $\S$ which means
that $\S_{12}=-S_{12}$ if both particles are fermions and 
$\S_{12}=S_{12}$ otherwise.
Again, both properties (ii) and (iii) are consequences of 
the {\bf crossing formulae}, which, for the general case 
of bosons or fermions, reads 
\be{3.18}\ba{ccl}
\lefteqn{_{\bar1}\la p_1\mi\O\mi p_2,\dots, p_n\ra^{in}_{2\dots n}}\\[2mm]
&=&\sigma_{\O1}\left\{\sum\limits_{j=2}^n\,{}_{\bar1}\la p_1\mi p_j\ra_j\,
\f_{2\dots\hat j\dots n}\,\S_{2j}\cdots\S_{j-1j}
+{\bf C}^{\bar11}\f_{12\dots n}(\t_1+i\pi_-,\dots,\t_n)\right\}\\[2mm]
&=&\sum\limits_{j=2}^n\,{}_{\bar1}\la p_1\mi p_j\ra_j\,
\f_{2\dots\hat j\dots n}\,\S_{jn}\cdots\S_{jj+1}
+\f_{2\dots n1}(\dots,\t_n,\t_1-i\pi_-)\,{\bf C}^{1\bar1}.
\ea\ee
replacing eq.~(\ref{3.14}).
The proof of these relations are also given in appendix \ref{sb}.

The appearance of $\S$ is natural in the context of
factorizing S-matrices.
See for example the general Yang-Baxter relation (\ref{2.5})
which is essential if  transitions as fermion + anti-fermion
$\to$ boson + anti-boson are possible.

\setcounter{equation}{0}
\section{Solution for the sine-Gordon alias massive Thirring model Model}
\label{s4}
We will now provide a constructive and systematic way of how to
solve the properties i)-v) for the co-vector valued function $f$ once
the scattering matrix is given. To capture the vectorial structure 
of the form factors we will employ the techniques of the 
``off-shell'' Bethe ansatz \cite{Hrachik1,Hrachik2} which we now
explain briefly. 

\subsection{The general formula}
As usual in the context of algebraic Bethe ansatz we define the monodromy
matrix
\be{4.1}
T_{1\dots n,0}(\ut,\t_0)=\S_{10}(\t_1-\t_0)\,\S_{20}(\t_2-\t_0)\cdots
\S_{n0}(\t_n-\t_0)=
\ba{c}
\unitlength3mm
\bp(10,4)
\put(0,2){\line(1,0){10}}
\put(2,0){\line(0,1){4}}
\put(4,0){\line(0,1){4}}
\put(8,0){\line(0,1){4}}
\put(1,0){$1$}
\put(3,0){$ 2$}
\put(7,0){$ n$}
\put(9,.8){$ 0$}
\put(5,1){$\dots$}
\ep
\ea
\ee
as a matrix acting in the tensor product of the ``quantum space"
$V_{1\dots n}=V_1\otimes\cdots\otimes V_n$ and the ``auxiliary space" $V_0$
(all $V_i\cong{\bf C}^2$ = soliton-antisoliton space).
The Yang-Baxter algebra relations yield
\be{4.2}
T_{1\dots n,a}(\ut,\t_a)\,T_{1\dots n,b}(\ut,\t_b)\,\S_{ab}(\t_a-\t_b)
=\S_{ab}(\t_a-\t_b)\,T_{1\dots n,b}(\ut,\t_b)\,T_{1\dots n,a}(\ut,\t_a)
\ee
which in turn implies  the basic algebraic properties of the sub-matrices
$A,B,C,D$ with respect to  the auxiliary space defined by
\be{4.3}
\def\pT#1#2{
\unitlength3mm
\bp(5,2.3)
\put(0,1){\line(1,0){5}}\put(1,0){\line(0,1){2}}
\put(4,0){\line(0,1){2}}\put(1.7,.5){$\dots$}
\put(.5,1){\vector(#1 1,0){.3}}\put(4.5,1){\vector(#2 1,0){.3}}
\ep}
T_{1\dots n,0}(\ut,\t)\equiv
\left(\matrix{A_{1\dots n}(\ut,\t)&B_{1\dots n}(\ut,\t)\cr
C_{1\dots n}(\ut,\t)&D_{1\dots n}(\ut,\t)}\right)\equiv
\left(\matrix{\pT{-}{-}&\pT{-}{}\cr\pT{}{-}&\pT{}{}}\right)~.
\ee
A Bethe ansatz co-vector in $V_{1\dots n}^\dagger$ is defined by
\be{4.4}
\ba{rcl}
\psi_{1\dots n}(\ut,u_1,\dots,u_m)&=&
\Omega_{1\dots n}C_{1\dots n}(\ut,u_1)\cdots C_{1\dots n}(\ut,u_m)\\
\ba{c}
\unitlength5mm
\bp(6,4)
\put(3,2){\oval(6,2)}
\put(3,2){\makebox(0,0){$\psi$}}
\put(1,0){\line(0,1){1}}
\put(5,0){\line(0,1){1}}
\put(0,0){$\t_1$}
\put(5.3,0){$\t_n$}
\put(2.5,.5){$\dots$}
\ep
\ea
&=&
\ba{c}
\unitlength5mm
\bp(6,4.5)
\put(0,1){\line(1,0){6}}
\put(0,1){\vector(1,0){.5}}
\put(6,1){\vector(-1,0){.5}}
\put(0,3){\line(1,0){6}}
\put(0,3){\vector(1,0){.5}}
\put(6,3){\vector(-1,0){.5}}
\put(1,0){\vector(0,1){4}}
\put(5,0){\vector(0,1){4}}
\put(0,0){$\t_1$}
\put(4,0){$\t_n$}
\put(5.4,.3){$u_m$}
\put(5.4,3.3){$u_1$}
\put(2.5,2){$\dots$}
\put(.3,1.7){$\vdots$}
\put(5.3,1.7){$\vdots$}
\ep
\ea
\ea
\ee
where $\Omega_{1\dots n}$ is the ``pseudo-vacuum'' co-vector 
consisting only of particles of highest weight.
When the monodromy matrix involves 
only the scattering matrix of soliton antisolitons it is given as
\be{4.5}
\Omega_{1\dots n}=\uparrow\otimes\cdots\otimes\uparrow
\ee
consisting only of solitons and fulfilling
\be{4.6}
\ba{rcl}
\Omega_{1\dots n}\,B_{1\dots n}(\ut,u)&=&0\\
\Omega_{1\dots n}\,A_{1\dots n}(\ut,u)
&=&\ds\prod_{i=1}^n\da(\t_i-u)\Omega_{1\dots n}\\
\Omega_{1\dots n}\,D_{1\dots n}(\ut,u)
&=&\ds\prod_{i=1}^n\db(\t_i-u)\Omega_{1\dots n}\,.
\ea
\ee
Here the eigenvalues of the matrices A and D, i.e. $\da$ and $\db$
are related to the amplitudes of the scattering matrix (refer 
(\ref{2.16})
via $\da=-a$ and $\db=-b$.

In the conventional Bethe ansatz  \cite{Betherev}, one is usually concerned
with the computation of the eigenvalues of the transfer matrix
\begin{equation}
\tau_{1 \ldots n}(\ut,u) = A_{1 \ldots n} (\ut,u) + D_{1 \ldots n} (\ut,u) 
\end{equation}
on a Bethe wave vector. Applying the transfer matrix to the co-vector
(\ref{4.4}) one obtains in general an equation of the form
\begin{eqnarray}
\label{4.8}
\lefteqn{
\psi_{1\dots n}(\ut,u_1,\dots,u_m)\,\tau_{1 \ldots n} (\ut,u)~=~
\Lambda(u | u_1,\dots | \ut)\,\psi_{1\dots n}(\ut,u_1,\dots,u_m)}\\ 
&&~~~~~~~~~~~~~~~~~-~\sum_{j=1}^m  
\Lambda_j(u_1,\dots,u_m | \ut)\,
\psi_{1\dots n}^j(\ut | u_1,\dots,u_{j-1},u,u_{j+1},\ldots,u_m)
\nonumber
\end{eqnarray}
where $\Lambda(u | u_1,\dots | \ut) $ and 
$\Lambda_j(u_1,\dots,u_m | \ut)$
are some complex valued functions. The co-vectors
$\psi_{1\dots n}^j(\ut | u_1,\dots,u_{j-1},u,u_{j+1},\ldots,u_m)$
are not proportional to the Bethe ansatz vectors
$\psi_{1\dots n}(\ut,u_1,\dots,u_m)$.
Hence in general, that is for an arbitrary set
of the spectral parameter, the Bethe ansatz vector is not
an eigenvector of the transfer matrix. To achieve this one 
usually imposes the validity of the Bethe ansatz equations,
i.e.  $\Lambda_j(u_1,\dots,u_m | \ut)=0~(j=1,\dots,m)$ such that the
so-called ``unwanted terms" vanish and one obtains
a genuine eigenvalue equation for the transfer matrix with
eigenvalue $\Lambda(u | u_1,\dots | \ut)$. In analogy to the 
one particle situation one may refer to such Bethe vectors 
as being ``on-shell" in contrast to the generic situation 
(\ref{4.8})
which is referred to as ``off-shell" \cite{Hrachik1,Hrachik2}.
In order to construct solutions to the properties 
(i)-(v) we shall employ the Bethe vector  (\ref{4.4})
in its ``off-shell" version.

Let us now consider the auxiliary form factor function given by
\be{4.9}
\f_\ua(\t_1,\dots,\t_n)=
\la0\mi\O\mi p_1,\dots,p_n\ra^{in}_\ua~,~~~{\rm for}~~\t_1>\dots>\t_n.
\ee
where the indices $\ua$ refer to solitons and antisolitons.

\begin{figure}[hbt]
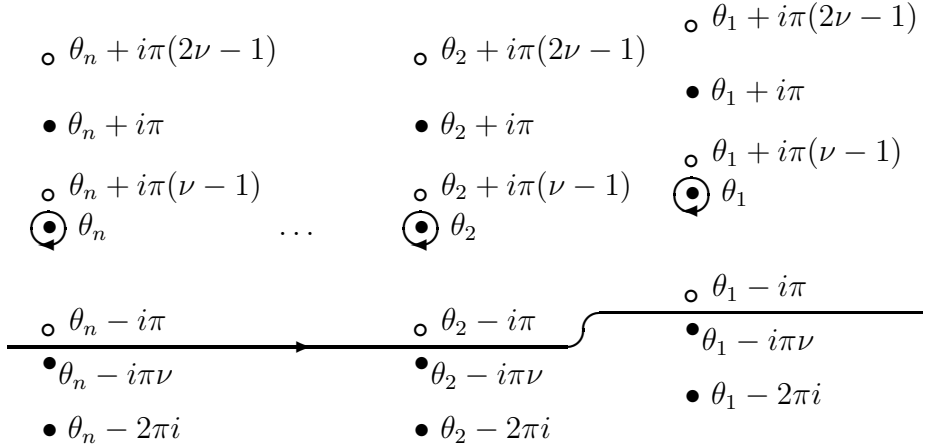

\def\poles#1{
\put(0,0){$\bullet~#1-2\pi i$}
\put(0,2){$\bullet$}\put(.5,1.6){$#1-i\pi\nu$}
\put(.19,3.2){\circle{.3}~$#1-i\pi$}
\put(0,6){$\bullet~~#1$}
\put(.2,6.2){\oval(1,1)}\put(-.1,5.71){\vector(-1,0){0}}
\put(.19,7.2){\circle{.3}~$#1+i\pi(\nu-1)$}
\put(0,9){$\bullet~#1+i\pi$}
\put(.19,11.2){\circle{.3}~$#1+i\pi(2\nu-1)$}
}
$$
\unitlength4.5mm
\bp(27,13)
\thicklines
\put(1,0){\poles{\t_n}}\put(8,6){\dots}\put(12,0){\poles{\t_2}}
\put(20,1){\poles{\t_1}}
\put(9,2.7){\vector(1,0){0}}
\put(0,3.2){\oval(34,1)[br]}\put(27,3.2){\oval(20,1)[tl]}
\ep
$$
\caption{\label{f5.1}\it
The integration contour $\cal C_\ut$ (for the repulsive case $\nu>1$). The
bullets belong to poles of the integrand resulting from
$a(\t_i-u_j)\,\phi(\t_i-u_j)$ and the small open circles belong to poles
originating from $b(\t_i-u_j)$ and $c(\t_i-u_j)$.
}
\end{figure}

\begin{theo}\label{t5}
The co-vector valued function $\f_{1\dots n}(\ut)$ fulfills the conditions
{\rm (i), (ii)} and {\rm (iii)} of section \ref{s3}
(see eqs.~(\ref{3.10}-\ref{3.12}))
if it is  represented by the following generalized Bethe ansatz \cite{BKZ}
\be{4.10}
\f_{1\dots n}(\ut)=N^\O_n
\int_{\cal C_\ut}du_1\cdots\int_{\cal C_\ut}du_m\,g(\ut,\u)\,
\Omega_{1\dots n}C_{1\dots n}(\ut,u_1)\cdots C_{1\dots n}(\ut,u_m)
\ee
with a normalization constant $N^\O_n$ and the scalar function
\be{4.11}
g(\ut,\u)=
\prod_{1\le i<j\le n}F(\t_{ij})
\prod_{i=1}^n\prod_{j=1}^m\phi(\t_i-u_j)
\prod_{1\le i<j\le m}\tau(u_i-u_j)\,
e^{\pm\tilde s\Big(2\sum u_j-\sum\t_i\Big)}
\ee
where  $\tilde s=s/q$ and $s$ is the ``spin" (c.f.~eq.~(\ref{3.15})) and
$q=n-2m$ is the charge of the operator $\O$.
The number $\tilde s$ is assumed to fulfil $\exp(2\pi i\tilde s)=(-1)^n$.
The function $F(\t)$ (see (\ref{4.16})) is a
soliton-soliton form factor fulfilling Watson's equations
\be{4.12}
F(\t)=-F(-\t)\,a(\t)=F(2\pi i-\t)
\ee
with the soliton-soliton scattering amplitude $a(\t)$ (see \ref{2.16}).
The scalar functions $\phi(u)$ and $\tau(u)$ are defined as
\be{4.13}
\phi(u)=\frac1{F(u)\,F(u+i\pi)}~~,~~~
\tau(u)=\frac1{\phi(u)\,\phi(-u)}~.
\ee
The integration contour $\cal C_\ut$ consists of several pieces
(see figure~\ref{f5.1}):
\begin{itemize}
\item[a)]
A line from $-\infty$ to $\infty$ avoiding all poles such that
$\Im\t_i-\pi-\epsilon<\Im u_j<\Im\t_i-\pi$.
\item[b)] Clock wise oriented circles around all poles (of the
$\phi(\t_i-u_j)$) at $u_j=\t_i$.
\end{itemize}
In addition we assume that the number of particles 
involved, i.e. $n$, to be odd. 
\end{theo}
This theorem is proven in appendix \ref{sc}. 

\subsubsection*{Remarks:}
\begin{itemize}
\item
The minus sign in eq.~(\ref{4.12})  is due to the fermionic statistics of the
solitons.
\item
A solution of the Watson's equations (\ref{4.12}) is
\be{4.16}
F(\t)=-i\sinh\h1\t\,f^{min}_{ss}(\t)
\ee
where the 'minimal' soliton-soliton form factor function is given as
$$
f_{ss}^{min}(\t)=\exp\int_0^\infty\frac{dt}t\frac{\sinh\frac12(1-\nu)t}
{\sinh\frac12\nu t\,\cosh\frac12t}\frac{1-\cosh t(1-\t/(i\pi))}{2\sinh t}.
$$
The corresponding functions $\phi(u)$ and $\tau(u)$ are (see appendix \ref{sd})
$$
\phi(u)=const.~\frac1{\sinh u}\exp\int_0^\infty\frac{dt}t
\frac{\sinh\frac12(1-\nu)t\,\Big(\cosh t(\h1-u/(i\pi))-1\Big)}
{\sinh\frac12\,\nu t\sinh t}
$$
$$\tau(u)=const.~\sinh u\sinh u/\nu$$
\item
Using Watson's equations (\ref{4.12}) for $F(u)$, crossing (\ref{2.17})
and unitarity (\ref{2.18}) for the sine-Gordon amplitudes one derives the
following identities for the scalar functions $\phi(u)$ and $\tau(u)$
\be{4.14}
\phi(u)=\phi(i\pi-u)=-\frac1{b(u)}\,\phi(u-i\pi)
=\frac{a(u-2\pi i)}{b(u)}\,\phi(u-2\pi i)~,
\ee
\be{4.15}
\tau(u)=\tau(-u)=
\frac{b(u)}{a(u)}\,\frac{a(2\pi i-u)}{b(2\pi i-u)}\,\tau(u-2\pi i)
\ee
where $b(u)$ is the soliton-antisoliton scattering amplitude related to
$a(u)$ by crossing $b(u)=a(i\pi-u)$.
\item
The number of C-operators $m$ depends on the charge $q=n-2m$ of the operator
$\O$, e.g. $m=(n-1)/2$ for the soliton field $\psi(x)$ with charge $q=1$.
\item
The integrals in eq.~(\ref{4.11}) converge if
$\h1(1+1/\nu)q\mp2\tilde s+2/\nu+1>0$.
\item
Note that other sine-Gordon form factors can be calculated from the general
formula (\ref{4.10}) using the bound state formula (\ref{3.13}).

\end{itemize}
We shall now  apply the general formula (\ref{4.10}) 
to an explicit example and exploit the fact that the
properties (i)-(iv) relate several different form factors
to each other. This will permit us to carry out various consistency checks.

\subsection{The two particle form factors}
We repeat some well known results (see for example \cite{KW,K2}).
According to equation (\ref{3.17}), the auxiliary function for the two 
particle form factor $f_{\alpha\beta}^{\cal O}(\t_{1},\t_{2})$ has to satisfy
$$
\f_{\alpha\beta}(\t_1,\t_2)=\sum_{\alpha'\beta'}
\f_{\beta'\alpha'}(\t_2,\t_1)\,
\S^{\beta'\alpha'}_{\alpha\beta}(\t_{12})
=\f_{\beta\alpha}(\t_2,\t_1-2\pi i)\,\sigma_{{\cal O}1}.
$$
These matrix equations may be  solved easily
by diagonalization of the S-matrix.
If there are only bosons involved we have to 
solve the scalar ``Watson's equations''
\be{4.17}
\f_e(\t)=\f_e(-\t)\,S_e(\t)=\f_e(2\pi i-\t)
\ee
where $S_e(\t)$ are the eigenvalues of the S-matrix given by eq. (\ref{2.11}).
In \cite{KW} it was shown that the general solution of these equations is of
the form
\be{4.18}
\f_e(\t)=\N_e K_e(\t)\,f^{min}_e(\t)
\ee
where $\N_e$ is a normalization factor,
$f^{min}_e(\t)$ is the minimal solution of Watson's equations without
any poles or zeroes in the physical strip $0\le$ Im $\t\le\pi$ and $K_e(\t)$
is an even periodic function with period $2\pi i$.
If the S-matrix eigenvalue is given by
\be{4.19}
S_e(\t)=\exp\int_0^\infty dt\,f(t)\sinh t\t/i\pi
\ee
the minimal solution of Watson's equations is given as
\be{4.20}
f_e^{min}(\t)=\exp\int_0^\infty dt\,f(t)\,
\frac{1-\cosh t(1-\t/i\pi)}{2\sinh t}\,.
\ee
If there are also fermions involved Watson's equations (\ref{4.17}) hold for
the invariant form factors (c.f.~(\ref{3.16})).
For the full matrix elements the representation (\ref{4.18}) holds with
additional factors $\exp(\pm \t_i/2)$ on the right hand side for all fermions.

The poles of $\f_e(\t)$ in the physical strip are determined by the
one-particle states in the channel corresponding to the S-matrix eigenvalue.
In \cite{KW} the minimality assumption was made, meaning 
that there are only these
poles and no zeroes in $0<\Im\,\t<\pi$.
This implies that
$$
K_e(\t)=\prod_{k=1}^L\frac1{\sinh\frac12(\t-\t_k)\sinh\frac12(\t+\t_k)}~,~~
(\Re\,\t_k=0,~0<\Im\,\t_k<\pi)\,.
$$
For several examples this assumption was checked against perturbation theory.

\subsection*{Examples}
We present two-particle form factors for several local operators and several
particle states of the sine-Gordon quantum field theory.
Some of them were already calculated in \cite{KW} (see also \cite{nankai}).
Up to normalizations the problem is solved by eqs.~(\ref{4.18}-\ref{4.20})
since the sine-Gordon S-matrix (\ref{2.15}) is diagonal except of the
soliton-anti-sector where the eigenvalues are given by eqs. (\ref{2.16})
and  (\ref{2.20}).

\subsubsection{The two-breather form factor}
{\def\O{\phi^2}
The simplest sine-Gordon form factor is that for a scalar operator
$\O(x)={\cal N}\phi^2(x)$
connecting the two-particle lowest breather state to the vacuum
$$
\f_{bb}(\t_{12})=\la0\mi\O\mi p_1,p_2\ra^{in}_{bb}
=\N_{bb}\,K_{bb}(\t_{12})\,f_{bb}^{min}(\t_{12})\,.
$$
According to (\ref{4.19}) and (\ref{4.20})
the minimal form factor function combined with (\ref{2.21}) reads
$$
f_{bb}^{min}(i\pi x)=-i\sinh\h1\t\,
\exp\int_0^\infty\frac{dt}t\,
2\frac{\cosh(\frac12-\nu)t}{\cosh\frac12t}\,
\frac{1-\cosh t(1-x)}{2\sinh t}\,.
$$
The ``minimality assumption'' implies that the 'pole function'
$$
K_{bb}(\t)=\frac1{\sinh\frac12(\t-i\pi\nu)\sinh\frac12(\t+i\pi\nu)}
$$
only possesses the pole corresponding to the second breather $b_2$ as a bound
state of two lowest breathers $b$.
The normalization constant can be calculated by means of the asymptotic
behavior \cite{KW}. Weinberg's power counting implies that in the limit of
infinite momentum transfer the form factor tends to its free value
$$
\la0\mi\O\mi p_1,p_2\ra^{in}_{bb}\to\la0\,|:\!\O\!:|\,p_1,p_2\ra^{free}_{bb}
=2\,Z^\phi~~{\rm as}~~(p_1+p_2)^2\to-\infty~.
$$
Here $Z^\phi$ is the wave-function renormalization constant of the fundamental
sine-Gordon field which has been calculated in
\cite{KW}
\be{4.21}
Z^{\phi}=(1+\nu)\frac{\h\pi\nu}{\sin\h\pi\nu}\,E(\nu)~.
\ee
We introduce the function
\ban{4.22}
E(x)&=&\exp\left(-\frac{1}{\pi}\int_{0}^{\pi x}\frac{t}{\sin t}dt\right)\\
&=&\exp\left\{-\frac1{\pi}\bigg(i\,{\rm L}\left(e^{i\pi x}\right)
+i\,{\rm L}\Big(e^{i\pi x}+1\Big)
-x\ln\Big(e^{i\pi x}+1\Big)+\frac{i\pi}{12} \bigg)\right\}\nn
\ean
where $L(x)$ $=\sum\limits_{n=1}^{\infty }\frac{x^{n}}{n^{2}}+\frac{1}{2}\ln
x\ln (1-x)$ denotes the Rogers dilogarithm \cite{Lewin}.
Notice, that this wave function renormalization constant satisfies
(compare figure~\ref{f5}) $0\leq Z^{\phi}\leq 1$, which is a general
consequence of positivity \cite{Le} (see also e.g.~\cite{IZ} p.~204). For the
free boson case, that is $\beta =0,~\nu=0$, we have $Z^{\phi}=1$. For the
free soliton case, i.e.~$g=0,~\nu=1$, where the breather decays
into soliton-antisoliton pairs, we have $Z^{\phi}=0$. 
\begin{figure}[hbt]
$$
\setlength{\unitlength}{0.240900pt}
\ifx\plotpoint\undefined\newsavebox{\plotpoint}\fi
\begin{picture}(850,540)(0,0)
\font\gnuplot=cmr10 at 10pt
\gnuplot
\sbox{\plotpoint}{\rule[-0.200pt]{0.400pt}{0.400pt}}%
\put(220.0,113.0){\rule[-0.200pt]{148.394pt}{0.400pt}}
\put(220.0,113.0){\rule[-0.200pt]{0.400pt}{97.324pt}}
\put(220.0,113.0){\rule[-0.200pt]{4.818pt}{0.400pt}}
\put(198,113){\makebox(0,0)[r]{0}}
\put(816.0,113.0){\rule[-0.200pt]{4.818pt}{0.400pt}}
\put(220.0,194.0){\rule[-0.200pt]{4.818pt}{0.400pt}}
\put(198,194){\makebox(0,0)[r]{0.2}}
\put(816.0,194.0){\rule[-0.200pt]{4.818pt}{0.400pt}}
\put(220.0,275.0){\rule[-0.200pt]{4.818pt}{0.400pt}}
\put(198,275){\makebox(0,0)[r]{0.4}}
\put(816.0,275.0){\rule[-0.200pt]{4.818pt}{0.400pt}}
\put(220.0,355.0){\rule[-0.200pt]{4.818pt}{0.400pt}}
\put(198,355){\makebox(0,0)[r]{0.6}}
\put(816.0,355.0){\rule[-0.200pt]{4.818pt}{0.400pt}}
\put(220.0,436.0){\rule[-0.200pt]{4.818pt}{0.400pt}}
\put(198,436){\makebox(0,0)[r]{0.8}}
\put(816.0,436.0){\rule[-0.200pt]{4.818pt}{0.400pt}}
\put(220.0,517.0){\rule[-0.200pt]{4.818pt}{0.400pt}}
\put(198,517){\makebox(0,0)[r]{1}}
\put(816.0,517.0){\rule[-0.200pt]{4.818pt}{0.400pt}}
\put(220.0,113.0){\rule[-0.200pt]{0.400pt}{4.818pt}}
\put(220,68){\makebox(0,0){0}}
\put(220.0,497.0){\rule[-0.200pt]{0.400pt}{4.818pt}}
\put(343.0,113.0){\rule[-0.200pt]{0.400pt}{4.818pt}}
\put(343,68){\makebox(0,0){0.2}}
\put(343.0,497.0){\rule[-0.200pt]{0.400pt}{4.818pt}}
\put(466.0,113.0){\rule[-0.200pt]{0.400pt}{4.818pt}}
\put(466,68){\makebox(0,0){0.4}}
\put(466.0,497.0){\rule[-0.200pt]{0.400pt}{4.818pt}}
\put(590.0,113.0){\rule[-0.200pt]{0.400pt}{4.818pt}}
\put(590,68){\makebox(0,0){0.6}}
\put(590.0,497.0){\rule[-0.200pt]{0.400pt}{4.818pt}}
\put(713.0,113.0){\rule[-0.200pt]{0.400pt}{4.818pt}}
\put(713,68){\makebox(0,0){0.8}}
\put(713.0,497.0){\rule[-0.200pt]{0.400pt}{4.818pt}}
\put(836.0,113.0){\rule[-0.200pt]{0.400pt}{4.818pt}}
\put(836,68){\makebox(0,0){1}}
\put(836.0,497.0){\rule[-0.200pt]{0.400pt}{4.818pt}}
\put(220.0,113.0){\rule[-0.200pt]{148.394pt}{0.400pt}}
\put(836.0,113.0){\rule[-0.200pt]{0.400pt}{97.324pt}}
\put(220.0,517.0){\rule[-0.200pt]{148.394pt}{0.400pt}}
\put(745,455){\makebox(0,0){$Z^\phi$}}
\put(528,23){\makebox(0,0){$\nu$}}
\put(220.0,113.0){\rule[-0.200pt]{0.400pt}{97.324pt}}
\put(220,517){\usebox{\plotpoint}}
\put(282,515.67){\rule{7.227pt}{0.400pt}}
\multiput(282.00,516.17)(15.000,-1.000){2}{\rule{3.613pt}{0.400pt}}
\put(312,514.67){\rule{7.468pt}{0.400pt}}
\multiput(312.00,515.17)(15.500,-1.000){2}{\rule{3.734pt}{0.400pt}}
\put(343,513.17){\rule{6.300pt}{0.400pt}}
\multiput(343.00,514.17)(17.924,-2.000){2}{\rule{3.150pt}{0.400pt}}
\put(374,511.17){\rule{6.300pt}{0.400pt}}
\multiput(374.00,512.17)(17.924,-2.000){2}{\rule{3.150pt}{0.400pt}}
\multiput(405.00,509.94)(4.429,-0.468){5}{\rule{3.200pt}{0.113pt}}
\multiput(405.00,510.17)(24.358,-4.000){2}{\rule{1.600pt}{0.400pt}}
\multiput(436.00,505.94)(4.283,-0.468){5}{\rule{3.100pt}{0.113pt}}
\multiput(436.00,506.17)(23.566,-4.000){2}{\rule{1.550pt}{0.400pt}}
\multiput(466.00,501.93)(2.751,-0.482){9}{\rule{2.167pt}{0.116pt}}
\multiput(466.00,502.17)(26.503,-6.000){2}{\rule{1.083pt}{0.400pt}}
\multiput(497.00,495.93)(2.013,-0.488){13}{\rule{1.650pt}{0.117pt}}
\multiput(497.00,496.17)(27.575,-8.000){2}{\rule{0.825pt}{0.400pt}}
\multiput(528.00,487.92)(1.590,-0.491){17}{\rule{1.340pt}{0.118pt}}
\multiput(528.00,488.17)(28.219,-10.000){2}{\rule{0.670pt}{0.400pt}}
\multiput(559.00,477.92)(1.210,-0.493){23}{\rule{1.054pt}{0.119pt}}
\multiput(559.00,478.17)(28.813,-13.000){2}{\rule{0.527pt}{0.400pt}}
\multiput(590.00,464.92)(0.888,-0.495){31}{\rule{0.806pt}{0.119pt}}
\multiput(590.00,465.17)(28.327,-17.000){2}{\rule{0.403pt}{0.400pt}}
\multiput(620.00,447.92)(0.740,-0.496){39}{\rule{0.690pt}{0.119pt}}
\multiput(620.00,448.17)(29.567,-21.000){2}{\rule{0.345pt}{0.400pt}}
\multiput(651.00,426.92)(0.574,-0.497){51}{\rule{0.559pt}{0.120pt}}
\multiput(651.00,427.17)(29.839,-27.000){2}{\rule{0.280pt}{0.400pt}}
\multiput(682.58,398.76)(0.497,-0.548){59}{\rule{0.120pt}{0.539pt}}
\multiput(681.17,399.88)(31.000,-32.882){2}{\rule{0.400pt}{0.269pt}}
\multiput(713.58,364.34)(0.497,-0.678){59}{\rule{0.120pt}{0.642pt}}
\multiput(712.17,365.67)(31.000,-40.668){2}{\rule{0.400pt}{0.321pt}}
\multiput(744.58,321.54)(0.497,-0.920){57}{\rule{0.120pt}{0.833pt}}
\multiput(743.17,323.27)(30.000,-53.270){2}{\rule{0.400pt}{0.417pt}}
\multiput(774.58,265.89)(0.497,-1.119){59}{\rule{0.120pt}{0.990pt}}
\multiput(773.17,267.94)(31.000,-66.945){2}{\rule{0.400pt}{0.495pt}}
\multiput(805.58,195.87)(0.497,-1.429){59}{\rule{0.120pt}{1.235pt}}
\multiput(804.17,198.44)(31.000,-85.436){2}{\rule{0.400pt}{0.618pt}}
\put(220.0,517.0){\rule[-0.200pt]{14.936pt}{0.400pt}}
\end{picture}
\setlength{\unitlength}{0.240900pt}
\ifx\plotpoint\undefined\newsavebox{\plotpoint}\fi
\begin{picture}(920,540)(0,0)
\font\gnuplot=cmr10 at 10pt
\gnuplot
\put(176.0,113.0){\rule[-0.200pt]{158.994pt}{0.400pt}}
\put(176.0,113.0){\rule[-0.200pt]{0.400pt}{97.324pt}}
\put(176.0,113.0){\rule[-0.200pt]{4.818pt}{0.400pt}}
\put(154,113){\makebox(0,0)[r]{0}}
\put(816.0,113.0){\rule[-0.200pt]{4.818pt}{0.400pt}}
\put(176.0,194.0){\rule[-0.200pt]{4.818pt}{0.400pt}}
\put(154,194){\makebox(0,0)[r]{0.2}}
\put(816.0,194.0){\rule[-0.200pt]{4.818pt}{0.400pt}}
\put(176.0,275.0){\rule[-0.200pt]{4.818pt}{0.400pt}}
\put(154,275){\makebox(0,0)[r]{0.4}}
\put(816.0,275.0){\rule[-0.200pt]{4.818pt}{0.400pt}}
\put(176.0,355.0){\rule[-0.200pt]{4.818pt}{0.400pt}}
\put(154,355){\makebox(0,0)[r]{0.6}}
\put(816.0,355.0){\rule[-0.200pt]{4.818pt}{0.400pt}}
\put(176.0,436.0){\rule[-0.200pt]{4.818pt}{0.400pt}}
\put(154,436){\makebox(0,0)[r]{0.8}}
\put(816.0,436.0){\rule[-0.200pt]{4.818pt}{0.400pt}}
\put(176.0,517.0){\rule[-0.200pt]{4.818pt}{0.400pt}}
\put(154,517){\makebox(0,0)[r]{1}}
\put(816.0,517.0){\rule[-0.200pt]{4.818pt}{0.400pt}}
\put(176.0,113.0){\rule[-0.200pt]{0.400pt}{4.818pt}}
\put(176,68){\makebox(0,0){0}}
\put(176.0,497.0){\rule[-0.200pt]{0.400pt}{4.818pt}}
\put(308.0,113.0){\rule[-0.200pt]{0.400pt}{4.818pt}}
\put(308,68){\makebox(0,0){0.2}}
\put(308.0,497.0){\rule[-0.200pt]{0.400pt}{4.818pt}}
\put(440.0,113.0){\rule[-0.200pt]{0.400pt}{4.818pt}}
\put(440,68){\makebox(0,0){0.4}}
\put(440.0,497.0){\rule[-0.200pt]{0.400pt}{4.818pt}}
\put(572.0,113.0){\rule[-0.200pt]{0.400pt}{4.818pt}}
\put(572,68){\makebox(0,0){0.6}}
\put(572.0,497.0){\rule[-0.200pt]{0.400pt}{4.818pt}}
\put(704.0,113.0){\rule[-0.200pt]{0.400pt}{4.818pt}}
\put(704,68){\makebox(0,0){0.8}}
\put(704.0,497.0){\rule[-0.200pt]{0.400pt}{4.818pt}}
\put(836.0,113.0){\rule[-0.200pt]{0.400pt}{4.818pt}}
\put(836,68){\makebox(0,0){1}}
\put(836.0,497.0){\rule[-0.200pt]{0.400pt}{4.818pt}}
\put(176.0,113.0){\rule[-0.200pt]{158.994pt}{0.400pt}}
\put(836.0,113.0){\rule[-0.200pt]{0.400pt}{97.324pt}}
\put(176.0,517.0){\rule[-0.200pt]{158.994pt}{0.400pt}}
\put(506,23){\makebox(0,0){$\nu$}}
\put(176.0,113.0){\rule[-0.200pt]{0.400pt}{97.324pt}}
\put(706,452){\makebox(0,0)[r]{$Z^{\phi^2}$}}
\put(176,113){\usebox{\plotpoint}}
\multiput(176.58,113.00)(0.495,0.888){31}{\rule{0.119pt}{0.806pt}}
\multiput(175.17,113.00)(17.000,28.327){2}{\rule{0.400pt}{0.403pt}}
\multiput(193.58,143.00)(0.494,0.849){29}{\rule{0.119pt}{0.775pt}}
\multiput(192.17,143.00)(16.000,25.391){2}{\rule{0.400pt}{0.388pt}}
\multiput(209.58,170.00)(0.495,0.738){31}{\rule{0.119pt}{0.688pt}}
\multiput(208.17,170.00)(17.000,23.572){2}{\rule{0.400pt}{0.344pt}}
\multiput(226.58,195.00)(0.494,0.689){29}{\rule{0.119pt}{0.650pt}}
\multiput(225.17,195.00)(16.000,20.651){2}{\rule{0.400pt}{0.325pt}}
\multiput(242.58,217.00)(0.495,0.588){31}{\rule{0.119pt}{0.571pt}}
\multiput(241.17,217.00)(17.000,18.816){2}{\rule{0.400pt}{0.285pt}}
\multiput(259.58,237.00)(0.494,0.529){29}{\rule{0.119pt}{0.525pt}}
\multiput(258.17,237.00)(16.000,15.910){2}{\rule{0.400pt}{0.263pt}}
\multiput(275.00,254.58)(0.607,0.494){25}{\rule{0.586pt}{0.119pt}}
\multiput(275.00,253.17)(15.784,14.000){2}{\rule{0.293pt}{0.400pt}}
\multiput(292.00,268.58)(0.669,0.492){21}{\rule{0.633pt}{0.119pt}}
\multiput(292.00,267.17)(14.685,12.000){2}{\rule{0.317pt}{0.400pt}}
\multiput(308.00,280.58)(0.860,0.491){17}{\rule{0.780pt}{0.118pt}}
\multiput(308.00,279.17)(15.381,10.000){2}{\rule{0.390pt}{0.400pt}}
\multiput(325.00,290.59)(1.395,0.482){9}{\rule{1.167pt}{0.116pt}}
\multiput(325.00,289.17)(13.579,6.000){2}{\rule{0.583pt}{0.400pt}}
\multiput(341.00,296.60)(2.382,0.468){5}{\rule{1.800pt}{0.113pt}}
\multiput(341.00,295.17)(13.264,4.000){2}{\rule{0.900pt}{0.400pt}}
\multiput(374.00,298.94)(2.382,-0.468){5}{\rule{1.800pt}{0.113pt}}
\multiput(374.00,299.17)(13.264,-4.000){2}{\rule{0.900pt}{0.400pt}}
\multiput(391.00,294.93)(1.022,-0.488){13}{\rule{0.900pt}{0.117pt}}
\multiput(391.00,295.17)(14.132,-8.000){2}{\rule{0.450pt}{0.400pt}}
\multiput(407.00,286.92)(0.712,-0.492){21}{\rule{0.667pt}{0.119pt}}
\multiput(407.00,287.17)(15.616,-12.000){2}{\rule{0.333pt}{0.400pt}}
\multiput(424.58,273.72)(0.494,-0.561){29}{\rule{0.119pt}{0.550pt}}
\multiput(423.17,274.86)(16.000,-16.858){2}{\rule{0.400pt}{0.275pt}}
\multiput(440.58,255.14)(0.495,-0.738){31}{\rule{0.119pt}{0.688pt}}
\multiput(439.17,256.57)(17.000,-23.572){2}{\rule{0.400pt}{0.344pt}}
\multiput(457.58,229.37)(0.494,-0.977){29}{\rule{0.119pt}{0.875pt}}
\multiput(456.17,231.18)(16.000,-29.184){2}{\rule{0.400pt}{0.438pt}}
\multiput(473.58,197.68)(0.495,-1.189){31}{\rule{0.119pt}{1.041pt}}
\multiput(472.17,199.84)(17.000,-37.839){2}{\rule{0.400pt}{0.521pt}}
\multiput(490.58,156.50)(0.494,-1.554){29}{\rule{0.119pt}{1.325pt}}
\multiput(489.17,159.25)(16.000,-46.250){2}{\rule{0.400pt}{0.663pt}}
\put(358.0,300.0){\rule[-0.200pt]{3.854pt}{0.400pt}}
\end{picture}
$$
\caption{\label{f5}\it
The wave function renormalization constants $Z^{\phi}$ and $Z^{\O}$ as a
function of the coupling $\nu=\frac{\beta^2}{8\pi-\beta^2}=\frac1{1+2g/\pi}$.
}
\end{figure}
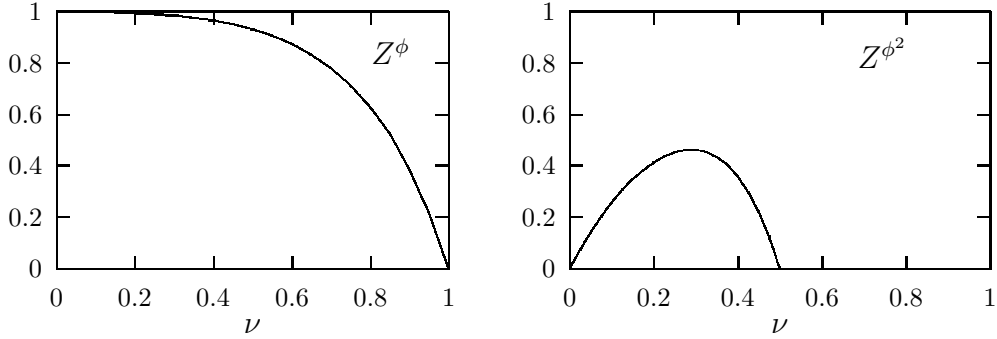
Using the asymptotic formula  
the normalization has been calculated in \cite{KW}
$$
\N_{bb}=-2(1+\nu)\frac\pi 2\nu\cot\frac\pi 2\nu.
$$

In addition we may now employ this result and compute a further
renormalization constant by means of the
bound state formula (\ref{3.13}). With $\varphi^{bb}_{b_2}=1$
(see eq.~(\ref{2.22})) we calculate the wave function renormalization 
constant $Z^{\O}$ via
$$
\Res_{\t=\t_0}\f_{bb}(\t)\Big(2i\Res_{\t=\t_0}S_{bb}(\t)\Big)^{-1/2}
=\f_{b_2}=\la0\mi\O\mi p_1+p_2\ra^{in}_{b_2}=\sqrt{Z^{\O}}
$$
where the fusing angle is $\t_0=i\pi\nu$.
The wave function renormalization constant turns out to be
\be{4.23}
Z^{\O}=\Big(Z^{\phi}\Big)^2\,\frac{\cos\pi\nu}{\cos^2\h\pi\nu}\,E(1-2\nu).
\ee
Again we have $0\leq Z^{\O}\leq 1$ and now $Z^{\O}$ vanishes at $\nu=0$
and $ \nu=1/2$, which is to be expected since at these values 
(compare for instance with the mass formula (\ref{2.14}))
the second breather decays into the two lowest breathers
or a soliton-antisoliton pair, respectively.
}

\subsubsection{The breather-soliton form factor}
We now choose  $\O(x)$ to be the fermi-field  $\psi(x)$ 
of the massive Thirring model
which annihilates the soliton.
{\def\O{\psi}
We assume the  breather-soliton form factor related to this field
$$\f_{bs}(\t_1,\t_2)=\la0\mi\O\mi p_1,p_2\ra^{in}_{bs}$$
 to acquire the following  form
\be{4.24}
\f_{bs}(\t_1,\t_2)=\N_{bs}\,
\Big(1+\N_5\,\gamma^5\,\coth\h1\t_{12}\Big)u(p_2)\,K_{bs}(\t_{12})\,
f_{bs}^{min}(\t_{12})\,,
\ee
that is consisting out of a scalar and a pseudoscalar coupling part. 
For our conventions concerning spinors and the $\gamma$-matrices see
section~5.
Upon employing  (\ref{4.19}) and (\ref{4.20}),
the minimal form factor 
function reads  together with equation (\ref{2.21})
$$
f_{bs}^{min}(i\pi x)=\sin\h\pi x\,\exp\int_0^\infty \frac{dt}t\,
2\frac{\cosh\frac12\nu t}{\cosh\frac12t}\,
\frac{1-\cosh t(1-x)}{2\sinh t}\,.
$$
Extracting explicitly the expected pole structure the ``pole function'' reads
$$
K_{bs}(\t)=\frac1{\sinh\frac12(\t-i\pi\frac{1+\nu}2)\,
\sinh\frac12(\t+i\pi\frac{1+\nu}2)}\,.
$$
Once more we may use the bound state formula (\ref{3.13}) in order to 
compute the normalization constants, with $\varphi^{bs}_s=1$ and 
$\t_0=i\h\pi(\nu+1)$ (see eq.~(\ref{2.22})) to obtain
$$
\Res_{\t_{12}=\t_0}\f_{bs}(\t_1,\t_2)\,
\Big(2i\Res_{\t_{12}=\t_0}S_{bs}(\t_{12})\Big)^{-1/2}
=\f_s=\la0\mi\O\mi p_1+p_2\ra^{in}_{s}=u(p_1+p_2)\,.
$$
This determines the normalization constant to be
$$
\N_{bs}=\frac{\cos^2\frac{\pi}2\nu}{E\Big(\frac12(1-\nu)\Big)}
\sqrt{\frac{E(\nu)}{\sin\frac{\pi}2\nu}}
$$
and the ratio of pseudo-scalar and scalar coupling to be
$$\N_5=-\tan\h\pi\nu\,\tan\mbox{$\frac\pi 4$}(1+\nu).$$
Note that formula (\ref{4.24}) may alternatively also be written as
\bea
\f_{bs}(\t_1,\t_2)
&=&\N_{bs}\,\cos^2\mbox{$\frac{\pi}4$}(1-\nu)\,\frac i{\gamma(p_1+p_2)-m}
\Big(1+\gamma^5\,\coth\h1\t_{12}\Big)u(p_2)\,
f_{bs}^{min}(\t_{12})\\
&=&\N_{bs}\,\frac{\sin\frac{\pi}4(1+\nu)}{\cos^2\frac{\pi}2\nu}
\left(\frac{e^{-\frac12i\pi\nu\gamma^5}}{\sinh\h1(\t_{12}+\t_0)}
+\frac{e^{\frac12i\pi\nu\gamma^5}}{\sinh\h1(\t_{12}-\t_0)}\right)
\frac{u(p_2)\,f_{bs}^{min}(\t_{12})}{\sinh\h1\t_{12}}\,.
\eea
}

\subsubsection{Soliton-antisoliton form factors}
Having an in-state, which  involves a soliton and an antisoliton, we have
several options for the operator $\O(x)$ such
that  the two-partice form factor is non vanishing.
\begin{itemize}
\item[a)]
Let  $\O(x)=j^\mu(x)={\cal N}\bar\psi\gamma^\mu\psi(x)$ the electromagnetic
current \cite{KW}
{\def\O{j^\mu}
$$
\f_{s\bar s}(\t_1,\t_2)=\la0\mi\O\mi p_1,p_2\ra^{in}_{s\bar s}
=\bar v(p_2)\gamma^\mu u(p_1)\,f_-(\t_{12}).
$$
The function $f_-(\t)$ fulfills Watsons equations with the negative $C$-and
$P$-parity S-matrix eigenvalue $S_-(\t)=
-\frac{\cosh\frac12(\t+i\pi)/\nu}{\cosh\frac12(\t-i\pi)/\nu}\,a(\t)$
(see eqs.~(\ref{2.16}-\ref{2.20})).
Taking the singularity structure into account we obtain \cite{KW},
with the help of (\ref{4.19}), (\ref{4.20}) and (\ref{2.16})
$$
f_-(\t)=\frac{\cosh\frac12(i\pi-\t)}{\cosh\frac12(i\pi-\t)/\nu}
\,f_{ss}^{min}(\t)
$$
and
$$
f_{ss}^{min}(i\pi x)=\exp\int_0^\infty\frac{dt}t\frac{\sinh\frac12(1-\nu)t}
{\sinh\frac12\nu t\,\cosh\frac12t}\frac{1-\cosh t(1-x)}{2\sinh t}\,.
$$
}

\item[b)]
Let  $\O(x)=\phi(x)$ the fundamental sine-Gordon field which correspond to the
lowest breather \cite{KW}
{\def\O{\phi}
$$
\f_{s\bar s}(\t_1,\t_2)=\la0\mi\O\mi p_1,p_2\ra^{in}_{s\bar s}
=\N_{s\bar s}\,\bar v(\t_2)u(\t_1)\,\frac1{\sinh\t_{12}}f_-(\t_{12})\,.
$$
The function $f_-(\t)$ is the same as in a).
Since Coleman's correspondence \cite{Co} relates the field 
$\phi$ and the current $j^\mu$ by
\be{4.25}
\epsilon^{\mu\nu}\partial_\nu\phi=-\frac{2\pi}\b j^\mu
\ee
the normalization constant turns out to be  \cite{KW}
$$\N_{s\bar s}=\frac{2\pi i}{\b M}\,.$$
We may now carry out a consistency check and compute once more
the wave function renormalization constant now starting, however, from a
different form factor. For this purpose
we use once again the bound state formula (\ref{3.13}) with
$\varphi^{s\bar s}_b=-\varphi^{\bar ss}_b=1/\sqrt{2}$
(see eq.~(\ref{2.19})) and calculate
$$
\Res_{\t=\t_0}\f_{s\bar s}(\t)\,\sqrt{2}\Big(2i\Res_{\t=\t_0}S_-(\t)
\Big)^{-1/2}=\f_b=\la0\mi\O\mi p\ra^{in}_b=\sqrt{Z^{\O}}
$$
where the fusing angle is $\t_0=i\pi(1-\nu)$. This computation leads to the
value for wave function renormalization constant 
of the previous subsection (\ref{4.21}) which has been obtained  
in \cite{KW} by slightly different arguments.
}
\item[c)]
Let $\O(x)={\cal N}\phi^2(x)$
{\def\O{\phi^2}
$$
\f_{s\bar s}(\t_1,\t_2)=\la0\mi\O\mi p_1,p_2\ra^{in}_{s\bar s}
=\N_{s\bar s}\,\bar v(\t_2)u(\t_1)\,f_+(\t_{12})\,.
$$
The function $f_+(\t)$ fulfills Watsons equations with the positive $C$-and
$P$-parity S-matrix eigenvalue $S_+(\t)=
-\frac{\sinh\frac12(\t+i\pi)/\nu}{\sinh\frac12(\t-i\pi)/\nu}\,a(\t)$
(see eqs.~(\ref{2.16}-\ref{2.20})). With (\ref{4.19}) and (\ref{4.20}) 
we obatin \cite{KW}  together with the explicit expression
for the integral representation of this amplitude of the 
scattering matrix (\ref{2.16})
$$
f_+(\t)=\frac{\sinh\frac12(i\pi-\t)}{\sinh\frac12(i\pi-\t)/\nu}
\,f_{ss}^{min}(\t)\,.
$$
We still have to fix the normalization constant $\N_{s\bar s}$, which may 
be achieved by employing the bound state formula (\ref{3.13}). Taking
$\varphi^{s\bar s}_{b_2}=\varphi^{\bar ss}_{b_2}=1/\sqrt{2}$
(see eq.~(\ref{2.19})) we calculate the wave function renormalization
constant to be
$$
\Res_{\t=\t_0}\f_{s\bar s}(\t)\,\sqrt{2}\,\Big(2i\Res_{\t=\t_0}S_+(\t)
\Big)^{-1/2}=\f_{b_2}=\la0\mi\O\mi p\ra^{in}_{b_2}=\sqrt{Z^{\O}}
$$
where $\t_0=i\pi(1-2\nu)$.
The wave function renormalization constant $Z^{\O}$
was calculated in the previous subsection (\ref{4.23}), such that 
we obtain the normalization constant
$$\N_{s\bar s}=\frac{(1+\nu)\,\pi}{8M\,\sin^3\h\pi\nu}\,.$$
}
\end{itemize}

\subsection{Three particle form factors}
We shall now analyse the general expression proposed in theorem 4.1 for an
explicit example.
First we recall the three breather form factor which was already
calculated in \cite{KW} and apply a consistency checks using (iii).
Furthermore we calculate the three soliton form factor using the general
formula (\ref{4.10}) and apply some consistency checks using (iii) and (iv).
\subsubsection{The three breather form factor}
We choose the operator $\O(x)$ to be the fundamental sine-Gordon field 
$\phi(x)$ which corresponds to the
lowest breather and consider the form factor
{\def\O{\phi}
$$
\f_{bbb}(\t_1,\t_2,\t_3)=\la0\mi\O\mi p_1,p_2,p_3\ra^{in}_{bbb}.
$$
The minimality assumption suggests the proposal \cite{KW}
$$
\f_{bbb}(\t_1,\t_2,\t_3)=\N_{bbb}\,
K_{bbb}(\t_1,\t_2,\t_3)\,
f_{bb}^{min}(\t_{12})\,f_{bb}^{min}(\t_{13})\,f_{bb}^{min}(\t_{23})
$$
with the ``pole function"
$$
K_{bbb}(\t_1,\t_2,\t_3)=\frac1{\cosh\frac12\t_{12}\,\cosh\frac12\t_{13}\,
\cosh\frac12\t_{23}}\,K_{bb}(\t_{12})\,K_{bb}(\t_{13})\,K_{bb}(\t_{23})~.
$$
The ``two-breather pole function"  $K_{bb}(\t)$ and 
the minimal form factor $f_{bb}^{min}(\t)$ were already provided above.
We use property (iii), i.e.  the recursion relation (\ref{3.12}), and
calculate
$$
\Res_{\t_{12}=i\pi}\f_{bbb}(\t_{12},\t_{13},\t_{23})
=2i\sqrt{Z^{\O}}\,(1-S_{bb}(\t_{23}))
$$
which determines the normalization constant \cite{KW}
$$\N_{bbb}=\h1\pi^2\nu^2(1+\nu)^2
\cot\h\pi\nu\,\cos^4\h\pi\nu\,\Big(Z^\phi\Big)^{-3/2}.
$$
}

\subsubsection{The three (anti)-soliton form factor}
We now choose $\O(x)$ to be the fermi-field $\psi^\pm(x)$ of the massive
Thirring model which annihilates the soliton. We consider the form factor
{\def\O{\psi^\pm}
$$
\f_{123}(\t_1,\t_2,\t_3)=\la0\mi\O\mi p_1,p_2,p_3\ra^{in}_{123}\,.
$$
Here $\pm$ refers to the first or second component of the spinor, 
respectively. 
Nonvanishing matrix elements contain two solitons and one antisoliton.
Taking in the general formula (\ref{4.10}) for $n=3$ and $m=1$ we obtain
\be{4.26}
\f_{123}(\ut)=\N_3\prod_{1\le i<j\le 3}F(\t_{ij})
\int_{\cal C}du\prod_{i=1}^3\phi(\t_i-u)\,
e^{\pm\big(\sum u-\sum\t_i/2\big)}\,\Omega_{123}\,C_{123}(\ut,u)\,.
\ee
Here the function $F(\t)$, which
fulfills Watsons equations
$$F(\t)=-F(-\t)\,a(\t)=F(2\pi i-\t)\,,$$
is closely related to the minimal form factor
which was computed above 
$$F(\t)=-i\sinh\h1\t\,f^{min}_{ss}(\t)\,.$$
The scalar function $\phi(u)$ reads
$$
\phi(u)=\frac1{F(u)F(u+i\pi)}\,.
$$
We now use property (iii), i.e.~the recursion relation (\ref{3.12}) 
and calculate
$$
\Res_{\t_{12}=i\pi}\f_{123}(\ut)=2i\,C_{12}\,\f_3\,(1-S_{23}(\t_{23}))
$$
which determines the normalization constant
\be{4.27}
\N_3=\pm i\frac{\sqrt{M}}{4\pi}\,\Big({f^{min}_{ss}}(0)\Big)^2.
\ee
Note that this follows also from the general recursion relation (\ref{B.7}).
The form factor is now fixed with all its constants. However, we also
expect  the bound state formula (\ref{3.13}) to hold  and we may
employ it now as a consistency check. We calculate with $\varphi^{12}_-$ 
given by eq.~(\ref{2.19}) and the fusing angle given by $\t_0=i\pi(1-\nu)$
$$
\Res_{\t_{12}=\t_0}\f_{123}(\t_1,\t_2,\t_3)\,\varphi^{12}_-\,
\Big(2i\Res_{\t=\t_0}S_-(\t)\Big)^{-1/2}
=\f_{b3}(\t_{(12)},\t_3)\,.
$$
The result of this computation coincides with the 
form factor  proposed in  (\ref{4.24}). Having convinced ourselves
of the mutual consistency of  several solutions we shall
now carry out an additional check and compare the 
results with conventional perturbation theory. 
}

\section{Perturbation theory: Massive Thirring model}\label{sa}
In order to check the three particle form factor of the fundamental fermi
field corresponding to the soliton in perturbation theory we calculate the
four point vertex function. We start with the Lagrangian
$${\cal L}^{MTM}
=\bar\psi(i\gamma\partial-M)\psi-\frac12g(\bar\psi\gamma^\mu\psi)^2~.$$
The fermi field $\psi(x)$ annihilates a soliton and creates an antisoliton
with the following normalisation
\be{5.28}
\la0\mi\psi(x)\mi p\ra_{\a}=\delta_{\a s}\,e^{-ipx}\,u(p)~,~~~
\la0\mi\bar\psi(x)\mi p\ra_{\a}=\delta_{\a\bar s}\,e^{ipx}\,\bar v(p).
\ee
We use the following conventions for the $\gamma$-matrices
\be{5.29}
\gamma^0=\left(\matrix{0&1\cr 1&0}\right)~,\quad
\gamma^1=\left(\matrix{0&1\cr -1&0}\right)~,\quad
\gamma^5=\gamma^0\gamma^1=\left(\matrix{-1&0\cr 0&1}\right)
\ee
and for the spinors
\be{5.30}
u(p)=\sqrt{M}\left(\matrix{e^{-\t/2}\cr e^{\t/2}}\right)~,\quad
v(p)=\sqrt{M}\,i\left(\matrix{e^{-\t/2}\cr -e^{\t/2}}\right)\quad
{\rm with}\quad p^\mu
=M\left(\matrix{\sinh\t\cr \cosh\t}\right)~.
\ee
We also employ the formulae
$$\{\gamma^\mu,\gamma^\nu\}=2g^{\mu\nu}~,\quad
[\gamma^\mu,\gamma^\nu]=2\epsilon^{\mu\nu}\gamma^5~,\quad
(\epsilon^{\mu\nu}=-\epsilon^{\nu\mu}~,~~\epsilon^{01}=1)
$$
$$
\gamma^\mu\gamma^\nu\gamma_\mu=0~,\quad
\epsilon^{\mu\rho}\epsilon^{\nu\sigma}=g^{\mu\sigma}g^{\nu\rho}-
g^{\mu\nu}g^{\rho\sigma}~,\quad
\gamma^5\gamma^\mu=\epsilon^{\mu\rho}\gamma_\rho~.
$$
The Lagrangian implies the Feynman rules of figure~\ref{fa1}
\begin{figure}[hbt]
$$
\ba{c}
\unitlength3mm
\bp(4,4)
\put(0,0){\line(1,1){4}}
\put(0,0){\vector(1,1){1}}
\put(2,2){\vector(1,1){1.5}}
\put(4,0){\line(-1,1){4}}
\put(4,0){\vector(-1,1){1}}
\put(2,2){\vector(-1,1){1.5}}
\put(2,2){\makebox(0,0){$\bullet$}}
\ep
\ea
=-ig\gamma^\mu\otimes\gamma_\mu~,~~~
\ba{c}
\unitlength4mm
\bp(6,1)
\put(0,0){\line(1,0){6}}
\put(2,0){\vector(-1,0){0}}
\put(3,.5){$k^\mu$}
\ep
\ea
=\frac i{\gamma k-M}.
$$
\caption{\label{fa1}\it
The Feynman rules for the massive Thiring model.
}
\end{figure}

The three particle matrix element of the fermi field up to order $g$ turns out to be
\bea
\lefteqn{\la0\mi\psi(0)\mi p_1,p_2,p_3\ra_{\bar sss}^{in}
~=~~~
\ba{c}
\unitlength3mm
\bp(4,5)
\put(0,1){\line(1,1){2}}
\put(1,2){\vector(-1,-1){.3}}
\put(2,1){\line(0,1){4}}
\put(2,2){\vector(0,1){.3}}
\put(2,4){\vector(0,1){.3}}
\put(4,1){\line(-1,1){2}}
\put(3,2){\vector(-1,1){.3}}
\put(2,3){\makebox(0,0){$\bullet$}}
\put(2,5){\makebox(0,0){$\bullet$}}
\put(-.6,0){$p_1$}
\put(1.7,0){$p_2$}
\put(4,0){$p_3$}
\ep
\ea}\\
&=&-ig\,\frac i{\gamma(p_1+p_2+p_3)-M}
\,\Big(\gamma^\mu u(p_2)\,\bar v(p_1)\gamma_\mu u(p_3)-
\gamma^\mu u(p_3)\,\bar v(p_1)\gamma_\mu u(p_2)\Big)+O(g^2)\\
&=&-ig\,\sinh\h1\t_{23}\,\frac{u(p_2)\cosh \h1\t_{12}+u(p_3)\cosh \h1\t_{13}}
{\cosh\h1\t_{12}\,\cosh\h1\t_{13}\,\cosh\h1\t_{23}}+O(g^2)\,.
\eea
Note that for the soliton-soliton scattering amplitude this implies
$$
a(\t)=1-ig\,\tanh\h1\t+O(g^2)
$$
in agreement with eq.~(\ref{2.16}).

To calculate the exact form factor up to this order we start from the general
formula (\ref{4.26})
{\def\O{\psi^\pm}
$$
\f_{\bar sss}(\t_1,\t_2,\t_3)=
\int_{{\cal C}_\ut} du\,I(\ut,u)
$$
with the integrand
$$I(\ut,u)=\N_3\prod_{1\le i<j\le 3}F(\t_{ij})\prod_{i=1}^3\phi(\t_i-u)\,
e^{\pm\big(\sum u-\sum\t_i/2\big)}\,\Big(\Omega C(\ut,u)\Big)_{\bar sss}.
$$
Using the residue theorem the integral may be written as
\bea
\int_{{\cal C}_\ut} du\,I(\ut,u)
&=&2\pi i\left(\Res_{\t_1-i\pi}-
\h1\bigg(\Res_{\t_1}+\Res_{\t_2}+\Res_{\t_3}
-\Res_{\t_1+i\pi(\nu-1)}\bigg)\right)I(\ut,u)\\
&&+~\h1\int_{{\cal C}_0} du\,\Big(I(\ut,u)+I(\ut,u+i\pi)\Big)
\eea
where the integration contour ${\cal C}_0$ is a line from
$-\infty$ to $\infty$ avoiding all poles such that
$\Im\t_i+\pi(\nu-2)<\Im u<\Im\t_i$ (for $\nu>1,~\nu=1/(1+2g/\pi)\approx 1$).
The integral on the right hand side is of higher order in $g$ and the residues
give
$$
\f_{\bar sss}(\t_1,\t_2,\t_3)=
\N_3\,\frac{\,\mp 4\pi g}{\sqrt{M}}
\,\sinh\h1\t_{23}\,\frac{u^\pm(p_2)\cosh\h1\t_{12}+u^\pm(p_3)\cosh\h1\t_{13}}
{\cosh\h1\t_{12}\,\cosh\h1\t_{13}\,\cosh\h1\t_{23}}+O(g^2)
$$
which is consistent with the result of the Feynman graph calculation because of
equation (\ref{4.27}) and $f^{min}_{ss}(0)=1+O(g)$. Hence we obtain mutual
consistency between the solutions of the form factor equations and 
conventional perturbation theory.
}
\section{Conclusions}
We have outlined in detail the so-called ``form factor program".
Using only the ``maximal analyticity assumption"
and the validity of the LSZ formalism we have derived general properties 
of form factors.
The properties are expressed in terms of the equations (i)--(v).
We provide a solution for these equations in a closed form, which
captures the vectorial structure by means of the ``off-shell" Bethe
ansatz  and the singularity structure in term of certain contour integrals.
The validity of this solution has been checked by constructing
various explicit two and three particle form factors. We have compared
our solution for the three particle form factor of the fundamental
fermi field with the expressions obtained from
perturbation theory in the massive Thirring model. We find
complete aggrement between these two approaches and we take this
as a further indication for the validity of the ``form factor program"
formalism. 

The vectorial nature of the form factors we present, 
is encapsulated  in the ``off-shell" Bethe ansatz and the singularities
are encoded in certain contour integrals.
We assume that this structure is of a universal nature
and will allow to construct further solutions of other integrable theories.
It will be highly interesting to work out such solutions explicitly. 
This task and the detailed study of the correlation functions 
obtained from these solutions is left to future investigations.  

We have applied the general formula (\ref{4.10}) 
to an explicit example and exploit the fact that the
properties (i)--(iv) relate several different form factors
to each other. This permits us to carry out various consistency
checks. We have for instance the following relations
$$
\ba[t]{lclcl}
f_{b_2s}^{\psi}&\stackrel{(iv)}{\longleftarrow}&
f_{\bar sss}^{\psi }&\stackrel{(iv)}{\longrightarrow }&f_{bs}^{\psi }\\
\downarrow\!{\s(iii)}\!\!\!\!\!&&\downarrow\!{\s(iii)}\!\!\!\!\!&&
\downarrow\!{\s(iv)}\\
f_{s}^{\psi }&=&f_{s}^{\psi }&=&f_{s}^{\psi }
\ea
~~~~~~~
\ba[t]{lclcr}
f_{bbb}^{\phi}&\stackrel{(iii)}{\longrightarrow}&
f_{b}^{\phi }&=&f_{b}^{\phi }\,\\
\downarrow\!{\s(iv)}\!\!\!\!\!&&\uparrow\!{\s(iii)}\!\!\!\!\!&&
\!\!\!\!\!\!\!\!\!{\s(iv)}\!\uparrow~\\
f_{b_2b}^{\phi }&\stackrel{(iv)}{\longleftarrow}&f_{s\bar sb}^{\phi }&
\stackrel{(iv)}{\longrightarrow}&f_{s\bar s}^{\phi }\\
&&&&\!\!\!\!\!\!\!\!\!\!\!\mbox{\scriptsize(\ref{4.25})}\,{\updownarrow}~\\
&&&&f_{s\bar s}^{j^\mu}
\ea
~~~~~~~
\ba[t]{lcr}
f_{s\bar sb}^{\phi^2}&\stackrel{(iv)}{\longrightarrow}&
f_{bb}^{\phi^2}\\
\downarrow\!{\s(iv)}\!\!\!\!\!&&\!\!\!\!{\s(iv)}\!\downarrow~\\
f_{s\bar s}^{\phi^2}&\stackrel{(iv)}{\longrightarrow}&f_{b_2}^{\phi^2}
\ea
$$
Several of these form factor relations and consistency checks have been
presented in this paper. The proof of further relations will be published
elsewhere.
\\[5mm]
{\bf Acknowledgment:} We would like to thank  A.A. Belavin, R. Schrader,
L.H. Kauffman, B. Schroer, F. Smirnov
for useful discussions and comments. H.B. is grateful to the VW project
``Cooperation with scientists from CIS" for financial support.
A.F., M.K. and A.Z. are grateful to the Deutsche
Forschungsgemeinschaft (Sfb288) for support.

\renewcommand{\theequation}{\mbox{\Alph{section}.\arabic{equation}}}
\appendix\section*{Appendix}\setcounter{section}{0}
\setcounter{equation}{0}
\section{Derivation of properties of generalized form factors}\label{sb}
In this appendix we derive the formulae for form factors of section \ref{s3}.
We use LSZ techniques \cite{LSZ} (see e.g. \cite{IZ}) and assume in addition
``maximal analyticity" which means that all
singularities originate from  physical intermediate states.
For simplicity we consider only particles with the same mass $m$.
Generalizations to the case of particles with different masses are obvious.
\subsection{Properties of generalized form factors for the pure bosonic case}
As usual we write the in-field as
\be{A.1}
\phi^{in}_\a(x)=\int\frac{dp}{2\pi2\omega}\left(a^{in}_\a(p)\,
e^{-ipx}+a^{in\,\dagger}_{\bar\a}(p)\,e^{ipx}\right).
\ee
It fulfills the Klein-Gordon equation $(\partial^2+m^2)\,\phi(x)=0$ and
when acting on states of the form (\ref{2.1}) it 
creates anti-particles and annihilates particles.
The commutation rules of the creation and annihilation operators are
\ban{A.2}
[\,a^{in}_{\a'}(p')\,,\,a^{in}_\a(p)]&=&0\\~
[\,a^{in}_{\a'}(p')\,,\,a^{in\,\dagger}_\a(p)]&=&\delta_{\a'\a}\,
2\omega\,2\pi\,\delta(p'-p)=\delta_{\a'\a}\,4\pi\,\delta(\t'-\t).
\ean
Corresponding formulae hold also for the out-field.

For the matrix elements of a local scalar operator $\O=\O(0)$ we have the
LSZ-reduction formulae \cite{LSZ}
\ban{A.4}
\lefteqn{_{\dots\bar\a'_1}^{~out}\la\dots,p'_1\mi\O\mi p_1,\dots\ra^{in}
_{\a_1\dots}}\\
&=&_{\dots\bar\a'_1}^{out}\la\dots,p'_1\mi a^{out\,\dagger}_{\a_1}(p_1)\,
\O\mi\dots\ra^{in}_{\dots}
+i\int d^2x~_{\dots\bar\a'_1}^{~out}\la\dots,p'_1\mi T
\left[ \O\,j^\dagger_{\a_1}(x) \right]
\mi\dots\ra^{in}_{\dots}\,e^{-ip_1x}\nn\\\nn
&=&^{out}_{~\dots}\la\dots\mi\O\,a^{in}_{\a'_1}(p'_1)
\mi p_1,\dots\ra^{in}_{\a_1\dots}
+i\int d^2x~_{~\dots}^{out}\la\dots\mi T
\left[ \O\,j^\dagger_{\a'_1}(x)  \right] \mi
p_1,\dots\ra^{in}_{\a_1\dots}\,e^{ip'_1x}
\ean
where $T$ is the time ordering operator and the source term $j(x)=
(\partial^2+m^2)\,\phi(x)$ is given by the interpolating field $\phi(x)$.
If $p_1$ ($p'_1$) corresponds to an
anti-particle (a particle) $j^\dagger$ has to be replaced by $j$.
On further reductions and combined with the assumption of maximal analyticity
the LSZ-formulae imply the crossing formula (\ref{3.4}) for the connected 
part of the matrix element.
We call a contribution to a matrix element 
$^{out}\la\dots\mi\O\mi p_1,\dots\ra^{in}$
\begin{itemize}
\item {\em disconnected} with respect to $p_1$, if its support 
as a distribution  with respect to $p_1$ is point like and
\item {\em connected} with respect to $p_1$ if it is a boundary value of an
analytic function of the Mandelstam variables $s_{1j}$.
\end{itemize}
With this notation the first terms in eqs.~(\ref{A.4})
are disconnected and the second ones are connected with respect to $p_1$
or $p_1'$, respectively.

If we interchange in eqs.~(\ref{A.4}) $in$ and $out$ the
time ordering is replaced by anti-time ordering.
Comparing eq.~(\ref{3.3}) with (\ref{3.2}) this means that $s+i\epsilon$
is replaced by $s-i\epsilon$. Combined with the completeness of the in- and
out-states we obtain the general Watson's equations (\ref{3.8}).
However, from integrability follows a stronger formula. To show this, we
consider the branch point $s_{12}=(m_1\pm m_2)^2$ separately. For simplicity
we assume $m_1=m_2$.
\begin{lemm}\label{lb1}
Let the S-matrix factorize as denoted in (\ref{3.7}), 
and let $s_{12}$ be in a neighbourhood of $4m^2$
or $0$ and all other $s_{ij}$ away from $4m^2$, then
\ban{A.5}
\F_\ua(s_{12}+i\epsilon,s_{ij}+i\epsilon)
&=&\F_{\ua'}(s_{12}-i\epsilon,s_{ij}+i\epsilon)
\,S^{\a'_2\a'_1}_{\a_1\a_2}(s_{12})~~~{\rm for}~~s_{12}\approx4m^2\\
\F_\ua(s_{12}+i\epsilon,s_{ij}+i\epsilon)
&=&\F_{\ua'}(s_{12}-i\epsilon,s_{ij}+i\epsilon)
~~~{\rm for}~~s_{12}\approx0
\label{A.6}
\ean
with $\ua=(\a_1,\a_2,\dots,\a_n)$ and $\ua'=(\a'_1,\a'_2,\dots,\a_n)$
and $2<i<j\le n$.
Corresponding formulae hold for all other branch points $s_{ij}=4m^2$.
\end{lemm}
{\bf Proof:}
By means of formula (\ref{3.4}) we may cross all particles except 1 and 2
to the left hand side. Using again LSZ we have for the full matrix element
\ban{A.7}
\lefteqn{^{out}\la p_3,\dots,p_n\mi\O\mi p_1,p_2\ra^{in}}\\
&=&^{out}\la p_3,\dots,p_n\mi a^{out\,\dagger}(p_1)\,\O\mi
p_2\ra+i\int d^2x~^{out}\la p_3,\dots,p_n\mi T
\left[\O\,j^\dagger(x) \right]
\mi p_2\ra\,e^{-ip_1x}\nn\\\label{A.8}
\lefteqn{^{out}\la p_3,\dots,p_n\mi\O\mi p_1,p_2\ra^{out}}\\
&=&^{out}\la p_3,\dots,p_n\mi a^{in\,\dagger}(p_1)\,\O\mi
p_2\ra-i\int d^2x~^{out}\la p_3,\dots,p_n\mi T^* \left[ \O\,
j^\dagger(x) \right] \mi p_2\ra\,e^{-ip_1x}~~~~~~\nn
\ean
where we have omitted the indices $\a$ and $T^*$ means anti-time ordering.
The term $^{out}\la p_3,\dots,p_n\mi a^{out\,\dagger}(p_1)\,
\O\mi p_2\ra$ is disconnected as in eq.~(\ref{A.4}), whereas\\
$^{out}\la p_3,\dots,p_n\mi a^{in\,\dagger}(p_1)\,
\O\mi p_2\ra$ in general contains also connected contributions.
However, for factorizing S-matrices this term is given by
$$^{out}\la p_3,\dots,p_n\mi a^{in\,\dagger}(p_1)
\mi q_1,\dots,q_m\ra^{in}\,^{in}\la q_m,\dots,q_1\mi
\O\mi p_2\ra$$
which is disconnected with respect to $p_1$.
Therefore, if we take the connected parts of eqs.~(\ref{A.7}) and
(\ref{A.8}) we obtain as in equations (\ref{3.2}) and (\ref{3.3})
\ban{A.9}
^{out}\la p_3,\dots,p_n\mi\O\mi p_1,p_2\ra^{in}_{conn.}
=\F_\ua(s_{12}+i\epsilon,(t_{rs})_{(1 \le r\le2<s\le n)},
(s_{kl}+i\epsilon)_{(2<k<l\le n)})\\
^{out}\la p_3,\dots,p_n\mi\O\mi p_1,p_2\ra^{out}_{conn.}
=\F_\ua(s_{12}-i\epsilon,(t_{rs})_{(1 \le r\le2<s\le n)},
(s_{kl}+i\epsilon)_{(2<k<l\le n)})
\ean
which implies the first claim.
Moreover, crossing in equations (\ref{A.7}) and (\ref{A.8}) in addition also
particle 2 to the left hand side, by the same arguments we confirm  the second
claim, since $\mi p_1\ra^{in}=\mi p_1\ra^{out}$. 

As a consequence of this lemma and the bose statistics of the particles we
have the property (i) (c.f. eq.~(\ref{3.10}))
\be{A.11}
\f_{\dots ij\dots}(\dots,\t_i,\t_j,\dots)
=\f_{\dots ji\dots}(\dots,\t_j,\t_i,\dots)\,
S_{ij}(\t_i-\t_j).
\ee
Iterating this formula we find that for $\t_1<\dots<\t_n$ the auxiliary
function $\f_\ua(\t)$ yields the matrix element for an $out$-state
\be{A.12}
\f_{\a_1\dots\a_n}(\t_1,\dots,\t_n)
=\f_{\a'_n\dots\a'_1}(\t_n,\dots,\t_1)\,
S^{\a'_n\dots\a'_1}_{\a_1\dots\a_n}(\ut)
=\la0\mi\O\mi p_1,\dots,p_n\ra^{out}_\ua.
\ee

We obtain property (ii) (see eq.~(\ref{3.11})) by comparing the analytic parts
of the crossing relations of the following lemma.
\begin{lemm}\label{lb2}
In terms of the auxiliary functions crossing for the full matrix elements reads
{\rm
\ban{A.13}
\lefteqn{
_{\bar 1}\la p_1\mi\O\mi p_2,\dots, p_n\ra^{in}_{2\dots n}}\\\nn
&=\cases{
\ba{l}\!\!
_{\bar1}\la p_1\mi p_2\ra_2
\,\f_{3\dots n}(\t_3,\dots,\t_n)\\
\hspace{2cm}+~{\bf C}^{\bar11}\,\f_{1\dots n}(\t_1+i\pi_-,\dots,\t_n)
\ea
&for $\t_1\ge\t_2>\dots>\t_n$
\cr
\ba{l}\!\!
_{\bar1}\la p_1\mi p_n\ra_n\,\f_{2\dots n-1}(\t_2,\dots,\t_{n-1})\\
\hspace{2cm}+~\f_{2\dots n1}(\t_2,\dots,\t_n,\t_1-i\pi_-)\,{\bf C}^{1\bar1}
\ea
&for $\t_2>\dots>\t_n\ge\t_1$}
\ean}
where $\pi_-=\pi-\epsilon$.
\end{lemm}
Together with property (i) this lemma implies the general crossing formulae
(\ref{3.14}) for arbitrary ordering of the rapidities.\\[2mm]
{\bf Proof:}
The disconnected contributions in eq.~(\ref{A.13}) follow directly 
from the LSZ - formulae (\ref{A.4}). Moreover
the LSZ - formulae imply that crossing of particle 1 means
$p_1\to-p_1$. In terms of the Mandelstam variables $s$ or the
rapidity differences (see also figure~\ref{f32} for the analytic properties) 
this means $s_{1j}+i\epsilon\to t_{1j}-i\epsilon$ 
or $\t_{1j}\to i\pi-\t_{1j}$. However, since there is no branch cut separating
region II and IV (see (\ref{A.6})), this is equivalent to considering 
$s_{1j}+i\epsilon \to t_{1j}+i\epsilon$ 
or $\t_{1j} \to i\pi+\t_{1j}$.
Hence, because of $\t_{1j}=|\t_1-\t_j|$ we have the equivalences
\bea
\t_{1j}\to i\pi+\t_{1j}&\Leftrightarrow&\t_1\to\t_1+i\pi~~~{\rm for}~~
\t_1>\t_j\\
\t_{1j}\to i\pi+\t_{1j}&\Leftrightarrow&\t_1\to\t_1-i\pi~~~{\rm for}~~
\t_1<\t_j
\eea
which imply the claim.

\begin{figure}[htb]
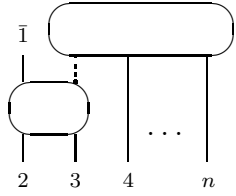

$$
\unitlength3.5mm
\bp(9,7)
\put(2,3){\oval(3,2)}
\put(5.5,6){\oval(7,2)}
\put(1,1){\line(0,1){1}}
\put(1,4){\line(0,1){1}}
\put(3,1){\line(0,1){1}}
\put(5,1){\line(0,1){4}}
\put(8,1){\line(0,1){4}}
\thicklines
\put(3,4){\dashbox{.15}(0,1){}}
\put(5.7,2){$\dots$}
\put(.8,5.5){$\s\bar1$}
\put(.8,0){$\s2$}
\put(2.8,0){$\s3$}
\put(4.8,0){$\s4$}
\put(7.8,0){$\s n$}
\ep
$$
\caption{\label{fb}\it
A graph contributing to a pole of a form factor. The dashed lines denotes an
off-shell line with the propagator $\frac i{(p_2+p_3-p_1)^2-m^2}$.
}
\end{figure}

The form factors have poles determined by one-particle states in any
subchannel\\ $(\a_i,\dots,\a_j)\subset(\a_1,\dots,\a_n)$, if the square of
the total momentum of all particles in the subchannel equals the one-particle
mass squared. We follow the arguments of \cite{KW} and in particular of
\cite{K2} (see also \cite{nankai,YZ}).
A particular type of poles is always present, even if there are no boundstates.
These poles are often refered to as kinematic poles.
If for instance in (\ref{A.13})
particle 1 is the anti-particle of 2, then one-particle intermediate
states with the quantum numbers of all other particles j ($2<j\le n$) yield
contributions to a pole at $p_1\approx p_2$, since then
$(p_2+p_j-p_1)^2\approx m_j^2$ in eq.~(\ref{A.13}), (see figure~\ref{fb}).
The residue of this pole is given by property (iii) of eq.~(\ref{3.12}).
This can be seen as follows.
By (ii) we have for $\t_1\approx\t_2$ and $\t_1\ne\t_j,~(j=3,\dots,n)$
\bea
\f_{12\dots n}(\t_1+i\pi_-,\dots)&\approx&
\frac1{\t_1-i\epsilon-\t_2}\,{\bf C}_{12}\,g(\t_2,\dots,\t_n)\\
\f_{2\dots n1}(\dots,\t_1-i\pi_-)&\approx&
\frac1{\t_1+i\epsilon-\t_2}\,{\bf C}_{12}\,g(\t_2,\dots,\t_n)
\eea
for some function $g(\t_2,\dots,\t_n)$. Employing now the well known
identity $\frac{1}{a \pm i \epsilon} = \frac{P}{a} \mp i \pi
\delta(a)$, with $P$ denoting the principal value, the general crossing
relations (\ref{3.14}) imply for the full matrix element for
$\t_1\approx\t_2$ and $\t_1\ne\t_j,~(j=3,\dots,n)$)
\bea
_{\bar1}\la p_1\mi\O\mi p_2,\dots, p_n\ra^{in}_{2\dots n}
&\approx&_{\bar1}\la p_1\mi p_2\ra_2\,\f_{3\dots n}+
\left(\frac P{\t_1-\t_2}+i\pi\delta(\t_1-\t_2)\right){\bf C}_{12}\,g\\
&\approx&_{\bar1}\la p_1\mi p_2\ra_2\,\f_{3\dots n}\,S_{2n}\cdots S_{23}+
\left(\frac P{\t_1-\t_2}-i\pi\delta(\t_1-\t_2)\right){\bf C}_{12}\,g.
\eea
Comparing the delta-function parts we obtain
$$
g(\t_2,\dots,\t_n)=2i\,\f_{3\dots n}(\t_3,\dots\t_n)\,\Big({\bf 1}-
S_{2n}\cdots S_{23}\Big)
$$
which yields (iii).

If there are also bound states there are additional poles \cite{KW} and 
we have also property (iv). The latter is obvious from figure~\ref{f30} and
eq.~(\ref{2.10}) up to a normalization. The normalization follows from
the following argument developed in \cite{KW}. Let us consider a model with
a bound state of type $c$ of two particles of type $a$ and $b$ such that the
attractive region
is connected analytically (by a coupling constant) to a repulsive region,
where the bound state decays.
For simplicity we consider first
the two-particle form factor
\be{A.14}
\f_{ab}(\t_{ab})=\la0\mi\O(0)\mi p_a,p_b\ra^{in}_{ab}~,~~(\t_a>\t_b)
\ee
such that the (scalar hermitian) operator $\O$ connects
the bound state $c$ with the vacuum
\be{A.15}
\f_{c}(\t_c)=\la0\mi\O(0)\mi p_c\ra_c=\sqrt{Z^\O}\ne0.
\ee
Then in the atractive region of the coupling the two-point Wightman function
reads
\be{A.16}
\la0\mi\O(x)\O(y)\mi0\ra=Z^\O\Delta_+(x-y,m^2_c)+\mbox{contributions from other
masses}
\ee
or the time ordered two-point function in momentum space fulfills
\be{A.17}
\la0\mi\,\tilde\O(k)\O(0)\mi0\ra\approx Z^\O
\frac i{k^2-m_c^2+i\epsilon}~~{\rm at}~~k^2\approx m_c^2
\ee
where $Z^\O$ is a wave function renormalization function.
In the repulsive region the contribution from the two-particle intermediate
states $ab$ is given by
\ban{A.18}
\la0\mi\O(x)\O(y)\mi0\ra
&=&\int\frac{dp_a}{4\pi\omega_a}\frac{dp_b}{4\pi\omega_b}
\frac12\,\la0\mi\O(x)\mi p_a,p_b\ra^{in}_{ab}
~^{in}\la p_a,p_b\mi\O(y)\mi0\ra^{ab}+\dots\nn\\
&=&\frac1{8\pi}\int_{-\infty}^{\infty}d\t\,
\f_{ab}(\t){\f}^{ab}(\t)\Delta_+(x-y,s_{ab})+\dots
\ean
where summation over the multiplets $a$ and $b$ is assumed with
$s_{ab}=m_a^2+m_b^2+2m_am_b\cosh\t$ and 
$\Delta_+(x,m^2)=(2\pi)^{-2}\int d^2ke^{ikx}\Theta(k_0)2\pi\delta(k^2-m^2)$.
In the repulsive region the functions $\f_{ab}(\t)$ and ${\f}^{ab}(\t)$
have poles in the unphysical
sheet at $\pm\t^c_{ab}$ (Im $\t^c_{ab}<0$), respectively. If we move to the
attractive region these poles will cross the integration path and by
analytic continuation we get
\ban{A.19}
\la0\mi\O(x)\O(y)\mi0\ra
&=&\frac1{8\pi}\left\{-\oint_{\t^c_{ab}}+\oint_{-\t^c_{ab}}+
\int_{-\infty}^{\infty}\right\}
d\t\,\f_{ab}(\t){\f}^{ab}(\t)\Delta_+(x-y,s_{ab})+\dots\nn\\
&=&\frac1{4i}\left(\Res_{\t=\t^c_{ab}}-\Res_{\t=-\t^c_{ab}}\right)
\f_{ab}(\t){\f}^{ab}(\t)\Delta_+(x-y,m_c^2)+\dots
\ean
Both residues give the same contribution, because
\bea
\Res_{\t=\t^c_{ab}}\f_{ab}(\t){\f}^{ab}(\t)
&=&\Res_{\t=\t^c_{ab}}\f_{ba}(-\t)\,S_{ab}(\t){\f}^{ab}(\t)=
\f_{ba}(-\t^c_{ab})\varphi^{ba}_cR_c\,\varphi_{ab}^c{\f}^{ab}(\t^c_{ab})\\
\Res_{\t=-\t^c_{ab}}\f_{ab}(\t){\f}^{ab}(\t)
&=&\Res_{\t=-\t^c_{ab}}\f_{ab}(-\t)\,S_{ba}(-\t){\f}^{ba}(-\t)=
-\f_{ab}(-\t^c_{ab})\varphi^{ab}_cR_c\,\varphi_{ba}^c{\f}^{ba}(\t^c_{ab})
\eea
where property (i) eq. (\ref{3.10}) and the residue formulae for the
S-Matrix (\ref{2.10}) and (\ref{2.11}) have been used.
Using this and comparing eqs. (\ref{A.16}) and (\ref{A.19}) we obtain
\bea
Z^\O=\f_{c}(\t_c)\,{\f}^{c}(\t_c)
&=&\frac1{4i}\left(\Res_{\t=\t^c_{ab}}-\Res_{\t=-\t^c_{ab}}\right)
\f_{ab}(\t){\f}^{ab}(\t)\nn\\
&=&
\left(\Res_{\t=\t^c_{ab}}\f_{ab}(\t)\varphi^{ab}_c\frac1{\sqrt{2iR_c}}\right)
\left(\frac1{\sqrt{2iR_c}}\Res_{\t=-\t^c_{ab}}\varphi_{ab}^c{\f}^{ab}(\t)
\right)
\eea
which agrees with (iv).
The general case may be proven similarly.


\subsection{Properties of form factors for the general case}
We now consider the case in which the particles are taken to be
fermions and the operators may be of fermionic or bosonic nature.
Again we use LSZ techniques \cite{LSZ} and ``maximal analyticity".
The two component fermionic in-field is
\be{A.20}
\psi^{in}_\a(x)=\int\frac{dp}{2\pi2\omega}\left(a^{in}_\a(p)\,u(p)\,
e^{-ipx}+a^{in\,\dagger}_{\bar \a}(p)\,v(p)\,e^{ipx}\right)
\ee
and fulfills the Dirac equation $(i\gamma\partial-m)\,\psi(x)=0$.
The anti-commutation relations are
\be{A.21}\ba{ccl}
\{\,a^{in}_{\a'}(p')\,,\,{a^{in}}_\a(p)\}&=&0\\
\{\,a^{in}_{\a'}(p')\,,\,a^{in\,\dagger}_\a(p)\}&=&\delta_{\a'\a}\,
2\omega\,2\pi\,\delta(p'-p)=\delta_{\a'\a}\,4\pi\,\delta(\t'-\t).
\ea\ee
Corresponding formulae hold for the out-field.
We use the conventions for the $\gamma$-matrices and the spinors
of (\ref{5.29}) and (\ref{5.30}).

The LSZ-reduction formulas for fermions read
\ban{A.22}\nn
_{\dots\bar\a'_1}^{~out}
\lefteqn{\la\dots,p'_1\mi\O\mi p_1,\dots\ra^{in}_{\a_1\dots}}\\
&=&\sigma_{\O\a_1}~_{\dots\bar\a'_1}^{out}\la\dots,p'_1\mi
a^{out\,\dagger}_{\a_1}(p_1)\,\O\mi\dots\ra^{in}_{\dots}\nn\\
&&\hspace{1cm}
+~i\int d^2x~_{\dots\bar\a'_1}^{~out}\la\dots,
p'_1\mi T \left[ \O\,\bar \jmath_{\a_1}(x) \right]
\mi\dots\ra^{in}_{\dots}\,u(p_1)\,e^{-ip_1x}\\
&=&\sigma_{\O\a'_1}~^{out}_{~\dots}\la\dots\mi\O\,a^{in}_{\a'_1}(p'_1)
\mi p_1,\dots\ra^{in}_{\a_1\dots}\nn\\
&&\hspace{1cm}-~i\,\sigma_{\O\a'_1}\int d^2x~_{~\dots}^{out}\la\dots\mi
T \left[ \O\,\bar \jmath_{\a'_1}(x) \right]
\mi p_1,\dots\ra^{in}_{\a_1\dots}\, v(p'_1)\,e^{ip'_1x}\label{A.23}
\ean
where
$\bar \jmath (x)=\bar\psi(x)\,
(i\gamma{\ds\mathop\partial^{\leftarrow}}
+m)$ and $\sigma_{\O\a}=-1$ if $\O$ is fermionic and $\sigma_{O\a}=1$
otherwise. Obviously $\sigma_{\O\a}=(-1)^n$ if n is the total
number of fermions in the states. A similar formula holds if we interchange
particles and anti-particles. The invariant form factors
$G^{(l)\,\O}(s_{ij}+i\epsilon)$ defined by eq.~(\ref{3.13})
are again boundary values of analytic functions. Again, if we interchange
$in$ and $out$ time ordering is replaced by anti-time ordering, which
means again that $s_{ij}+i\epsilon$ is replaced by $s_{ij}-i\epsilon$.
The crossing relation for the connected part of the matrix element reads
\ban{A.24}
\lefteqn{^{out}_{\bar\a_1\dots\a_m}\la p_1,\dots,p_m\mi\O\mi
p_{m+1},\dots,p_n\ra^{in~{\rm conn.}}_{\a_{m+1}\dots\bar\a_n}}\nn\\
&=&(-1)^m\prod_{i=1}^m\sigma_{\O\a_i}\,\sum\bar v(p_n)\cdots\bar u(p_m)\,
\Gamma_{\mu_1\dots\mu_k}\,u(p_{m+1})\cdots v(p_1)\,
p^{\mu_1}_{i_1}\dots p^{\mu_k}_{i_k}\\\nn&&
G^\O_{\a_1\dots\bar\a_n}(s_{ij}+i\epsilon,t_{rs}-i\epsilon,s_{kl}+i\epsilon)
\ean
Watson's equations for the
invariant form factor functions $G$ acquire the same form
 as those for $F$ for the bosonic
case (\ref{3.8}). Also Lemma \ref{lb1} holds for the invariant form factors
$G$.

Analogously to the bosonic case it is convenient to introduce the
vector valued auxiliary function $\f_\ua(\ut)$ which is
considered as an analytic function of the rapidities of the particles.
Its components coincide again with the physical matrix elements
for a particular order of the rapidities.
\be{A.25}
\f_\ua(\t_1,\dots,\t_n):=\la0\mi\O\mi p_1,\dots,p_n\ra^{in}_\ua
~,~~~{\rm for}~~\t_1>\dots>\t_n.
\ee
In the other sectors the function $\f_\ua(\ut)$ is again given by 
analytic continuation.
Again, as a consequence of Lemma \ref{lb1} and the fermi statistics of the
particles we have now the property (i) in the form (c.f. eq.~(\ref{3.17}))
\be{A.26}
\f_{\dots ij\dots}(\dots,\t_i,\t_j,\dots)
=\f_{\dots ji\dots}(\dots,\t_j,\t_i,\dots)\,
(-S)_{ij}(\t_i-\t_j).
\ee

The LSZ-formulae (\ref{A.22}) and (\ref{A.23}) imply the general crossing
formulae (\ref{3.18}). Note that some signs in these formulae depend on the
choice of the relative phases of the $u$- and the $v$-spinors taken in
eq.~(\ref{5.30}). The crossing formulae again, as for the bosonic case,
implies the properties (ii) and (iii) as given by (\ref{3.17}), where we have
used the fact that $\sigma_{\O1}=\sigma_{2n}\cdots\sigma_{23}$, if
particle 1 has the same statistics as particle 2.

\setcounter{equation}{0}
\section{Proof of the theorem \ref{t5}}
\label{sc}
In the proof of theorem \ref{t5} we follow \cite{BKZ} (see also \cite{Resh}).

\subsubsection*{Proof that eq.~(\ref{4.10}) fulfills (i):}
Property (i) (c.f.~eq.~(\ref{3.17})) follows directly from the Yang-Baxter
equations, the definitions
of the soliton-soliton S-matrix (\ref{2.15}) and the pseudo-ground state
$\Omega$ and Watson's equations for $F(\t)$
\bea
F(\t_{ji})\,\Omega_{\dots ji\dots}\,
C_{\dots ji\dots}({\dots\t_j,\t_i\dots})\,\S_{ij}(\t_{ij})
&=&F(\t_{ji})\,\Omega_{\dots ji\dots}\,\S_{ij}(\t_{ij})\,
C_{\dots ij\dots}({\dots\t_i,\t_j\dots})\\
&=&-F(\t_{ji})\,a(\t_{ij})\,\Omega_{\dots ij\dots}\,
C_{\dots ij\dots}({\dots\t_i,\t_j\dots})\\
&=&F(\t_{ij})\,\Omega_{\dots ij\dots}\,
C_{\dots ij\dots}({\dots\t_i,\t_j\dots}).
\eea
The minus sign is due to the fermi statistics of the solitons
(c.f. (\ref{3.1}))

\subsubsection*{Proof that eq.~(\ref{4.10}) fulfills (ii):}
Using (i) the property (ii) (c.f.~eq.~(\ref{3.17})) may be rewritten as a
difference equation
\be{B.1}
\f_{1\dots n}(\ut)=\f_{1\dots n}(\ut')\,Q_{1\dots n}(\ut)
\ee
where $\ut'=(\t_1,\dots,\t_n'=\t_n-2\pi i)$ and $Q(\ut)$ is the trace of
the monodromy matrix (\ref{4.1}) over the auxiliary space for the specific
value of the spectral parameter $\t_0=\t_n$
\be{B.2}
Q_{1\dots n}(\ut)={\rm tr}_0T^Q_{1\dots n}(\ut)~,~~{\rm with}~~
T^Q_{1\dots n}(\ut)=T_{1\dots n}(\ut,\t_n)
\ee
since $S_{n0}(0)=P_{n0}$ is the permutation operator.
This may be depicted as
$$
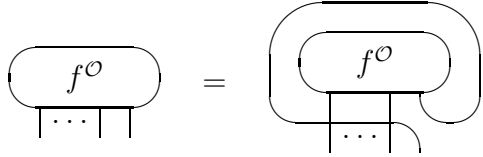
\ba{c}
\unitlength4mm
\bp(5,4)
\put(2.5,2){\oval(5,2)}
\put(2.5,2){\makebox(0,0){$\f$}}
\put(1,0){\line(0,1){1}}
\put(3,0){\line(0,1){1}}
\put(4,0){\line(0,1){1}}
\put(1.4,.5){$\dots$}
\ep
\ea
~~=~~
\ba{c}
\unitlength4mm
\bp(7,5)
\put(6,2){\oval(2,2)[b]}
\put(3.5,2){\oval(7,6)[t]}
\put(3.5,2){\oval(7,2)[lb]}
\put(3.5,0){\oval(3,2)[rt]}
\put(3.5,3){\oval(5,2)}
\put(3.5,3){\makebox(0,0){$\f$}}
\put(2,0){\line(0,1){2}}
\put(4,0){\line(0,1){2}}
\put(2.4,.5){$\dots$}
\ep
\ea
$$
In the following we will suppress the indices $1\dots n$.
The Yang-Baxter relations (\ref{4.2}) imply the well known commutation rules
for the matrices $A,C,D$ defined in eq.~(\ref{4.3})
\ban{B.3}
C(\ut,u)C(\ut,v)&=&C(\ut,v)C(\ut,u)\nn\\
C(\ut,u)A(\ut,\t)&=&\frac{a(\t-u)}{b(\t-u)}A(\ut,\t)C(\ut,u)
-\frac{c(\t-u)}{b(\t-u)}A(\ut,u)C(\ut,\t)\\\nn
C(\ut,u)D(\ut,\t)&=&\frac{a(u-\t)}{b(u-\t)}D(\ut,\t)C(\ut,u)
-\frac{c(u-\t)}{b(u-\t)}D(\ut,u)C(\ut,\t)
\ean
In addition there are commutation rules where also the matrices $A^Q,C^Q,D^Q$
defined by
$$T^Q(\ut)=\left(\matrix{A^Q(\ut)&B^Q(\ut)\cr C^Q(\ut)&D^Q(\ut)}\right)$$
are involved \cite{BKW}
\ban{B.4}
C(\ut',u)A^Q(\ut)&=&\frac{a(\t_n'-u)}{b(\t_n-u)}A^Q(\ut)C(\ut,u)
-\frac{c(\t_n-u)}{b(\t_n-u)}A(\ut',u)C^Q(\ut)\\\nn
C(\ut',u)D^Q(\ut)&=&\frac{a(u-\t_n)}{b(u-\t_n')}D^Q(\ut)C(\ut,u)
-\frac{c(u-\t_n')}{b(u-\t_n')}D(\ut',u)C^Q(\ut)\,.
\ean
To analyze the right hand side of eq.~(\ref{B.1}) we proceed as
follows:
We apply the trace of $T^Q(\ut)$ to the co-vector
$\f(\ut')$ as given by eq.~(\ref{4.10}) and the Bethe ansatz (\ref{4.4}).
In the contribution from $A^Q(\ut)$
$$
\Omega\,C(\ut',u_1)\cdots C(\ut',u_m)\,A^Q(\ut)=
\ba{c}
\unitlength5mm
\bp(8,5.5)
\put(0,2){\line(1,0){8}}
\put(0,2){\vector(1,0){.5}}
\put(8,2){\vector(-1,0){.5}}
\put(0,4){\line(1,0){8}}
\put(0,4){\vector(1,0){.5}}
\put(8,4){\vector(-1,0){.5}}
\put(1,0){\vector(0,1){5}}
\put(6,0){\vector(0,1){5}}
\put(7.6,5){\oval(1.2,8)[lb]}
\put(1,1){\vector(-1,0){1}}
\put(7,4){\vector(0,1){1}}
\put(0,0){\oval(14,2)[rt]}
\put(8,1.){\vector(-1,0){.5}}
\put(1.3,0){$\t_1$}
\put(4.3,0){$\t_{n-1}$}
\put(7.2,0){$\t_{n}$}
\put(7.4,2.3){$u_m$}
\put(7.4,4.3){$u_1$}
\put(3,3){$\dots$}
\put(.3,2.7){$\vdots$}
\put(7.8,1.1){$\t_n'$}
\ep
\ea
$$
because of charge conservation only the amplitudes $a(\t_n'-u_j)$
appear in the S-matrices $\dot S(\t_n'-u_j)$ which are constituents of the
C-operators. Therefore we may shift all $u_j$-integration
contours ${\cal C}_{\ut'}$ to ${\cal C}_{\ut}$ without changing the values of
the integrals, because the functions $a(\t_n'-u_j)\phi(\t_n'-u_j)$ are
holomorphic inside ${\cal C}_{\ut'}-{\cal C}_{\ut}$. 

We now proceed as usual in the algebraic Bethe ansatz and push the $A^Q(\ut)$
and $D^Q(\ut)$ through all the C-operators using
the commutation rules (\ref{B.4}) and obtain
\be{B.5}
C(\ut',u_1)\cdots C(\ut',u_m)\,A^Q(\ut)=\prod_{j=1}^m
\frac{a(\t_n'-u_j)}{b(\t_n-u_j)}\,A^Q(\ut)\,C(\ut,u_1)\cdots C(\ut,u_m)
+uw_A\,,
\ee
\be{B.6}
C(\ut',u_1)\cdots C(\ut',u_m)\,D^Q(\ut)=\prod_{j=1}^m
\frac{a(u_j-\t_n)}{b(u_j-\t_n')}\,D^Q(\ut)\,C(\ut,u_1)\cdots C(\ut,u_m)
+uw_D\,.
\ee
The ``wanted terms" written explicitly originate from the first term in the
commutations rules (\ref{B.4}); all other contributions
yield the so-called ``unwanted terms".
If we insert these equations into the representation (\ref{4.10}) of
$f(\ut')$ we find that the desired contribution from $A^Q$ already gives
the  result we are looking for. The wanted contribution from $D^Q$ applied to $\Omega$
gives zero. The unwanted contributions cancel after integration over the $u_j$.
All these three facts can be seen as follows. We have
$$
\Omega\,A^Q(\ut)=\prod_{i=1}^n\da(\t_i-\t_n)\,\Omega~, ~~~\Omega\,D^Q(\ut)=0
$$
which follow from eq.~(\ref{4.6}).

The relations (\ref{4.14}) for $\phi(u)$ and (\ref{4.12}) for $F(\t)$ imply
that the wanted term from $A^Q$ yields $f(\ut)$.
The commutation relations (\ref{B.3}) and (\ref{B.4}) imply that
the unwanted terms are proportional to a product of $C$-operators, where
exactly one $C(\ut,u_j)$ is replaced by $C^Q(\ut)$.
Because of the commutativity of the $C$s  it is sufficient to
consider only the unwanted terms for $j=m$ which are denoted by $uw_A^m(\u)$
and $uw_D^m(\u)$.
They come from the second term in (\ref{B.4}) when $A^Q(\ut)$ is commuted
with $C(\ut,u_m)$. Then the resulting $A(\ut,u_m)$ pushed through the
other $C$s and taking only the first terms in (\ref{B.3}) into account
and correspondingly for $D^Q(\ut;u_m)$ we arrive at
$$
uw_A^m(\u)=-\frac{c(\t_n-u_m)}{b(\t_n-u_m)}\,
\prod_{j<m}\frac{a(u_m-u_j)}{b(u_m-u_j)}\,A(\ut',u_m)\,
C(\ut',u_1)\dots C^Q(\ut)
$$
$$
uw_D^m(\u)=-\frac{c(u_m-\t_n')}{b(u_m-\t_n')}\,
\prod_{j<m}\frac{a(u_j-u_m)}{b(u_j-u_m)}\,D(\ut',u_m)\,
C(\ut',u_1)\dots C^Q(\ut).
$$
Using again (\ref{4.6}) and the relations (\ref{4.14}) and (\ref{4.15}) for
$\phi(u)$ and $\tau(u)$ we obtain
$$
g(\ut',\u)\,\Omega\,uw_D^m(\u)=-g(\ut',\u')\,\Omega\,uw_A^m(\u')
$$
where also $c(u)/b(u)=-c(-u)/b(-u)$ has been used and $\u'=(u_1,\dots,u_m'=
u_m+2\pi i)$.
Therefore after integration of the A-unwanted term along ${\cal C}_\ut$
and the D-unwanted term along ${\cal C}_{\ut'}$ both cancel.


\subsubsection*{Proof that eq.~(\ref{4.10}) fulfills (iii):}
We will prove that eq.~(\ref{4.10}) fulfills (iii) (see eq.~(\ref{3.17}))
in the form of
$$
\Res_{\t_{1n}=i\pi}\f_{1\dots n}(\t_1,\dots,\t_n)=-\sigma_{\O n}\,2i\,
{\bf C}_{1n}\,\f_{2\dots n-1}(\t_2,\dots,\t_{n-1})\Big(
{\bf 1}_{2\dots n-1}-S_{2n}\cdots S_{n-1n}\Big)
$$
which is equivalent to eq.~(\ref{3.17}) due to (i).
We consider the $n$-particle form factor function given by eq.~(\ref{4.10})
\bea
\f_{1\dots n}(\ut)
=\prod_{i=1}^m\left(\int_{{\cal C}_\ut}du_i\right)\,g_{n}(\ut,\u)\,
\Omega_{1\dots n}C_{1\dots n}(\ut,u_1)\cdots C_{1\dots n}(\ut,u_m)
\eea
with the scalar function
\bea
g_{n}(\ut,\u)&=&\frac{N^\O_n}{N^\O_{n-2}}\,
g_{n-2}(\tilde\ut,\tilde\u)\\
&&\times F(\t_1-\t_n)\,\prod_{i=2}^{n-1}\Big(F(\t_1-\t_i)\,F(\t_i-\t_n)\Big)
\prod_{i=1}^{n}\phi(\t_i-u_m)\\
&&\times\prod_{j=1}^{m-1}\Big(\phi(\t_1-u_j)\,\phi(\t_n-u_j)\,
\tau(u_j-u_m)\Big)\ e^{\pm\tilde s(2u_m-\t_1-\t_n)}
\eea
where $\tilde\ut=\t_2,\dots,\t_{n-1}$ and $\tilde\u=u_1,\dots,u_{m-1}$.
We calculate the residue of this function
at $\t_1=\t_n+i\pi$. It consists of three terms
$$
\Res_{\t_1=\t_n+i\pi}\f_{1\dots n}(\ut)=R_1+R_2+R_3
$$
This is because each of the
$m$ integration contours will be ``pinched" at three points:
\begin{itemize}
\item[(1)] $u_j=\t_n=\t_1-i\pi$, 
\item[(2)] $u_j=\t_n+i\pi=\t_1$
\item[(3)] $u_j=\t_n-i\pi=\t_1-2i\pi$.
\end{itemize}
Due to symmetry it is sufficient to determine the contribution from the
$u_m$-integration and multiply the result by $m$.

The contribution of (1) is given by $u_m$-integration along
the small circle around $u_m=\t_n$ (see figure~\ref{f5.1}). The S-matrix
$S(\t_n-u_m)$ yields the permutation operator $S(0)=P$ and
$S(\t_1-u_m)$ the annihilation-creation operator $S(i\pi)=K$ 
$$
S_{\a\b}^{\delta\gamma}(0)=\delta_{\a\delta}\,\delta_{\b\gamma}=
\ba{c}
\unitlength1.5mm
\bp(8,8)
\put(2,2){\oval(4,4)[rt]}
\put(6,6){\oval(4,4)[lb]}
\put(3.5,0){$\a$}
\put(7,3.5){$\b$}
\put(3.5,7){$\gamma$}
\put(0,3.5){$\delta$}
\ep
\ea~~,~~~~
S_{\a\b}^{\delta\gamma}(i\pi)=
{\bf C}_{\a\b}\,{\bf C}^{\delta\gamma}=
\delta_{\a\bar\b}\,\delta_{\delta\bar\gamma}=
\ba{c}
\unitlength1.5mm
\bp(8,8)
\put(2,6){\oval(4,4)[rb]}
\put(6,2){\oval(4,4)[lt]}
\put(3.5,0){$\a$}
\put(7,3.5){$\b$}
\put(3.5,7){$\gamma$}
\put(0,3.5){$\delta$}
\ep~.
\ea
$$
Therefore we have for $u_m=\t_n=\t_1-i\pi$
\bea
\Omega_{1\dots n}\,C_{1\dots n}(\ut,u_1)\cdots
C_{1\dots n}(\ut,\t_n)
=
\ba{c}
\unitlength6mm
\bp(8,5.5)
\put(0,2){\line(1,0){8}}
\put(0,2){\vector(1,0){.5}}
\put(1,2){\vector(1,0){.5}}
\put(7,2){\vector(-1,0){.8}}
\put(8,2){\vector(-1,0){.8}}
\put(0,4){\line(1,0){8}}
\put(0,4){\vector(1,0){.5}}
\put(1,4){\vector(1,0){.5}}
\put(7,4){\vector(-1,0){.8}}
\put(8,4){\vector(-1,0){.5}}
\put(2,0){\vector(0,1){5}}
\put(6,0){\vector(0,1){5}}
\put(4,0){\oval(6,2)[t]}
\put(.4,5){\oval(1.2,8)[rb]}
\put(1,4){\vector(0,1){1}}
\put(0,1){\vector(1,0){.5}}
\put(7.6,5){\oval(1.2,8)[lb]}
\put(7,4){\vector(0,1){1}}
\put(8,1){\vector(-1,0){.5}}
\put(.3,0){$\t_1$}
\put(2.2,0){$\t_2$}
\put(4.7,0){$\t_{n-1}$}
\put(7.2,0){$\t_n$}
\put(7.4,1.2){$u_m$}
\put(7.4,4.3){$u_1$}
\put(2.6,1.2){$u_m=\t_n$}
\put(7.4,1.2){$v$}
\put(3.5,3){$\dots$}
\put(.3,2.7){$\vdots$}
\ep
\ea~~~~~~~~~~~~~~~\\
=\prod_{j=1}^{m-1}\Big(\db(\t_1-u_j)\da(\t_n-u_j)\Big)\,
{\bf C}_{1n}\,\Omega_{2\dots n-1}C_{2\dots n-1}(\tilde\ut,u_1)\cdots\\
\cdots C_{2\dots n-1}(\tilde\ut,u_{m-1})\,\S_{2n}\cdots\S_{n-1n} 
\eea
where ${\bf C}_{1n}$ is the charge conjugation matrix with ${\bf C}_{\a\b}=
\delta_{\bar\a\b}$. We have  used the fact that because of charge conservation
the amplitude $b(\cdot)$ only contributes to the S-matrices
$S(\t_1-u_j)$ and $a(\cdot)$ to the S-matrices $S(\t_n-u_j)$.

We combine this with the scalar function $g_n$ and after having performed
the remaining $u_j$-integrations we obtain
\bea
R_{1\dots n}^{(1)}
&=&{\bf C}_{1n}\,\f_{2\dots n-1}(\tilde\ut)\,\S_{2n}\cdots\S_{n-1n}\\
&&\times m\,\frac{N^\O_n}{N^\O_{n-2}}\,
\Res_{\t_1=\t_n+i\pi}(-2\pi i)\Res_{u_m=\t_n}\da(\t_n-u_m)\phi(\t_n-u_m)\,
\db(\t_1-u_m)\phi(\t_1-u_m)\\
&&\times F(i\pi)\,\prod_{i=2}^{n-1}\Big(F(\t_1-\t_i)F(\t_i-\t_n)
\phi(\t_i-u_m)\Big)\\
&&\times\prod_{j=1}^{m-1}\Big(\db(\t_1-u_j)\phi(\t_1-u_j)
\da(\t_n-u_j)\phi(\t_n-u_j)
\tau(u_j-u_m)\Big)\,e^{\pm\tilde s(2u_m-\t_1-\t_n)}\\
&=&2i\,\sigma_{\O n}\,{\bf C}_{1n}\,\f_{2\dots n-1}(\tilde\ut)\,
S_{2n}\cdots S_{n-1n}
\eea
if we relate the normalization constants by the recursion relation
\be{B.7}
N^\O_n=N^\O_{n-2}e^{\pm i\pi\tilde s}\frac{\Big(f^{min}_{ss}(0)\Big)^2}{4\pi m}~.
\ee
We have also used that $\sigma_{\O n}=(-1)^n$ and
$$F(u)\,F(u+i\pi)\,\phi(u)=1 ~~~{\rm and}~~~~
b(u+i\pi)\,\phi(u+i\pi)\,a(u)\,\phi(u)\,\tau(-u)=1$$
which follows from the definitions (\ref{4.12}) and (\ref{4.13}).
Finally we have used the normalization $F(i\pi)=1$ and
$\Res_{u=0}\da(u)\phi(u)=\Res_{u=i\pi}\db(u)\phi(u)=-2i/f^{min}_{ss}(0)$
because of eq.~(\ref{4.16}).
Note also that the signs from the $\da$s and $\db$s cancel and
$\sigma_{\O n}=(-1)^{n-2}$, since all particles are fermions.

The remaining contribution to (iii) is due to $R_2$ and $R_3$
$$
R_{1\dots n}^{(2)}+R_{1\dots n}^{(3)}
=2i\,{\bf C}_{1n}\,\f_{2\dots n-1}(\tilde\ut).
$$
If both particles at 1 and n are solitons both vanish. If one particle at 1 or
n is an soliton
and the other an anti-soliton one term gives the desired expression and the
other vanish. If both particles at 1 and n
are anti-solitons both terms cancel. These fact can be proven as follows.

The contribution of (2) is given by the $u_m$-integration along
the small circle around $u_M=\t_1$ (see again figure~\ref{f5.1}). Now
$S(\t_1-u_m)$ yields the  permutation operator $S(0)=P$ and
the co-vector part of this contribution for $u_m=\t_1=\t_n+i\pi$ is
\ban{B.8}
\lefteqn{\Omega_{1\dots n}\,C_{1\dots n}(\ut,u_1)\cdots
C_{1\dots n}(\ut,u_m=\t_1)\,P_n(s)}\nn\\
&&=
\ba{c}
\unitlength6mm
\bp(8,5.5)
\put(8,2){\line(-1,0){8}}
\put(8,2){\vector(-1,0){.5}}
\put(8,4){\line(-1,0){8}}
\put(8,4){\vector(-1,0){0.5}}
\put(0,2){\vector(1,0){0.6}}
\put(0,4){\vector(1,0){0.6}}
\put(7,0){\vector(0,1){5}}
\put(6,0){\vector(0,1){5}}
\put(2,0){\vector(0,1){5}}
\put(1,4){\vector(0,1){1}}
\put(.4,0.00){\oval(1.2,2)[rt]}
\put(0,1){\vector(1,0){.5}}
\put(8,5){\oval(14,8)[lb]}
\put(8,1){\vector(-1,0){.5}}
\put(.3,0){$\t_1$}
\put(1.3,0){$\t_2$}
\put(4.7,0){$\t_{n-1}$}
\put(7.4,4.3){$u_1$}
\put(7.6,1.2){$u_m$}
\put(3.6,1.2){$u_m$}
\put(3.5,3){$\dots$}
\put(.3,2.7){$\vdots$}
\put(7.3,-.1){$\t_n$}
\ep
\ea
~~~=~~~
\ba{c}
\unitlength6mm
\bp(8,5.5)
\put(0,2){\line(1,0){8}}
\put(0,2){\vector(1,0){.5}}
\put(0,3.5){\line(1,0){8}}
\put(0,3.5){\vector(1,0){.5}}
\put(8,2){\vector(-1,0){.6}}
\put(8,3.5){\vector(-1,0){.6}}
\put(6,0){\vector(0,1){5}}
\put(5,0){\vector(0,1){5}}
\put(2,0){\vector(0,1){5}}
\put(1,4.6){\vector(0,1){.4}}
\put(.4,0){\oval(1.2,2)[rt]}
\put(0,1){\vector(1,0){.5}}
\put(7.6,3){\oval(1.2,4)[lb]}
\put(8,1){\vector(-1,0){.5}}
\put(4,3){\oval(6,2.2)[rt]}
\put(4,4.6){\oval(6,1)[lb]}
\put(.3,0){$\t_1$}
\put(7.1,2.5){$\t_1$}
\put(1.3,0){$\t_2$}
\put(3.7,0){$\t_{n-1}$}
\put(7.4,3.8){$u_1$}
\put(7.6,1.2){$u_m$}
\put(3,4.4){$u_m$}
\put(3,2.7){$\dots$}
\put(.3,2.5){$\vdots$}
\put(6.3,-.1){$\t_n$}
\ep
\ea\nn\\
&&=\prod_{i=1}^n\da(\t_i-u_m)\,
\prod_{j=1}^{m-1}\Big(\da(\t_1-u_j)\da(\t_n-u_j)\Big)\nn\\
&&~~~~~\times{\bf C}_{1n}\,\Omega_{2\dots n-1}
C_{2\dots n-1}(\tilde\ut,u_1)\cdots C_{2\dots n-1}(\tilde\ut,u_{m-1}) 
\,P_1(\bar s)\,P_n(s)
\ean
where the Yang-Baxter relation (\ref{2.8}) has been used iteratively.
$P_1(\bar s)$ and $P_n(s)$ project onto the components where the particle at
1 is a soliton and at $n$ is an anti-soliton, respectively.
The remaining components if both particles at 1 and n are
anti-solitons are calculated below.

We combine this with the scalar function $g_n$ and after having performed
the remaining $u_j$-integrations we obtain
\bea
\lefteqn{R_{1\dots n}^{(2)}\,P_n(s)
\ =\ {\bf C}_{1n}\,\f_{2\dots n-1}(\tilde\ut)\,P_1(\bar s)\,P_n(s)}\\
&&\times m\,\frac{N^\O_n}{N^\O_{n-2}}\,
\Res_{\t_1=\t_n+i\pi}(-2\pi i)\Res_{u_m=\t_1}\da(\t_1-u_m)\phi(\t_1-u_m)\,
\da(\t_n-u_m)\phi(\t_n-u_m)\\
&&\times F(i\pi)\,\prod_{i=2}^{n-1}\Big(F(\t_1-\t_i)F(\t_i-\t_n)
\da(\t_i-u_m)\phi(\t_i-u_m)\Big)\\
&&\times\prod_{j=1}^{m-1}\Big(\da(\t_1-u_j)\phi(\t_1-u_j)
\da(\t_n-u_j)\phi(\t_n-u_j)
\tau(u_j-u_m)\Big)\,e^{\pm\tilde s(2u_m-\t_1-\t_n)}\\
&=&-2i\,\sigma_{\O n}\,{\bf C}_{1n}\,\f_{2\dots n-1}(\tilde\ut)
\,P_1(\bar s)\,P_n(s)~.
\eea
if we apply the condition $\exp(2\pi\tilde s)=(-1)^n=\sigma_{\O n}$ and
if we relate the normalization constants as above.
We have used the identities
\be{B.9}
F(-u)\,F(u+i\pi)\,\da(u)\,\phi(u)=1~~{\rm and}~~
a(u)\,\phi(u)\,a(u-i\pi)\,\phi(u-i\pi)\,\tau(-u)=1.
\ee

The contribution of (3) is given by $u_m$-integration along
the small circle around $u_m=\t_n-i\pi$ (see again figure~\ref{f5.1}). Now
$S(\t_n-u_m)$ yields the  annihilation-creation operator $S(i\pi)=K$ and
the co-vector part of this contribution for $u_m=\t_1-2\pi i=\t_n-i\pi$ is
\ban{B.10}
\lefteqn{\Omega_{1\dots n}\,C_{1\dots n}(\ut,u_1)\cdots
C_{1\dots n}(\ut,u_m=\t_n-i\pi)\,P_1(s)}\nn\\
&&=
\ba{c}
\unitlength6mm
\bp(8,5.5)
\put(0,2){\line(1,0){8}}
\put(0,2){\vector(1,0){0.5}}
\put(0,4){\line(1,0){8}}
\put(0,4){\vector(1,0){0.5}}
\put(8,2){\vector(-1,0){0.6}}
\put(8,4){\vector(-1,0){0.6}}
\put(1,0){\vector(0,1){5}}
\put(2,0){\vector(0,1){5}}
\put(6,0){\vector(0,1){5}}
\put(7,4){\vector(0,1){1}}
\put(7.6,0){\oval(1.2,2)[lt]}
\put(8,1){\vector(-1,0){0.5}}
\put(0,5){\oval(14,8)[rb]}
\put(0,1){\vector(1,0){.5}}
\put(.3,0){$\t_1$}
\put(1.3,0){$\t_2$}
\put(4.7,0){$\t_{n-1}$}
\put(7.4,4.3){$u_1$}
\put(7.6,1.2){$u_m$}
\put(3.6,1.2){$u_m$}
\put(3.5,3){$\dots$}
\put(.3,2.7){$\vdots$}
\put(7.3,-.1){$\t_n$}
\ep
\ea
~~~=~~~
\ba{c}
\unitlength6mm
\bp(8,5.5)
\put(0,2){\line(1,0){8}}
\put(0,2){\vector(1,0){.5}}
\put(0,3.5){\line(1,0){8}}
\put(0,3.5){\vector(1,0){.5}}
\put(8,2){\vector(-1,0){.6}}
\put(8,3.5){\vector(-1,0){.6}}
\put(2,0){\vector(0,1){5}}
\put(3,0){\vector(0,1){5}}
\put(6,0){\vector(0,1){5}}
\put(7,4.6){\vector(0,1){.4}}
\put(7.6,0){\oval(1.2,2)[lt]}
\put(8,1){\vector(-1,0){.5}}
\put(.4,3){\oval(1.2,4)[rb]}
\put(0,1){\vector(1,0){.5}}
\put(4,3){\oval(6,2.2)[lt]}
\put(4,4.6){\oval(6,1)[rb]}
\put(1.3,0){$\t_1$}
\put(2.3,0){$\t_2$}
\put(4.7,0){$\t_{n-1}$}
\put(7.4,3.8){$u_1$}
\put(7.6,1.2){$u_m$}
\put(4,4.4){$u_m$}
\put(4,2.7){$\dots$}
\put(.3,2.5){$\vdots$}
\put(7.3,-.1){$\t_n$}
\put(1.1,2.5){$\t_n$}
\ep
\ea\nn\\
&&=\prod_{i=1}^n\db(\t_i-u_m)\,
\prod_{j=1}^{m-1}\Big(\db(\t_1-u_j)\db(\t_n-u_j)\Big)\nn\\
&&~~~~\times{\bf C}_{1n}\,\Omega_{2\dots n-1}
C_{2\dots n-1}(\tilde\ut,u_1)\cdots C_{2\dots n-1}(\tilde\ut,u_{m-1}) 
\,P_1(s)\,P_n(\bar s)
\ean
if the particle at 1 is a soliton and that at n an anti-soliton. This
contribution obviously vanishes if n is a soliton.
Again the remaining components if both particles at 1 and n are
anti-solitons are calculated below.

Again we combine this with the scalar function $g_n$ and after having performed
the remaining $u_j$-integrations we obtain
\bea
\lefteqn{R_{1\dots n}^{(3)}\,P_1(s)
\ =\ {\bf C}_{1n}\,\f_{2\dots n-1}(\tilde\ut)\,P_1(s)\,P_n(\bar s)}\\
&&\times m\,\frac{N^\O_n}{N^\O_{n-2}}\,
\Res_{\t_1=\t_n+i\pi}(-2\pi i)\Res_{u_m=\t_n-i\pi}\db(\t_n-u_m)\phi(\t_n-u_m)\,
\db(\t_1-u_m)\phi(\t_1-u_m)\\
&&\times F(i\pi)\,\prod_{i=2}^{n-1}\Big(F(\t_1-\t_i)F(\t_i-\t_n)
\db(\t_i-u_m)\phi(\t_i-u_m)\Big)\\
&&\times\prod_{j=1}^{m-1}\Big(\db(\t_1-u_j)\phi(\t_1-u_j)
\db(\t_n-u_j)\phi(\t_n-u_j)
\tau(u_j-u_m)\Big)\,e^{\pm\tilde s(2u_m-\t_1-\t_n)}\\
&=&-2i\,\sigma_{\O n}\,{\bf C}_{1n}\,\f_{2\dots n-1}(\tilde\ut)
\,P_1(s)\,P_n(\bar s)
\eea
provided that  we fix the normalization constants as above.
We have used the identities
\be{B.11}
\db(u)\,F(-u)\,F(u+i\pi)\,\phi(u)=1~~{\rm and}~~
a(u)\,\phi(u)\,a(u-i\pi)\,\phi(u-i\pi)\,\tau(-u)=1.
\ee


Finally we calculate $R_{1\dots n}^{(2)}+R_{1\dots n}^{(3)}$ for the case that
both particles at 1 and at n are anti-solitons.
Instead of eq.~(\ref{B.8}) we have now for $u_m=\t_1=\t_n+i\pi$
\bea
&&\prod_{i=1}^n\da(\t_i-u_m)\,
\prod_{j=1}^{m-1}\Big(\da(\t_1-u_j)\da(\t_n-u_j)\Big)
\frac{c(\t_n-u_{m-1})}{a(\t_n-u_{m-1})}\\
&&~~~\times\Omega_{2\dots n-1}
C_{2\dots n-1}(\tilde\ut,u_1)\cdots C_{2\dots n-1}(\tilde\ut,u_{m-2})\,
D_{2\dots n-1}(\tilde\ut,u_{m-1})\,P_1(\bar s)\,P_n(\bar s)+\dots\\
&&
\eea
and because of the Yang-Baxter relations (\ref{B.6})
\bea
&&C(\tilde\ut,u_1)\cdots C(\tilde\ut,u_{m-2})\,D(\tilde\ut,u_{m-1})\\
&&\hspace{2cm}=\prod_{j=1}^{m-2}\frac{a(u_j-u_{m-1})}{b(u_j-u_{m-1})}
\,D(\tilde\ut,u_{m-1})\,C(\tilde\ut,u_1)\cdots C(\tilde\ut,u_{m-2})+\dots
\eea
where the dots refer to similar terms with $D_{2\dots n-1}(\tilde\ut,u_j),
~(j<m-1)$. Because of symmetry with respect to the C-operators it is
sufficient to consider only this term.

Similarly we get instead of eq.~(\ref{B.10}) $u_m=\t_1-2\pi i=\t_n-i\pi$
\bea
&&\prod_{i=1}^n\db(\t_i-u_m)\,
\prod_{j=1}^{m-1}\Big(\da(\t_1-u_j)\da(\t_n-u_j)\Big)
\frac{c(\t_1-u_{m-1})}{b(\t_1-u_{m-1})}\\
&&~~~\times\Omega_{2\dots n-1}
C_{2\dots n-1}(\tilde\ut,u_1)\cdots C_{2\dots n-1}(\tilde\ut,u_{m-2})\,
A_{2\dots n-1}(\tilde\ut,u_{m-1})\,P_1(\bar s)\,P_n(\bar s)+\dots
\eea
and because of the Yang-Baxter relations (\ref{B.6})
\bea
&&C(\tilde\ut,u_1)\cdots C(\tilde\ut,u_{m-2})\,A(\tilde\ut,u_{m-1})\\
&&\hspace{2cm}=\prod_{j=1}^{m-2}\frac{a(u_{m-1}-u_j)}{b(u_{m-1}-u_j)}
\,A(\tilde\ut,u_{m-1})\,C(\tilde\ut,u_1)\cdots C(\tilde\ut,u_{m-2})+\dots
\eea
where the dots again refer to similar terms with $A_{2\dots n-1}(\tilde\ut,u_j),
~(j<m-1)$.
We apply $D_{2\dots n-1}(\tilde\ut,u_{m-1})$ and
$A_{2\dots n-1}(\tilde\ut,u_{m-1})$ to the pseudo-vacuum,
use as above the identities (\ref{B.9}) and (\ref{B.11}),
and find that the sum $R^{(2)}+R^{(3)}$ is proportional to (with $u=u_{m-1}$)
\bea
\int_{{\cal C}_{\tilde\ut}}du\left\{
\frac{c(\t_n-u)}{a(\t_n-u)}
\prod_{i=2}^{n-1}\db(\t_i-u)\phi(\t_i-u)\,
\prod_{j=1}^{m-2}\frac{a(u_j-u)}{b(u_j-u)}
\tau(u_j-u)\right.~~~~~~~~~~\\
+\left.
\frac{c(\t_1-u)}{b(\t_1-u)}
\prod_{i=2}^{n-1}\da(\t_i-u)\phi(\t_i-u)\,
\prod_{j=1}^{m-2}\frac{a(u-u_j)}{b(u-u_j)}
\tau(u_j-u)\right\}=I
\eea
Due to crossing we have
$$
\frac{c(\t_n-u_{m-1})}{a(\t_n-u_{m-1})}=
-\frac{c(\t_1-u_{m-1}-2\pi i)}{b(\t_1-u_{m-1}-2\pi i)}.
$$
In addition we use the quasi-periodicity properties (\ref{4.14}) and
(\ref{4.15}) of $\phi(\cdot)$ and $\tau(\cdot)$ and get
\bea
I&=&\left\{-\int_{{\cal C}_{\tilde\ut}+2\pi i}+
\int_{{\cal C}_{\tilde\ut}}\right\}du_{m-1}
\frac{c(\t_1-u_{m-1})}{b(\t_1-u_{m-1})}\\
&&\prod_{i=2}^{n-1}\da(\t_i-u_{m-1})\phi(\t_i-u_{m-1})\,
\prod_{j=1}^{m-2}\frac{a(u_{m-1}-u_j)}{b(u_{m-1}-u_j)}
\tau(u_j-u_{m-1})
\eea
which vanishes since the integrand is holomorphic inside the contour
${\cal C}_{\tilde\ut}-({\cal C}_{\tilde\ut}+2\pi i)$.

\setcounter{equation}{0}
\section{Some useful formulae}\label{sd}
In this appendix we provide some explicit formulae
(which partly may also be found elsewhere in the
literature)  for
typical scattering matrices, minimal form factors and
some auxiliary functions which we frequently employed
in the explicit computations. We state some typical
integral representation, which are very useful since via (\ref{4.19})
and (\ref{4.20}) they relate the scattering matrix
and the minimal form factors effortlessly. The infinite product
representations in terms of Gamma functions, obtained from  
the evaluation of the integrals or the direct solution of the
functional relations, make the singularity structure more
transparent. For numerical purposes it is often more 
useful to express the Gamma functions  with
the help of Euler's product representation in terms of
rational functions at the cost of an additional infinite
product.

A typical S-matrix eigenvalue is (for $a>0$)
$$
S(i\pi x,a)=\frac{a+x}{a-x}=
\exp\int_0^\infty\frac{dt}t\,2e^{-ta}\sinh tx~.
$$
According to (\ref{4.19}) and (\ref{4.20}) the corresponding 
minimal form factor function is therefore
\bea
f^{min}(i\pi x,a)
&=&\exp\int_0^\infty\frac{dt}t\,2e^{-ta}\frac{1-\cosh t(1-x)}{2\sinh t}\\
&=&\prod_{l=0}^\infty\frac{(2l+2+a-x)(2l+a+x)}{(2l+1+a)^2}
=\frac{\Gamma^2(\frac12+\frac a2)}
{\Gamma(1+\frac a2-\frac x2)\Gamma(\frac a2+\frac x2)}.
\eea
In particular, for $a=0$ we recover the scattering matrix of the Ising model
$$
S=-1~~\to~~f^{min}(i\pi x)=\sin\frac\pi2x
$$
For negative values of $a$ we use $S(\t,a)=1/S(\t,-a)$
and $f^{min}(\t,a)=1/f^{min}(\t,-a)$.
A further typical S-matrix eigenvalue\footnote{For instance
almost all diagonal scattering matrices related to
perturbation of certain conformal field theories may be built
out of these elementary blocks.} is (for $0<a<1$)
$$
S(i\pi x,a)=\frac{\sin\frac\pi 2(a+x)}{\sin\frac\pi 2(a-x)}=
\exp\int_0^\infty\frac{dt}t\,2\frac{\sinh t(1-a)}{\sinh t}\sinh tx~.
$$
with the corresponding minimal form factor function
\bea
f^{min}(i\pi x,a)
&=&\exp\int_0^\infty\frac{dt}t\,2\frac{\sinh t(1-a)}{\sinh t}\,
\frac{1-\cosh t(1-x)}{2\sinh t}\\
&=&\prod_{k=0}^\infty\prod_{l=0}^\infty\,
\frac{2l+2k+a+x}{2l+2k+2-a+x}\,\frac{2l+2k+2+a-x}{2l+2k+4-a-x}
\left(\frac{2l+2k+3-a}{2l+2k+1+a}\right)^2\\
&=&
\prod_{k=0}^\infty
\frac{\Gamma(k+1-\frac a2+\frac x2)}{\Gamma(k+\frac a2+\frac x2)}\,
\frac{\Gamma(k+2-\frac a2-\frac x2)}{\Gamma(k+1+\frac a2-\frac x2)}
\left(\frac{\Gamma(k+\frac12+\frac a2)}{\Gamma(k+\frac32-\frac a2)}\right)^2~.
\eea

The sine-Gordon soliton-soliton S-matrix  reads
\bea
&&S_{ss}(i\pi x)=a(i\pi x)=
\exp\int_0^\infty\frac{dt}t\,\frac{\sinh\frac12(1-\nu)t}
{\sinh\frac12\nu t\,\cosh\frac12t}\,\sinh tx\\
&=&\prod_{k=0}^\infty\prod_{l=0}^\infty
\frac{2l+\nu+k\nu+x}{2l+\nu+k\nu-x}\,
\frac{2l+2+k\nu+x}{2l+2+k\nu-x}\,
\frac{2l+1+\nu+k\nu-x}{2l+1+\nu+k\nu+x}\,
\frac{2l+1+k\nu-x}{2l+1+k\nu+x}\\
&=&\prod_{k=0}^\infty
\frac{\Gamma((\nu+k\nu-x)/2)}{\Gamma((\nu+k\nu+x)/2)}\,
\frac{\Gamma((2+k\nu-x)/2)}{\Gamma((2+k\nu+x)/2)}\,
\frac{\Gamma((1+\nu+k\nu+x)/2)}{\Gamma((1+\nu+k\nu-x)/2)}\,
\frac{\Gamma((1+k\nu+x)/2)}{\Gamma((1+k\nu-x)/2)}
\eea
with
$$a(i\pi(x+\nu))=-\cot\frac\pi 2x\,\cot\frac\pi 2(x+\nu)\,a(i\pi x).$$
Consequently, the  minimal form factor function is

\bea
\lefteqn{f_{ss}^{min}(i\pi x)
=\exp\int_0^\infty\frac{dt}t\frac{\sinh\frac12(1-\nu)t}
{\sinh\frac12\nu t\,\cosh\frac12t}\frac{1-\cosh t(1-x)}{2\sinh t}}\\
&=&\prod_{k=0}^\infty\prod_{l=0}^\infty\prod_{m=0}^\infty
\frac{2m+2l+2+\nu+k\nu-x}{2m+2l+3+\nu+k\nu-x}~
\frac{2m+2l+\nu+k\nu+x}{2m+2l+1+\nu+k\nu+x}\\
&&~~~~~~~~~~~\times
\frac{2m+2l+4+k\nu-x}{2m+2l+3+k\nu-x}~
\frac{2m+2l+2+k\nu+x}{2m+2l+1+k\nu+x}
\\
&&~~~~~~~~~~~\times
\left(\frac{2m+2l+2+\nu+k\nu}{2m+2l+1+\nu+k\nu}~
\frac{2m+2l+2+k\nu}{2m+2l+3+k\nu}\right)^2
\\
&=&\prod_{k=0}^\infty\prod_{l=0}^\infty
\frac{\Gamma\Big(l+\frac12(3+\nu+k\nu-x)\Big)}
{\Gamma\Big(l+1+\frac12(\nu+k\nu-x)\Big)}~
\frac{\Gamma\Big(l+\frac12(1+\nu+k\nu+x)\Big)}
{\Gamma\Big(l+\frac12(\nu+k\nu+x)\Big)}\\
&&~~~~~\times
\frac{\Gamma\Big(l+\frac12(3+k\nu-x)\Big)}
{\Gamma\Big(l+2+\frac12(k\nu-x)\Big)}~
\frac{\Gamma\Big(l+\frac12(1+k\nu+x)\Big)}
{\Gamma\Big(l+1+\frac12(k\nu+x)\Big)}\\
&&~~~~~\times
\frac{\Gamma^2\Big(l+\frac12(1+\nu+k\nu)\Big)}
{\Gamma^2\Big(l+1+\frac12(\nu+k\nu)\Big)}~
\frac{\Gamma^2\Big(l+\frac12(3+k\nu)\Big)}
{\Gamma^2\Big(l+1+\frac12k\nu\Big)}
\eea
with asymptotic behavior
for $|\Re \t|\to\infty$, ($|\Im\t-\pi|<\frac\pi2(3+\nu-|1-\nu|)$)
$$
f_{ss}^{min}(i\pi-\t)=c_{ss}\left(e^{\frac{1-\nu}{4\nu}|\t|}+o(1)\right)
$$
with the constant
$$
c_{ss}=\exp\frac12\int_0^\infty\frac{dt}t\left(\frac{\sinh\frac12(1-\nu)t}
{\sinh\frac12\nu t\,\cosh\frac12t\,\sinh t}-\frac{1-\nu}{\nu t}\right)\,.
$$
The corresponding functions $\phi(u)=\Big(F(u)F(i\pi+u)\Big)^{-1}$ and
$\tau(u)=\Big(\phi(u)\phi(-u)\Big)^{-1}$ with $F(i\pi x)=\sin(\frac\pi 2 x)
f_{ss}^{min}(i\pi x)$ are
\bea
\lefteqn{\phi(i\pi x)
=\frac1{F^2\Big(\frac{i\pi}2\Big)}
\frac1{\sin(\pi x)}\exp\int_0^\infty\frac{dt}t
\frac{\sinh\frac12(1-\nu)t\,\Big(\cosh t(\h1-x)-1\Big)}
{\sinh\frac12\,\nu t\sinh t}}\\
&=&\frac1{F^2\Big(\frac{i\pi}2\Big)}\prod_{k=0}^\infty\prod_{l=0}^\infty
\frac{2l+1+\nu+k\nu+x}{2l+k\nu+x}\frac{2l+2+\nu+k\nu-x}{2l+1+k\nu-x}
\left(\frac{2l+k\nu+\frac12}{2l+\nu+k\nu+\frac32}\right)^2\\
&=&\frac1{F^2\Big(\frac{i\pi}2\Big)}\prod_{k=0}^\infty
\frac{\Gamma\Big(\frac12(k\nu+x)\Big)}{\Gamma\Big(\frac12(1+\nu+k\nu+x)\Big)}\,
\frac{\Gamma\Big(\frac12(1+k\nu-x)\Big)}{\Gamma\Big(\frac12(2+\nu+k\nu-x)\Big)}
\frac{\Gamma^2\Big(\frac{\nu+k\nu}2+\frac34\Big)}
{\Gamma^2\Big(\frac{k\nu}2+\frac14\Big)}
\eea
with
$$\phi(i\pi(x+\nu))=\frac{\sin\frac\pi 2x}{\cos\frac\pi 2(x+\nu)}\,
\phi(i\pi x)$$
and
$$
\tau(i\pi x)=\frac{F^2(i\pi/2)F^2(-i\pi/2)}{\sin\frac\pi{2\nu}}~
\sin\pi x\,\sin(\pi x/\nu)\,.
$$


\end{document}